\documentclass[11pt]{amsart}
\usepackage[margin = 1in]{geometry}
\usepackage[many]{tcolorbox}
\usepackage{tgpagella}
\usepackage{amsmath,amssymb,amsthm}
\usepackage{graphicx,epsfig,epstopdf}
\usepackage{url,mathtools}
\usepackage{setspace}
\usepackage{colortbl,framed}
\usepackage{enumitem}
\usepackage{bbm, bm}
\usepackage{footmisc}
\usepackage{float}
\usepackage[style = authoryear, backend = bibtex, natbib = true]{biblatex}
\usepackage[foot]{amsaddr}
\usepackage{pagecolor,lipsum}
\usepackage{wallpaper}
\usepackage{graphicx}
\usepackage{mathrsfs}
\RequirePackage[colorlinks,citecolor=blue,urlcolor=blue]{hyperref}
\RequirePackage{hypernat}

\usepackage{cleveref}
\makeatletter
\renewcommand{\email}[2][]{  \ifx\emails\@empty\relax\else{\g@addto@macro\emails{,\space}}\fi  \@ifnotempty{#1}{\g@addto@macro\emails{\textrm{(#1)}\space}}  \g@addto@macro\emails{#2}}
\makeatother
\newtheorem{innercustomthm}{Condition}
\newenvironment{customthm}[1]
  {\renewcommand\theinnercustomthm{#1}\innercustomthm}
  {\endinnercustomthm}
\newtheorem{innercustomgeneric}{\customgenericname}
\providecommand{\customgenericname}{}

\renewcommand{\qed}{\hfill{\tiny \ensuremath{\blacksquare} }}

\def\sfz{\mathsf{z}}

\newcommand{\indep}{\mathop{\perp\!\!\!\!\perp}}

\renewcommand{\qed}{\hfill {\tiny {\ensuremath{\blacksquare}}}}
\newtheorem{theorem}{Theorem}[section]

\newtheorem{lemma}{Lemma}[section]
\newtheorem{proposition}{Proposition}[section]

\newtheorem{definition}{Definition}[section]
\newtheorem{remark}{Remark}[section]
\newtheorem{condition}{Condition}[section]
\theoremstyle{definition}

\numberwithin{remark}{section}
\newtheorem{example}{Example}
\numberwithin{equation}{section}
\numberwithin{theorem}{section}

\makeatletter
\def\@setthanks{\vspace{-\baselineskip}\def\thanks##1{\@par##1\@addpunct.}\thankses}
\makeatother

\renewcommand{\Pr}{{\mathrm{Pr}}}

\renewcommand{\(}{\left(}
\renewcommand{\)}{\right)}
\renewcommand{\hat}{\widehat}

\renewcommand{\hat}{\widehat}
\renewcommand{\tilde}{\widetilde}
\renewcommand{\leq}{\leqslant}
\renewcommand{\geq}{\geqslant}

\def\argmin{\mathsf{arg \; min}}

\renewcommand{\(}{\left(}
\renewcommand{\)}{\right)}
\renewcommand{\[}{\left[}
\renewcommand{\]}{\right]}

\RequirePackage[OT1]{fontenc}

\def\H{\mathsf{H}}
\def\NH{\mathsf{NH}}

\def\BP{\mathbb{P}}

\def\BL{\mathbb{L}}

\def\E{\mathsf{E}}
\def\P{\mathsf{P}}

\def\IIFF{\mathbb{IF}}
\def\IF{\mathsf{IF}}

\def\log{\mathsf{log}}

\def\var{\mathsf{var}}

\def\cov{\mathsf{cov}}

\def\Bias{\mathsf{Bias}}

\def\se{\mathsf{s.e.}}

\def\CI{\mathsf{CI}}

\def\EB{\mathsf{EB}}
\def\TB{\mathsf{TB}}

\def\tr{\mathsf{tr}}

\def\cf{\mathsf{cf}}

\def\Holder{\text{H\"{o}lder}}

\def\min{\mathsf{min}}
\def\max{\mathsf{max}}

\def\zbar{\bar{\mathsf{z}}}

\def\calR{\mathcal{R}}
\def\calE{\mathcal{E}}

\def\calH{\mathcal{H}}

\def\zkpsi{\tilde{\psi}_{k} (\theta)}

\def\CS{\mathsf{CS}}
\def\CSBias{\mathsf{CSBias}}

\newtheoremstyle{named}{}{}{\itshape}{}{\bfseries}{.}{.5em}{\thmnote{#3's }#1}
\theoremstyle{named}

\newcommand{\myreferences}{Master.bib}
\nocite{*}
\begin{document}
\newrefsection

\title[Assumption-lean higher-order testing]{\normalfont A\lowercase{ssumption-lean Falsification Tests of Rate Double-Robustness of Double-Machine-Learning Estimators}}
\author{$\text{Lin Liu}^{1}$}
\thanks{1. Institute of Natural Sciences, MOE-LSC, School of Mathematical Sciences, CMA-Shanghai, SJTU-Yale Joint Center for Biostatistics and Data Science, Shanghai Jiao Tong University; Shanghai Artificial Intelligence Laboratory \href{linliu@sjtu.edu.cn}{linliu@sjtu.edu.cn}}
\author{$\text{Rajarshi Mukherjee}^{2}$} 
\thanks{2. Department of Biostatistics, Harvard T. H. Chan School of Public Health, \href{ram521@mail.harvard.edu}{ram521@mail.harvard.edu}}
\author{$\text{James M. Robins}^{3}$}
\thanks{3. Department of Epidemiology and Department of Biostatistics, Harvard T. H. Chan School of Public Health, \href{robins@hsph.harvard.edu}{robins@hsph.harvard.edu}}
\thanks{This article is based on a talk given by the last author (JMR) in the ``\href{https://www.nber.org/sites/default/files/2020-08/CEME_Micro_2019.pdf}{2019 Conference for Celebrating Whitney Newey's Contributions to Econometrics}'' held at MIT, and is heavily inspired by works of Whitney Newey. The authors would like to thank three anonymous referees and \href{https://economics.yale.edu/people/xiaohong-chen}{Xiaohong Chen} for constructive comments and \href{https://gaofn.xyz/}{Fengnan Gao} (School of Mathematics and Statistics at the University College Dublin) for discussions that significantly improve the paper. LL was supported by NSFC Grants No.12101397 and 12090024, Shanghai Municipal Science and Technology Major Project No.2021SHZDZX0102, Shanghai Municipal Science and Technology Grants No.21ZR1431000 and 21JC1402900. RM was partially supported by NSF Grant EAGER-1941419. JMR was supported by the U.S. Office of Naval Research grant N000141912446, and National Institutes of Health (NIH) awards R01 AG057869 and R01 AI127271.}
\maketitle

\begin{abstract}
The class of doubly-robust (DR) functionals studied by Rotnitzky et al. (2021) is of central importance in economics and biostatistics. It strictly includes both (i) the class of mean-square continuous functionals that can be written as an expectation of an affine functional of a conditional expectation studied by Chernozhukov et al. (2022b) and (ii) the class of functionals studied by Robins et al. (2008). The present state-of-the-art estimators for DR functionals $\psi$ are double-machine-learning (DML) estimators (Chernozhukov et al., 2018). A DML estimator $\widehat{\psi}_{1}$ of $\psi$ depends on estimates $\widehat{p} (x)$ and $\widehat{b} (x)$ of a pair of nuisance functions $p(x)$ and $b(x)$, and is said to satisfy ``rate double-robustness'' if the Cauchy--Schwarz upper bound of its bias is $o (n^{- 1/2})$. Rate double-robustness implies that the bias is $o (n^{-1/2})$, but the converse is false. Were it achievable, our scientific goal would have been to construct valid, assumption-lean (i.e. no complexity-reducing assumptions on $b$ or $p$) tests of the validity of a nominal $(1 - \alpha)$ Wald confidence interval (CI) centered at $\widehat{\psi}_{1}$. But this would require a test of the bias to be $o (n^{-1/2})$, which can be shown not to exist. We therefore adopt the less ambitious goal of falsifying, when possible, an analyst's justification for her claim that the reported $(1 - \alpha)$ Wald CI is valid. In many instances, an analyst justifies her claim by imposing complexity-reducing assumptions on $b$ and $p$ to ensure ``rate double-robustness''. Here we exhibit valid, assumption-lean tests of $\H_{0}$: ``rate double-robustness holds'', with non-trivial power against certain alternatives. If $\H_{0}$ is rejected, we will have falsified her justification. However, no assumption-lean test of $\H_{0}$, including ours, can be a consistent test. Thus, the failure of our test to reject is not meaningful evidence in favor of $\H_{0}$.
\end{abstract}
{\footnotesize \textbf{Key words:} Econometrics, Causal Inference, Machine Learning, Doubly-Robust Functionals, Higher-Order $U$-Statistics, Higher-Order Influence Functions}

\allowdisplaybreaks

\section{Introduction}
\label{sec:introduction} 

Suppose a data analyst has constructed and published a nominal $1 - \alpha$ large sample Wald confidence interval (CI) for a mean-square continuous linear functional $\psi$ of a conditional expectation $b (x) = \E [Y | X = x]$ with the Wald CI centered by a doubly-robust Double Machine Learning (DML) estimator $\hat{\psi}_{1}$ of $\psi$ \citep{chernozhukov2018double} (see Section \ref{sec:dml} for a formal definition of $\hat{\psi}_{1}$). The estimator $\hat{\psi}_{1}$ will depend on estimates $\hat{b}(x)$ and $\hat{p}(x)$ of two functions of $x$: $b(x)$ itself and the function $p(x)$ occurring in the Riesz representer of the linear functional. Were it achievable, our goal would be to construct an assumption-lean (i.e. essentially assumption-free) empirical test, with non-trivial power against certain alternatives, of the \textit{null hypothesis} that the true asymptotic coverage for $\psi$ of the above nominal $1 - \alpha$ Wald CI is greater than or equal to $1 - \alpha$ under repeated sampling. By definition, assumption-lean tests make no complexity-reducing assumptions (such as smoothness or sparsity) on $b (x)$ or $p (x)$. If such a test rejects (with a very small p-value), we would have falsified (or more precisely, have strong evidence) that the true coverage of the published Wald CI is less than nominal. Henceforth, we will say a large sample Wald CI is valid if and only if the above null hypothesis is true. Unfortunately, following \citet{robins1997toward}, such tests do not exist; for intuition, see Section \ref{sec:prelude}.

We therefore adopt the following less ambitious, but \textit{partially} achievable goal: Our (new) goal is to construct an empirical assumption-lean test that can falsify an analyst's \textit{justification} for the claim that their nominal Wald CI centered at a DML estimator is valid. If falsified, the analyst should then retract any claim of validity. We refer to our goal as \textit{partially} achievable because our test can only falsify certain types of justification. To formally characterize which types, we first must review the properties of DML estimators.

A necessary and (essentially) sufficient condition for the validity of a Wald CI for $\psi$ centered at a DML estimator is that the (asymptotic) bias of the estimator is $o_{p}(n^{-1/2})$. \citet{chernozhukov2018double} showed that a sufficient (but not necessary) condition for validity is that the (weighted) $L_{2}(\mathsf{P})$ rate of convergence $n^{-\kappa_{b}}$ of $\hat{b}$ to $b$ multiplied by the rate of convergence $n^{-\kappa_{p}}$ of $\hat{p}$ to $p$ is $o_{p}(n^{-1/2})$ or, equivalently, $\kappa_{b} + \kappa_{p} > 1 / 2$. Conditions in similar spirit also appeared in earlier works in the econometrics and statistics literature, such as \citet{chen2008semiparametric, chen2015sieve}. This property was termed ``rate double-robustness'' in \citet{smucler2019unifying}.

The main technical result of our paper is that we construct an assumption-lean empirical test, with power against certain alternatives, of the null hypothesis that $\kappa_{b} + \kappa_{p} > 1 / 2$, or in words, ``that rate double-robustness \textit{is} true''. Thus, our test can potentially falsify the justification of any analyst who uses, either explicitly or implicitly, ``rate double-robustness'' to justify the validity of her Wald CI. An example would be an analyst who (i) makes explicit, restrictive assumptions on both the complexities of the functions $b$ and $p$ (e.g. in terms of smoothness or sparsity) and on the algorithms used in their fitting followed by (ii) an appeal to theorems that guarantee $\kappa_{b} + \kappa_{p} > 1 / 2$ under these restrictions. For instance, when $b$ and $p$ are fit by minimizing a (weighted) penalized empirical (squared $L_{2}$-) loss (see \eqref{loss_data}) using deep neural networks \citep{farrell2021deep, chen2020causal, xu2022deepmed}, $L_{2}(\mathsf{P})$-convergence rates satisfying $\kappa_{b} + \kappa_{p} > 1 / 2$ can be proved by assuming $b$ and $p$ live in sufficiently smooth \text{H\"{o}lder}{} spaces \citep{schmidt2020nonparametric}\footnote{If our test rejects the hypothesis that $\kappa_{b} + \kappa_{p} > 1 / 2$, it also rejects the hypothesis that any assumed collection of restrictions on complexity of and fitting algorithms for $b$ and $p$ that imply $\kappa_{b} + \kappa_{p} > 1 / 2$ are all true.}. A second example, especially common in the applied literature, is an analyst who reports a nominal Wald CI centered on a DML estimator without any explicit discussion of its validity, other than citing \citet{chernozhukov2018double}. We also regard such an analyst's implicit justification for her Wald CI as being by appeal to rate double-robustness.

On the other hand, there are ``justifications'' for Wald CI validity that cannot be falsified by our assumption-lean tests rejecting $\kappa_{b} + \kappa_{p} > 1 / 2$ with a very small p-value. Specifically, some recent papers have proposed novel DML estimators that, under very restrictive assumptions on both $b$ and $p$ and on the algorithms used in their estimation, have bias $o_{p} (n^{-1/2})$ and thus can center valid Wald CIs, even though $\kappa_{b} + \kappa_{p} < 1 / 2$ \citep{newey2018cross, kennedy2020towards}. Therefore an analyst who justifies the validity of her Wald CIs by appeal to these restrictive assumptions is not at all surprised to learn that the null hypothesis $\kappa_{b} + \kappa_{p} > 1 / 2$ is false. These novel estimators are discussed both later in the Introduction and in Section \ref{sec:nonstandard}.

In the remainder of the paper, we restrict the functional $\psi$ to the class of Mixed-Bias or Doubly-Robust (DR) functionals \citep{rotnitzky2021characterization}. This class strictly includes both (i) the class of mean-square continuous functionals that can be written as an expectation of an affine functional of a conditional expectation studied by \citet{chernozhukov2022automatic} and (ii) the class of functionals studied by \citet{robins2008higher}\footnote{The class of Mixed-Bias or DR functionals considered in this paper is itself contained in the class of functionals discussed in Section 5 of \citet{chernozhukov2022locally}, that constitutes the most general class of functionals for which there exist first-order DR estimators.}. Many DR functionals are of substantive scientific and economic interest, see the examples after Definition \ref{def:dr}. Our unified treatment of the entire DR functional class requires that we use rather abstract notation. In order to prevent such abstract notation from making the reader miss the forest for the trees, we shall complete the introduction by using a familiar functional to motivate both our goals and our methodology. Furthermore, detailed regularity conditions will be suppressed in the Introduction to facilitate the exposition.

The motivation for our paper is best described by the following inferential quandary faced many (perhaps dozens) of times daily by data analysts employed by large tech companies. The quandary arises when the analyst needs to estimate a population average effect $\E [Y (a = 1)] - \E [Y (a = 0)]$ of a dichotomous treatment $A$ on a response $Y$ from observational data. Here $Y (a)$ is the counterfactual outcome under treatment level $a$. Often, data on a very high-dimensional vector $X$ (with dimension $d$) of pretreatment covariates are available and deemed sufficient for ignorability $Y (a) \indep A | X$ to hold, thereby identifying\footnote{The minus sign plays no essential role and is added for notational convenience that will be made clear in Definition \ref{def:dr}.} $\psi^{a} \equiv -\E [Y (a)]$ as $- \E [b_{a} (X)]$ with $b_{a} (x) = \E [Y | X = x, A = a]$ when positivity holds. \citet{chernozhukov2018double} argued persuasively that in an effort to obtain valid inference (i.e. confidence intervals) for $\psi^{a}$ [and thus for the average treatment effect (ATE) $- \psi^{a = 1} + \psi^{a = 0}$] with very high dimensional $X$, one should use (cross-fitted) DML estimators $\hat{\psi}_{\mathsf{cf}, 1}^{a}$ to center nominal $1 - \alpha$ large sample Wald CI $\hat{\psi}_{\mathsf{cf}, 1}^{a} \pm z_{\alpha / 2} \hat{\mathsf{s.e.}} (\hat{\psi}_{\mathsf{cf}, 1}^{a})$ where $z_{\alpha / 2}$ is the $\alpha / 2$ upper-quantile of a standard normal random variable and $\hat{\mathsf{s.e.}}(\hat{\psi}_{\mathsf{cf}, 1}^{a})$ is the estimator of the standard error $\mathsf{s.e.}(\hat{\psi}_{\mathsf{cf}, 1}^{a})$ of $\hat{\psi}_{\mathsf{cf}, 1}^{a}$ given in Proposition \ref{thm:drml} below. As a result, DML estimators rapidly became the standard in high tech.

For DR functionals, DML estimators combine the benefits of cross-fitting (cf), double robustness, and machine learning of nuisance parameters \citep{chernozhukov2018double, chernozhukov2022locally}. Henceforth, for notational convenience, we remove the $a$ index by restricting to the case $a = 1$ so, for example, $\psi^{a=1}$ becomes $\psi$ and $b_{a}$ becomes $b$. Then, to compute a DML estimator, the data is randomly divided into two (or more) samples -- the estimation sample of size $n$ and the training or nuisance/training sample of size $n_{\mathsf{tr}} = N - n$ with $1 - c > n / N > c$ for some $c \in (0, 1)$. To simplify the exposition we take $c = 0.5$. Estimators $\hat{b} (\cdot)$ and $\hat{p} (\cdot)$ of $b (\cdot)$ and inverse propensity score $p (\cdot) \coloneqq 1 / \mathsf{E}[A|X = \cdot]$, where $p^{-1}$ is assumed to be strictly bounded between $(0, 1)$, are computed from the training sample data using modern black-box highly nonlinear machine learning algorithms, often deep neural networks. In semiparametric statistics literature, $b$ and $p$ are referred to as nuisance parameters/functions. Let $\mathsf{P}_{n}$ denote a sample average over the estimation sample. Then the doubly-robust one step estimator $\hat{\psi}_{1} = \hat{\psi} + \mathsf{P}_{n} [\hat{\mathsf{IF}}_{1, \psi}] = \mathsf{P}_{n} [- \hat{b} (X) - A \hat{p} (X) (Y - \hat{b} (X))]$ is constructed by adding to an initial plug-in estimator $\hat{\psi} = \mathsf{P}_{n} [-\hat{b} (X)]$ the sample average of an estimate of the first order influence function $\mathsf{IF}_{1, \psi} = - b (X) - A p (X) (Y - b (X)) - \psi$ of $\psi$\footnote{In the econometrics literature, $\hat{\psi}$ is referred to as a first stage estimator, $\hat{\psi}_{1}$ as the second stage estimator, and $\mathsf{P}_{n} [- A \hat{p} (X) (Y - \hat{b}(X))]$ is the debiasing term that makes $\hat{\psi}_{1}$ a doubly-robust estimator satisfying Neyman orthogonality \citep{chernozhukov2018double, chernozhukov2022locally}.}. The cross-fit DML estimator $\hat{\psi}_{\mathsf{cf}, 1}$ is the arithmetic average of $\hat{\psi}_{1}$ and $\bar{\hat{\psi}}_{1}$, where $\bar{\hat{\psi}}_{1}$ is computed like $\hat{\psi}_{1}$ but with the roles of training and estimation samples switched. We define a standard DML estimator to be a DML estimator where $\hat{b}$ and $\hat{p}$ are separately estimated, each using data from the entire training sample; see Section \ref{sec:dml}. This allows us to distinguish standard DML estimators from, for instance, ``DCDR estimators'' of \citet{newey2018cross} that estimate $b$ and $p$ from separate non-overlapping subsamples of the training sample.

The analyst's inferential quandary is how to justify the claim that $\hat{\psi}_{\mathsf{cf}, 1}\pm z_{\alpha /2}\hat{\mathsf{s.e.}}(\hat{\psi}_{\mathsf{cf}, 1})$ is a valid large sample $1-\alpha $ Wald CI. For it to be valid the following are necessary: (i) $\hat{\psi}_{\mathsf{cf}, 1}$ is asymptotically normal, (ii) $\hat{\mathsf{s.e.}} (\hat{\psi}_{\mathsf{cf}, 1}) / \mathsf{s.e.}(\hat{\psi}_{\mathsf{cf}, 1})$ converges to $1$ in probability and (iii) the (asymptotic) bias of $\hat{\psi}_{\mathsf{cf}, 1}$ is of smaller order than $\mathsf{s.e.} (\hat{\psi}_{\mathsf{cf}, 1})$. Since $\hat{\psi}_{\mathsf{cf}, 1}$ is the average of $\hat{\psi}_{1}$ and its ``twin'' $\bar{\hat{\psi}}_{1}$, it must be the case that $\hat{\psi}_{1} \pm z_{\alpha / 2} \hat{\mathsf{s.e.}} (\hat{\psi}_{1})$ is also a valid large sample $1 - \alpha$ Wald CI for $\psi$, therefore satisfying (i) - (iii) with $\hat{\psi}_{1}$ substituted for $\hat{\psi}_{\mathsf{cf}, 1}$. Since $\mathsf{s.e.} (\hat{\psi}_{1})$ is order $n^{-1/2}$, it is necessary that its bias is $o_{p} (n^{-1/2})$ to satisfy (iii) above. As discussed in the literature \citep{newey2018cross}, by far the most difficult of the three assumptions to satisfy and thus to justify is (iii). As mentioned above, for a standard DML estimator, if $\kappa_{b} + \kappa_{p} > 1 / 2$ holds, then (iii) holds. Specifically, that $\kappa_{b} + \kappa_{p} > 1 / 2$ implies the bias of $\hat{\psi}_{1}$ is $o_{p} (n^{-1/2})$ is a consequence of applying Cauchy-Schwarz (CS) inequality to upper bound the bias, as we show next. The exact conditional bias of $\hat{\psi}_{1}$ given the training sample, denoted as $\mathsf{Bias} (\hat{\psi}_{1})$, is $\mathsf{E} [A (\hat{b} (X) - b (X)) (\hat{p} (X) - p (X))] = \int p^{-1}(x) (\hat{b} (x) - b (x)) (\hat{p} (x) - p (x)) \mathrm{d} F (x)$, where $F$ denotes the distribution of $X$. Here and henceforth, unless stated otherwise, all expectations will be understood to be conditional on the training sample,  although that fact is suppressed in the notation for brevity. It then follows from the CS inequality that 
\begin{equation}
|\mathsf{Bias} (\hat{\psi}_{1})| \leq \{\mathsf{E} [p^{-1} (X) (\hat{b} (X) - b (X))^{2}]\}^{1 / 2} \{\mathsf{E} [p^{-1} (X) (\hat{p} (X) - p (X))^{2}]\}^{1 / 2}.
\label{cs-heuristic}
\end{equation}
We refer to the RHS of the above display as the (conditional on the training sample) Cauchy-Schwarz (CS) bias, denoted as $\mathsf{CSBias}(\hat{\psi}_{1})$, of $\hat{\psi}_{1}$.  
More generally, for any pair of positive functions $w = (w_{b}, w_{p})$ of $x$ strictly bounded from above and below, define  
\begin{equation*}
\mathsf{CSBias}^{w} (\hat{\psi}_{1})\equiv \{\mathsf{E} [w_{b} (X) (\hat{b} (X) - b(X))^{2}]\}^{1/2} \{\mathsf{E} [w_{p} (X) (\hat{p} (X) - p (X))^{2}]\}^{1/2}.
\end{equation*}
Then applying \text{H\"{o}lder}{} inequality, as in \eqref{holder} later in our paper, we have that $\mathsf{CSBias}^{w}(\hat{\psi}_{1})$ and $\mathsf{CSBias} (\hat{\psi}_{1})$ are equal up to a multiplicative positive constant, except that when $w_{b} = w_{p} = p^{-1}$, the equality is exact. Hence $\mathsf{CSBias}^{w} (\hat{\psi}_{1}) = o_{p} (n^{-1/2})$, if and only if $\mathsf{CSBias}(\hat{\psi}_{1})=o_{p}(n^{-1/2})$, if and only if rate double-robustness holds ($\kappa_{b} + \kappa_{p} > 1 / 2$, where we refer to $n^{-\kappa_{b}}$ and $n^{-\kappa_{p}}$ as the $p^{-1}$-weighted $L_{2} (\P)$ convergence rates of $\hat{b}$ and $\hat{p}$ to $b$ and $p$). Thus $\mathsf{CSBias}^{w}(\hat{\psi}_{1})=o_{p}(n^{-1/2})$ also implies $\mathsf{Bias}(\hat{\psi}_{1})=o_{p}(n^{-1/2})$.

\subsection*{Main technical contributions}

If we can empirically reject the null hypothesis $\NH_{0, \CS}: \mathsf{CSBias} (\hat{\psi}_{1}) = o_{p}(n^{-1/2})$ encoding rate double-robustness, then $\kappa_{b} + \kappa_{p} \leq 1 / 2$, and we falsify the justification of the validity of a Wald CI centered at the standard DML estimator $\hat{\psi}_{1}$. The main technical contribution of this paper is to construct a test of $\NH_{0, \CS}$.



The reader might rightfully complain at this point that finite sample tests of asymptotic hypotheses such as $\NH_{0, \CS}$ that concern rates of convergence cannot be constructed. To overcome this problem, we pair each asymptotic null hypothesis of interest with a natural non-asymptotic null hypothesis and then, by convention, formally declare the asymptotic hypothesis (not) rejected if the paired non-asymptotic null hypothesis is (not) rejected. Then we say that the asymptotic null hypothesis has been \textit{operationalized} by its paired non-asymptotic null hypothesis. In particular, we pair the asymptotic null hypothesis $\NH_{0, \CS}$ with the following non-asymptotic null hypothesis
\begin{equation}
\label{h0-cs}
\H_{0, \CS} (\delta): \CSBias (\hat{\psi}_{1}) \leq \delta \se (\hat{\psi}_{1})
\end{equation}
and construct an $\alpha^{\dag}$-level falsification test of $\H_{0, \CS} (\delta)$, where $\delta > 0$ is chosen by the analyst. With such an operationalized pairing
\begin{equation}
\label{pairing}
\begin{split}
\mathsf{NH}_{0, \CS}: \CSBias (\hat{\psi}_{1}) = o_{p} (n^{- 1 / 2}) \sim \H_{0, \CS} (\delta): \CSBias (\hat{\psi}_{1}) < \se (\hat{\psi}_{1}) \delta,
\end{split}
\end{equation}
we will, by convention, declare $\mathsf{NH}_{0, \CS}$ (not) rejected if $\H_{0, \CS} (\delta)$ is (not) rejected. The larger $\delta$ that one chooses, the stronger evidence that rejection of $\H_{0, \CS} (\delta)$ has against the asymptotic null hypothesis $\NH_{0, \CS}$, but the less power one has to reject $\H_{0, \CS} (\delta)$. Suppose the analyst insists that her justification is $\CSBias^{w} (\hat{\psi}_{1}) = o (n^{- 1 / 2})$ as defined in \eqref{equiv}, then, by the aforementioned equivalence between $\CSBias (\hat{\psi}_{1})$ and $\CSBias^{w} (\hat{\psi}_{1})$, we can use the same operationalized pairing \eqref{pairing} to empirically falsify $\CSBias_{\theta}^{w} (\hat{\psi}_{1}) = o (n^{- 1 / 2})$. We remark that one could, in principle, choose $\delta$ to be a diminishing sequence as a function of the sample size $n$.

Since we want our test to be assumption-lean, we make essentially no assumptions on the nuisance functions $b$ or $p$, their estimates $\hat{b}$ or $\hat{p}$, or the algorithms used to construct $\hat{b}$ and $\hat{p}$ from the training sample. To avoid relying on assumptions on $\hat{b}$ and $\hat{p}$, the falsification test and its properties are established by conditioning on the training sample, so the training sample is treated as fixed and statements such as $\CSBias (\hat{\psi}_{1}) = o_{p} (n^{-1/2})$ become $\CSBias (\hat{\psi}_{1}) = o (n^{-1/2})$. By arguing as in \citet{robins1997toward} or \citet{ritov2014bayesian}, in absence of complexity-reducing assumptions on $b$ or $p$, there is no uniformly consistent estimator of either $\Bias (\hat{\psi}_{1})$ or $\CSBias (\hat{\psi}_{1})$. However, we will exhibit a functional, denoted as $\Bias_{k} (\hat{\psi}_{1})$, that is uniformly consistently estimable using a third-order $U$-statistic [derived using the theory of Higher-Order Influence Functions (HOIFs) \citep{robins2008higher, liu2017semiparametric}] such that one can reject the ``$k$-projected'' null hypothesis
\begin{equation}
\label{h0k}
\H_{0, k} (\delta): |\Bias_{k} (\hat{\psi}_{1})| \leq \delta \se (\hat{\psi}_{1})
\end{equation}
with nontrivial power. Here $\Bias_{k} (\hat{\psi}_{1})$ is defined as
\begin{align*}
\Bias_{k} (\hat{\psi}_{1}) \coloneqq \E [\Pi [p^{-1 / 2} (b - \hat{b}) | p^{-1 / 2} \zbar_{k}] (X) \Pi [p^{-1 / 2} (p - \hat{p}) | p^{-1 / 2} \zbar_{k}] (X)],
\end{align*}
where for any $h \in L_{2} (\P_{F})$, $\Pi [h | p^{-1 / 2} \zbar_{k}]$ denotes the population projection of $h$ onto the linear span of a (user-selected) $k$-dimensional dictionary (or basis functions) $p^{-1 / 2} \zbar_{k} = p^{-1 / 2} (\sfz_{1}, \cdots, \sfz_{k})^{\top}$. We now show that $|\Bias_{k} (\hat{\psi}_{1})|$ is a lower bound for $\CSBias (\hat{\psi}_{1})$ as follows:
\begin{equation}
\label{csbias}
\begin{split}
& \vert \Bias_{k} (\hat{\psi}_{1}) \vert \leq \underbrace{\{\E [\Pi [p^{-1 / 2} (b - \hat{b}) | p^{-1 / 2} \zbar_{k}] (X)^{2}]\}^{1 / 2} \{\E [\Pi [p^{-1 / 2} (p - \hat{p}) | p^{-1 / 2} \zbar_{k}] (X)^{2}]\}^{1 / 2}}_{\eqqcolon \CSBias_{k} (\hat{\psi}_{1})} \leq \CSBias (\hat{\psi}_{1}),
\end{split}
\end{equation}
where the first inequality again follows from CS inequality and the second inequality follows from the fact that projection contracts norms\footnote{The reason why we do not focus on a different $k$-projected null hypothesis
\begin{equation}
\label{h0k-cs}
\H_{0, k, \CS} (\delta): \CSBias_{k} (\hat{\psi}_{1}) \leq \delta \se (\hat{\psi}_{1}).
\end{equation}
will be explained in Remark \ref{rem:temper} of Section \ref{sec:main}. In principle, though, one can also consider testing $\H_{0, k, \CS} (\delta)$.}\label{ft:temper}.
Hence if $\H_{0, k} (\delta)$ is false, then $\H_{0, \CS} (\delta)$ is false and if $\H_{0, \CS} (\delta)$ is true, then $\H_{0, k} (\delta)$ is true; but the converses of the above two clauses do not necessarily hold. In fact, no test, ours included, can be a consistent test of $\H_{0, \CS} (\delta)$ (that is, no test can have power against all alternatives to $\H_{0, \CS} (\delta)$) unless one makes further possibly incorrect complexity-reducing assumptions on the nuisance functions of $b$ and $p$ and their estimates $\hat{b}$ and $\hat{p}$. This again follows from the argument in \citet{robins1997toward} or \citet{ritov2014bayesian}. But we also provide an intuitive explanation in Section \ref{sec:prelude}.

To further illustrate our approach, we can rewrite $\Bias_{k} (\hat{\psi}_{1})$ as follows (see Appendix \ref{app:oracle}):
\begin{align*}
\Bias_{k} (\hat{\psi}_{1}) = \E [A (Y - \hat{b} (X)) \zbar_{k} (X)^{\top}] \Sigma_{k}^{-1} \E [\zbar_{k} (X) (A \hat{p} (X) - 1)]
\end{align*}
where $\Sigma_{k} \equiv \E [A \zbar_{k} (X) \zbar_{k} (X)^{\top}]$. When $\Sigma_{k}$ is known, $\Bias_{k} (\hat{\psi}_{1})$ can be unbiasedly estimated by the following second-order $U$-statistic:
\begin{align*}
\hat{\IIFF}_{22, k} (\Sigma_{k}^{-1}) \equiv \frac{1}{n (n - 1)} \sum_{1 \leq i_{1} \neq i_{2} \leq n} A_{i_{1}} (Y_{i_{1}} - \hat{b} (X_{i_{1}})) \zbar_{k} (X_{i_{1}})^{\top} \Sigma_{k}^{-1} \zbar_{k} (X_{i_{2}}) (A_{i_{2}} \hat{p} (X_{i_{2}}) - 1).
\end{align*}
$\hat{\IIFF}_{22, k} (\Sigma_{k}^{-1})$ is the second-order influence function of the functional $\Bias_{k} (\hat{\psi}_{1})$; see Appendix \ref{app:hoif} for a more precise definition. Since $\Sigma_{k}$ is generally unknown, we replace $\Sigma_{k}$ by an estimate $\Sigma_{k}$ from the training sample. However, $\hat{\IIFF}_{22, k} (\hat{\Sigma}_{k}^{-1})$ is no longer unbiased for $\Bias_{k} (\hat{\psi}_{1})$. As a consequence, to protect the level of our test, we replace $\hat{\IIFF}_{22, k} (\hat{\Sigma}_{k}^{-1})$ by a third-order $U$-statistic to correct for the additional bias; see Section \ref{sec:main} for more details.

A preliminary version of the above idea has appeared in \citet{liu2020nearly}, but the results therein are mainly for (1) the case when the distribution of the potentially high-dimensional covariates $X$ is known, or the so-called {\it semisupervised} setting, and (2) one special case of DR functionals, the expected conditional covariance. In this article, we advance the literature in the following regards.
\begin{itemize}
\item First, we extend the results in \citet{liu2020nearly} to the more realistic case where $\Sigma_{k}$ is unknown. In particular, we construct a test of the null hypothesis of rate double-robustness based on a third-order $U$-statistic (which is the estimated third-order influence function of $\Bias_{k} (\hat{\psi}_{1})$). 
\item Second, the results in this paper require much weaker assumptions\footnote{Note that both the assumptions in this paper and in \citet{liu2020nearly} are easily checkable by the analysts; see Conditions \ref{cond:w} and \ref{cond:sw}.} on the dictionary $\zbar_{k}$ than those in \citet{liu2020nearly}, allowing the proposed methodology to be used more broadly in substantive studies. 
\item Third, we extend the results of \citet{liu2020nearly} for expected conditional covariance to the entire DR functional class, which requires that we derive the HOIFs of DR functionals. This could be of independent interest, considering the role of HOIFs in constructing rate-optimal estimators for smooth/differentiable functionals or their generalizations \citep{tchetgen2008minimax, kennedy2022minimax}. In Appendix \ref{app:prox_hoif}, we construct the HOIFs for the average treatment effect in the presence of unmeasured confounding (or under endogeneity) in the setting of proximal causal inference \citep{miao2018identifying, miao2018confounding}. This part is built on \citet{cui2023semiparametric}'s derivation of the first-order influence function under the assumptions underlying proximal causal inference\footnote{Based on personal communication, \href{https://statistics.wharton.upenn.edu/profile/ett/}{Eric Tchetgen Tchetgen}, \href{https://sites.google.com/view/yifancui}{Yifan Cui}, and colleagues have also independently derived HOIFs of this parameter under the proximal causal learning framework in an unpublished manuscript.}. A more thorough discussion is deferred to Section \ref{sec:conclusion} and Appendix \ref{app:prox_hoif}.
\end{itemize}


\subsection*{Literature overview}
To the best of our knowledge, the assumption-lean falsification test as constructed in this paper is new in the literature (except for its precursor \citet{liu2020nearly}), and has different purposes from specification tests \citep{newey1985maximum} in the econometrics literature. Thus we first mention a subset of the fast-growing literature in statistics and econometrics on estimating and drawing statistical inference for (certain members of) DR functionals using standard DML estimators; due to space limitation, see \citet{farrell2015robust, chernozhukov2018double, smucler2019unifying, bradic2019sparsity, chernozhukov2022locally} and references therein. 

Some recent works also consider further refinement of standard DML estimators \citep{newey2018cross, kline2020leave, bradic2019minimax, mcgrath2022undersmoothing, kennedy2020towards}. The main distinction of these nonstandard DML estimators from the standard ones is to estimate $b$ and $p$ from separate subsamples of the training sample. Earlier in the introduction, we used the word ``novel'' rather than ``nonstandard'' in describing these DML estimators. Under very restrictive, specific complexity-reducing assumptions on $b$ and $p$ and on the algorithms used in their estimation, the estimators may have bias $o (n^{- 1 / 2})$ yet rate double-robustness fails to hold \citep{newey1990semiparametric, newey1994large}. In the absence of such restrictive, specific complexity-reducing assumptions and fitting algorithms, it is unclear if these nonstandard DML estimator still outperform standard one either in theory or in practice. In fact, a lower bound established recently in \citet{balakrishnan2023fundamental} shows that standard DML estimators are minimax optimal under an assumption-lean model that imposes no complexity-reducing assumptions on the nuisance functions $b$ or $p$. As a result, this article mainly focuses on the standard DML estimators but these nonstandard ones will also be considered briefly in Section \ref{sec:nonstandard}.

Last but not least, we remark that our work is also closely related to the literature on $\sqrt{n}$-consistent estimation and inference for low-dimensional parameters (implicitly) defined via (conditional) moment restrictions involving nonparametric nuisance functions; e.g., see \citet{ai2003efficient, ai2007estimation, ai2012semiparametric, chen2015sieve}, to name a few. Such parameters encompass the DR functionals studied in this paper, and can be applied to endogeneity settings \citep{ai2003efficient, angrist1996identification, tchetgen2020introduction}. Extending our framework (specifically the theory of higher-order influence functions) to such more complicated parameters is still an open problem. 

\subsection*{Organization of the paper}
The remainder of the paper is arranged as follows. In Section \ref{sec:review}, we describe the mathematical setup formally and review the definition and properties of DR functionals recently characterized in \citet{rotnitzky2021characterization}. In Section \ref{sec:dr}, we review the statistical properties of their standard DML estimators, based upon which we motivate and formally define our approach. 

The main result of this paper, Section \ref{sec:main}, is to construct a valid $\alpha^{\dag}$-level falsification test of $\H_{0, \CS} (\delta)$ based on a third-order $U$-statistic, denoted as $\hat{\IIFF}_{22 \rightarrow 33, k} (\hat{\Sigma}_{k}^{-1})$ (see \eqref{if2233}), which is the estimated third-order influence function of $\Bias_{k} (\hat{\psi}_{1})$ but with $\Sigma_{k}$ replaced by $\hat{\Sigma}_{k}$, following the notation used in \citet{robins2008higher}; also see Appendix \ref{app:hoif} for derivations.

In Section \ref{sec:nonstandard}, we study if the proposed falsification test could be also meaningful for nonstandard DML estimators, whose bias could be $o (n^{-1/2})$ even when $\kappa_{b} + \kappa_{p} \leq 1 / 2$ (rate double-robustness is violated). We first argue that the $k$-projected null hypothesis $\H_{0, k} (\delta)$ is a natural hypothesis to falsify. Then we construct a valid $\alpha^{\dag}$-level test of $\H_{0, k} (\delta)$ by modifying the falsification test in Section \ref{sec:main} using higher order $U$-statistics. Since higher order $U$-statistics are computational costly, we also propose an early-stopping strategy that takes the analyst's computational budget into account. In Section \ref{sec:simulation} we present results of simulation studies to evaluate the finite sample performance of our methods. Section \ref{sec:conclusion} concludes with a discussion of some open problems. Many of the technical details are deferred to the Appendix.

\section{Formal setup and an illustration of our approach}
\label{sec:review}
The formal setup is as follows. We observe $N$ i.i.d. copies of the data vector $O = (W, X)$ drawn from some unknown probability distribution $\P_{\theta}$ belonging to a locally nonparametric model 
\begin{equation*}
\mathcal{M} = \left\{ \P_{\theta}; \theta = (b, p, \theta \setminus \{b, p\}) \in \Theta = \mathcal{B} \times \mathcal{P} \times \Theta \setminus \{\mathcal{B}, \mathcal{P}\} \right\}
\end{equation*}
parameterized by the parameter $\theta = (b, p, \theta \setminus \{b, p\})$ with $b, p, \theta \setminus \{b, p\}$ variation independent. To stress
the dependence on $\theta$, from here on, we will attach $\theta$ to many symbols that have appeared, such as $\psi (\theta)$, $\E_{\theta}$, $\Bias_{\theta} (\hat{\psi}_{1})$, $\CSBias_{\theta} (\hat{\psi}_{1})$, $\Bias_{\theta, k} (\hat{\psi}_{1})$, etc. $X$ is a $d$-dimensional random vector with compact support (with distribution function $F$) whose density $f$ is bounded away from $0$ and $\infty$ on its support, and $d$ is allowed to increase with $N$. Here the maps $b: x \mapsto b(x) \in \mathcal{B}$ and $p: x \mapsto p(x) \in \mathcal{P}$ have range bounded and contained in $\mathbb{R}$. $\mathcal{M}$ is locally nonparametric in the sense that the tangent space for the model at each $\theta \in \Theta$ is equal to $L_{2} (\P_{\theta})$, e.g. when $b, p$ belong to \Holder{} balls with certain smoothness. 

To avoid extraneous technical issues, we assume that observed data $O$ is bounded with probability 1 (see Remark \ref{rem:w} for further discussion). We consider functionals (i.e. parameters) $\psi: \theta \mapsto \psi (\theta)$ that possess a (first order) influence function\footnote{The term ``influence function'' when used without further qualification is to be understood to be the first order influence function.} \citep{ichimura2022influence} $\IF_{1, \psi} (\theta) = \mathsf{if}_{1, \psi} (O; \theta)$ (and thus a positive and finite semiparametric variance bound \citep{newey1990semiparametric}) and are contained in the mixed bias or doubly-robust class of functionals of \citet{rotnitzky2021characterization} defined as follows. Under the locally nonparametric models defined above, the influence function $\IF_{1, \psi} (\theta)$ of $\psi (\theta)$ with respect to the tangent space $L_{2} (\P_{\theta})$ is unique.

\begin{definition}[Definition of a mixed bias or doubly-robust functional (DR functional) (Definition 1 of \citet{rotnitzky2021characterization})]
\label{def:dr} 
$\psi (\theta)$ is a doubly-robust functional if, for each $\theta \in \Theta$ there exists $b: x \mapsto b (x) \in \mathcal{B}$ and $p: x \mapsto p (x) \in \mathcal{P}$ such that (i) $\theta = (b, p, \theta \setminus \{b, p\})$ and $\Theta = \mathcal{B} \times \mathcal{P} \times \Theta \setminus \{\mathcal{B}, \mathcal{P}\}$ and (ii) for any $\theta, \theta'$ 
\begin{equation}
\psi (\theta) - \psi (\theta^{\prime}) + \E_{\theta} \left[ \IF_{1, \psi} (\theta^{\prime}) \right] = \E_{\theta} \left[ S_{bp} (b (X) - b^{\prime} (X)) (p (X) - p^{\prime} (X)) \right] \label{eq:drbias}
\end{equation}
where $S_{bp} \equiv s_{bp} (O)$ with $o \mapsto s_{bp} (o)$ a known function that does not depend on $\theta$ or $\theta'$ satisfying either $\P_{\theta} (S_{bp} \geq 0) = 1$ or $\P_{\theta} (S_{bp} \leq 0) = 1$. $b$ and $p$ are called nuisance functions in the semiparametric statistics literature, a terminology we also adopt in this paper. Finally, we let $$\lambda (x) \coloneqq \E_{\theta} [S_{bp} |X = x]$$ and assume $\lambda$ to be strictly bounded from above and below.
\end{definition}

To understand the implication of DR functionals, let $\P_{n}$ be the empirical mean operator over the estimation sample and suppose $\theta'$ were an estimate of $\theta$ from the training sample (regarded as fixed, i.e. non-random). It then follows that the one step estimator $\psi (\theta') + \P_{n} \left[ \IF_{1, \psi} (\theta') \right]$ is doubly-robust \citep{scharfstein1999rejoinder, robins2001comments, bang2005doubly}. That is, by \eqref{eq:drbias}, it is unbiased for $\psi (\theta)$ under $\P_{\theta}$ if either $b = b'$ or $p = p'$. Because of this fact, we will use the term DR functional in this paper, as is done in much of the current literature, instead of the ``mixed bias'' terminology employed in \citet{rotnitzky2021characterization}. 
To ease notation, we will restrict consideration to DR functionals for which $\P_{\theta} (S_{bp} \geq 0) = 1$. For a DR functional $\psi^{\dag} (\theta)$ of substantive interest for which $\P_{\theta} (S_{bp} \leq 0) = 1$, we will instead analyze $\psi (\theta) = - \psi^{\dag} (\theta)$.

Let $W = (Y, A)$. Below are some examples of DR functionals that are of substantive interest in economics and statistics.

\begin{enumerate}
\item The counterfactual mean of $Y$ when $\{0, 1\}$-valued $A$ is set to $1$ (under strong ignorability), $\E_{\theta} [Y (a = 1)]$ is identified from the distribution of $W = (A, A Y, X)$ by the DR functional $\psi^{\dag} (\theta) = \E_{\theta} [b (X)]$ with $b (x) = \E_{\theta} [Y | X = x, A = 1]$, $p (x) = 1 / \E_{\theta} [A | X = x]$, $S_{bp} = - A$, and $\lambda (x) = - p (x)^{-1}$. Here $p(x)$ is the inverse of the propensity score $\pi (x) = \E_{\theta} [A | X = x]$. Because $S_{bp} = - A$, we instead analyze the DR functional 
\begin{equation*}
\psi (\theta) = - \psi^{\dag} (\theta) = - \E_{\theta} [b (X)] = - \E_{\theta} [Y (a = 1)]
\end{equation*}
for which $S_{bp} = A$. We will use $\psi (\theta) = - \E_{\theta} [b (X)] = - \E_{\theta} [Y (a = 1)]$ as a running example below\footnote{\citet{chernozhukov2022automatic} used an alternative decomposition under which they included our $A, X$ in their $X$. See Example 1 of \citet{rotnitzky2021characterization}.}. Average treatment effect, one of the most intensively studied causal parameters, is thus a difference of two DR functionals.


\item The expected conditional covariance 
\begin{equation*}
\psi (\theta) = \E_{\theta} [(Y - b (X)) (A - p (X))]
\end{equation*}
with $b (x) = \E_{\theta} [Y | X = x], p (x) = \E_{\theta} [A | X = x]$ is a DR functional (equivalently, a MB functional) with $S_{bp} = \lambda (X) \equiv 1$. When $Y$ and $b (X)$ are replaced by $A$ and $p (X)$, $\psi (\theta)$ reduces to the expected conditional variance $\phi (\theta) = \E_{\theta} [(A - p (X))^{2}]$. The expected conditional covariance has been popularized recently by \citet{shah2020hardness} for the problem of conditional independence testing, under a different name ``generalized covariance measure''. The expected conditional covariance also relates to the regression coefficient $\tau$ in the following semiparametric regression: $Y = \tau A + b (X) + \varepsilon$ where $\varepsilon$ is mean zero conditional on $A$ and $X$ \citep{li2011higher}.
\end{enumerate}

To save space, we refer interested readers to \citet{rotnitzky2021characterization} for further examples, including Average Treatment Effect on the Treated (ATT) (Example 8 therein), ATE under sensitivity analysis models (Example 3 therein) and etc.

Before proceeding further, we collect some frequently used notation. We use $\E_{\theta} [\cdot], \var_{\theta} [\cdot]$ and etc. to denote the expectation, variance, and etc. with respect to $\P_{\theta}$. The data is randomly divided into an estimation sample and a training (equivalently, nuisance) sample of size $n = N / 2$. To avoid notational clutter, all expectations, variances, and probabilities are conditional on the training sample unless otherwise stated. For a (random) vector $V$, $\Vert V \Vert_{\theta} \equiv \E_{\theta} [V^{\otimes 2}]^{1/2} = \E_{\theta} [V^{\top} V]^{1/2}$ denotes its $L_{2} (\P_{\theta})$ norm conditioning on the training sample, $\Vert V \Vert \equiv (V^{\otimes 2})^{1/2} = (V^{\top} V)^{1/2}$ its $\ell_{2}$ norm and $\Vert V \Vert_{\infty}$ its $L_{\infty} (\P_{\theta})$ norm. For any matrix $M$, $\Vert M \Vert$ will be reserved for its operator norm. Given an integer $k$, and a random vector $\zbar_{k} = \zbar_{k} (X)$, $\Pi_{\theta} [\cdot | \zbar_{k}]$ denotes the population linear projection operator onto the linear space spanned by $\zbar_{k}$ conditioning on the training sample, and $\Pi_{\theta}^{\perp} [\cdot | \zbar_{k}] = \left( \mathsf{I} - \Pi_{\theta} \right) \left[ \cdot | \zbar_{k} \right]$ is the projection onto the ortho-complement of $\zbar_{k}$ in the Hilbert space $L_{2} (\P_{F})$ where $\P_{F}$ denotes the marginal law of $X$ under $\theta$. That is, for a random variable $V$, 
\begin{equation}
\Pi _{\theta} [V | \zbar_{k}] (x) = \zbar_{k}^{\top} (x) \{\E_{\theta} [\zbar_{k} (X) \zbar_{k} (X)^{\top}]\}^{-1} \E_{\theta} [\zbar_{k} (X) V], \Pi _{\theta}^{\perp} [V | \zbar_{k}] (x) = V - \Pi_{\theta} [V | \zbar_{k}] (x). \label{eq:projection}
\end{equation}

The following common asymptotic notations are used throughout the paper: $x \lesssim y$ (equivalently $x = O (y)$ or $y = \Omega (x)$) denotes that there exists some constant $C > 0$ such that $x \leq C y$, $x \asymp y$ (equivalently $x = \Theta (y)$) means there exist some constants $c_{1} > c_{2} > 0$ such that $c_{2} |y| \leq |x| \leq c_{1} |y|$. $x = o (y)$ or $y = \omega (x)$ or $y \gg x$ or $x \ll y$ is equivalent to $\lim_{x, y \rightarrow \infty} \frac{x}{y} = 0$. For a random variable $X_{n}$ with law $\P$ possibly depending on the sample size $n$, $X_{n} = O_{\P} (a_{n})$ denotes that $X_{n} / a_{n}$ is bounded in $\P$-probability, and $X_{n} = o_{\P} (a_{n})$ means that $\lim_{n \rightarrow \infty} \P (|X_{n} / a_{n} | \geq \epsilon) = 0$ for every positive $\epsilon$.

\subsection{Influence functions of DR functionals}\leavevmode
\label{sec:dr}

In this section, we generalize our discussion in the Introduction on $\psi (\theta) = - \E_{\theta} [Y (1)]$ under ignorability to the DR functionals and provide a more detailed exposition. We first review the influence functions of DR functionals (see Definition \ref{def:dr}) established in \citet{rotnitzky2021characterization}. Although the definition of DR functional is quite abstract, \citet{rotnitzky2021characterization} derived the (nonparametric) influence functions of DR functionals by directly leveraging the form of the product bias appeared in Definition \ref{def:dr}. We summarize their results below for the sake of completeness.

\begin{proposition}[A summary of Theorems 1 and 2 of \citet{rotnitzky2021characterization}]
\label{thm:if1} 
Suppose (1) $\psi (\theta)$ for $\theta = \left( b, p, \theta \setminus \{b, p\} \right) \in \Theta \equiv \mathcal{B} \times \mathcal{P} \times \Theta \setminus \{\mathcal{B}, \mathcal{P}\}$ is a DR functional (equivalently MB functional) according to Definition \ref{def:dr}; (2) the regularity Condition \ref{cond:r} in Appendix \ref{app:regular} (Condition 1 of \citet{rotnitzky2021characterization}) holds; (3) $b (X)$, $p (X)$, $\lambda (X) b (X)$ and $\lambda (X) p (X)$ belong to $L_{2} (\P_{F})$ for all $\theta \in \Theta$.

Then there exists a statistic $S_{0}$ and linear maps $h \mapsto m_{1} (O, h)$ for $h \in \mathcal{B}$ and $h \mapsto m_{2} (O, h)$ for $h \in \mathcal{P}$ independent of $\theta$ satisfying the following:

\begin{itemize}
\item The influence function of $\psi (\theta)$ is given by:
\begin{equation}  \label{eq:if1}
\IF_{1, \psi} (\theta) = \mathcal{H} (b, p) - \psi (\theta)
\end{equation}
where $\mathcal{H} (b, p) \coloneqq S_{bp} b (X) p (X) + m_{1} (O, b) + m_{2} (O, p) + S_{0}$;
\item Furthermore, the following first-order doubly-robust moment conditions hold:
\begin{align}  
\label{eq:mean_zero}
\left\{ \begin{array}{c}
\E_{\theta} [S_{bp} h (X) p (X) + m_{1} (O, h)] = 0 \text{ \ for all } h \in \mathcal{B}, \\ 
\E_{\theta} [S_{bp} b (X) h (X) + m_{2} (O, h)] = 0 \text{ \ for all } h \in \mathcal{P}.
\end{array} \right.
\end{align}
And $\psi (\theta) = \E_{\theta} [m_{1} (O, b) + S_{0}] = \E_{\theta} [m_{2} (O, p) + S_{0}] = \E_{\theta} [- S_{bp} b (X) p (X) + S_{0}]$.
\item Further suppose the maps $h \mapsto \E_{\theta} [m_{1} (O, h)]$ and $h \mapsto \E_{\theta} [m_{2} (O, h)]$ for $h \in L_{2} (\P_{F})$ are continuous and linear with Riesz representers $\mathcal{R}_{1}(X) \equiv \mathcal{R}_{1}(X; \theta)$ and $\mathcal{R}_{2}(X) \equiv \mathcal{R}_{2}(X; \theta)$\footnote{\label{fn:riesz}The Riesz representer $\calR (X)$ of a continuous linear functional $h \mapsto \E [m (O, h)]$ for $h \in L_{2} (\P_{F})$ is, by definition, the function of $X$ satisfying $\E [m (O, h)] = \E \left[ \calR (X) h (X) \right]$; see \citet{chen2007large, chen2015sieve, chernozhukov2022debiased, chernozhukov2022automatic, rotnitzky2021characterization}.} respectively. Then $b (X) = - \calR_{2} (X) / \lambda (X)$ and $p (X) = - \calR_{1} (X) / \lambda (X)$; moreover, $b$ and $p$ are the (global) minimizers of the following minimization problems 
\begin{equation}
\label{loss}
\begin{split}
b (\cdot) & = \mathsf{arg \; min}_{h \in L_{2} (\P_{F})} \E_{\theta} \left[ S_{bp} \frac{h (X)^{2}}{2} + m_{2} (O, h) \right], \\
p (\cdot) & = \mathsf{arg \; min}_{h \in L_{2} (\P_{F})} \E_{\theta} \left[ S_{bp} \frac{h (X)^{2}}{2} + m_{1} (O, h) \right].
\end{split}
\end{equation}
\end{itemize}
\end{proposition}

We first give some examples of the influence functions $\IF_{1,\psi} (\theta)$ and Riesz representers of DR functionals. Similar moment conditions such as \eqref{eq:mean_zero} can be traced back to \citet{newey1994asymptotic, newey1997convergence, newey1998undersmoothing, newey2004twicing, chernozhukov2022automatic}.

\begin{example}[Some examples of Riesz representers for DR functionals]
\label{eg:riesz}\leavevmode

\begin{enumerate}
\item $\psi (\theta) = - \E_{\theta} [b (X)] = - \E_{\theta} [Y (a = 1)]$. Then $\IF_{1, \psi} (\theta) = \mathcal{H} (b, p) - \psi (\theta)$ with 
\begin{equation*}
\mathcal{H} (b, p) = - \left( - A b (X) p (X) + b (X) + A Y p (X) \right).
\end{equation*}
Here $S_{bp} = A$ and $\lambda (X) = p (X)^{-1}$, $m_{1} (O, b) = - b(X)$, $m_{2} (O, p) = - A Y p (X)$ and $S_{0} = 0$. Then $\mathcal{R}_{1} (X) = -1$ and $\mathcal{R}_{2} (X) = - b (X) / p (X)$ since 
\begin{equation*}
\left\{ \begin{array}{l}
\E_{\theta} \left[ m_{1} (O, h) \right] = \E_{\theta} [- h (X)], \\ 
\E_{\theta} \left[ m_{2} (O, h) \right] = \E_{\theta} [- A Y h (X)] = \E_{\theta} \left[ - \left\{ p (X) \right\}^{-1} b (X) h (X) \right].
\end{array} \right.
\end{equation*}

\item $\psi (\theta) = \E_{\theta} \left[ (Y - b (X)) (A - p (X)) \right]$ -- expected conditional covariance between $A$ and $Y$ given $X$ with $b (X) = \E_{\theta} [Y | X], p (X) = \E_{\theta} [A | X]$: $\IF_{1, \psi} (\theta) = \mathcal{H} (b, p) - \psi (\theta)$ with $\mathcal{H} (b, p) = b (X) p (X) - A b (X) - Y p (X) + A Y$, $S_{bp} = 1$ and $\lambda (X) = 1$, $m_{1} (O, b) = - A b (X)$, $m_{2} (O, p) = - Y p (X)$, and $S_{0} = A Y$. Then $\mathcal{R}_{1} (X) = - p (X)$ and $\mathcal{R}_{2} (X) = - b(X)$ since 
\begin{equation*}
\left\{ \begin{array}{l}
\E_{\theta} \left[ m_{1} (O, h) \right] = \E_{\theta} [- A h(X)] = \E_{\theta} [- p(X) h(X)], \\ 
\E_{\theta} \left[ m_{2} (O, h) \right] = \E_{\theta} [- Y h(X)] = \E_{\theta} [- b(X) h(X)].
\end{array} \right.
\end{equation*}
\end{enumerate}
\end{example}

\allowdisplaybreaks
We now briefly comment on the three parts of Proposition \ref{thm:if1} as they are all important for future development of the paper. Eq. \eqref{eq:if1} exhibits the general formula of influence functions of DR functionals, which are the basic building blocks for standard DML estimators and most nonstandard DML estimators. Also, our framework heavily relies on higher order influence functions, which are derived from the (first-order) influence functions. Eq. \eqref{eq:mean_zero} and \eqref{loss} together suggest natural loss functions that could be used to fit the nuisance functions $b$ and $p$ from data. To see why, we first make the following important observation:
\begin{lemma}
\label{lem:loss}
The minimization problems given in \eqref{loss}, with the function class $L_{2} (\P_{F})$ replaced by some $\mathcal{F} \subset L_{2} (\P_{F})$
\begin{equation}
\label{loss_f}
\begin{split}
\tilde{b} (\cdot) = \mathsf{arg \; min}_{h \in \mathcal{F}} \E_{\theta} \left[ S_{bp} \frac{h (X)^{2}}{2} + m_{2} (O, h) \right], \tilde{p} (\cdot) = \mathsf{arg \; min}_{h \in \mathcal{F}} \E_{\theta} \left[ S_{bp} \frac{h (X)^{2}}{2} + m_{1} (O, h) \right]
\end{split}
\end{equation}
are equivalent to:
\begin{equation}
\label{equi_loss_f}
\begin{split}
\tilde{b} (\cdot) = \mathsf{arg \; min}_{h \in \mathcal{F}} \E_{\theta} \left[ \lambda (X) (b (X) - h (X))^{2} \right], \tilde{p} (\cdot) = \mathsf{arg \; min}_{h \in \mathcal{F}} \E_{\theta} \left[ \lambda (X) (p (X) - h (X))^{2} \right].
\end{split}
\end{equation}
\end{lemma}
The proof of Lemma \ref{lem:loss} can be found in Appendix \ref{app:loss}. Based on Lemma \ref{lem:loss}, one can often establish rates of convergence of $\hat{b}, \hat{p}$ to $b, p$ in $L_{2} (\P_{\theta})$ norm by solving the following minimization problem from the training sample:
\begin{equation}
\label{loss_data}
\begin{split}
\hat{b} = \underset{h \in \mathcal{F}}{\argmin} \P_{n_{\tr}} \left[ S_{bp} \frac{h^{2}}{2} + m_{2} (O, h) \right], \hat{p} = \underset{h \in \mathcal{F}}{\argmin} \P_{n_{\tr}} \left[ S_{bp} \frac{h^{2}}{2} + m_{1} (O, h) \right].
\end{split}
\end{equation}
where $\mathcal{F}$ is the set of functions computable by the machine learning algorithm. This is because the convergence properties of $\hat{b}$ and $\hat{p}$ are often established by excess risk bound that connects the empirical loss \eqref{loss_data} to the expected $\lambda$-weighted $L_{2} (\P_{\theta})$-loss \eqref{loss_f} under certain complexity-reducing assumptions. Importantly, given positive weight functions $w_{b}, w_{p}$ over $X$, that are strictly bounded from above and below, any $w_{b}$- (or $w_{p}$-) weighted $L_{2} (\P_{\theta})$-loss and $\lambda$-weighted $L_{2} (\P_{\theta})$-loss are equivalent up to constants under our assumption on $\lambda$ in Definition \ref{def:dr}:
\begin{equation}
\label{holder}
\inf_{x} \frac{\lambda (x)}{w_{b} (x)} \E_{\theta} [w_{b} (X) \{\hat{b} (X) - b(X)\}^{2}] \leq \E_{\theta} [\lambda (X) \{\hat{b} (X) - b(X)\}^{2}] \leq \sup_{x} \frac{\lambda (x)}{w_{b} (x)} \E_{\theta} [w_{b} (X) \{\hat{b} (X) - b(X)\}^{2}]
\end{equation}
and similarly for $\E_{\theta} [\lambda (X) \{\hat{p} (X) - p (X)\}^{2}]$. In Appendix \ref{app:loss}, we also point out several other possible choices, including unweighted $L_{2} (\P_{\theta})$-loss, cross-entropy loss, and adversarial loss minimization strategies, that are also equivalent to the $\lambda$-weighted $L_{2} (\P_{\theta})$-loss up to multiplicative or additive constants.

For DR functionals, $\CSBias_{\theta} (\hat{\psi}_{1})$, the Cauchy-Schwarz bias of $\hat{\psi}_{1}$, is defined as
\begin{equation}
\label{cs-general}
\CSBias_{\theta} (\hat{\psi}_{1}) \equiv \left\{ \E_{\theta} [\lambda (X) (\hat{b} (X) - b (X))^{2}] \E_{\theta} [\lambda (X) (\hat{p} (X) - p (X))^{2}] \right\}^{1 / 2}
\end{equation}
which is an upper bound of (the absolute value of) $\Bias_{\theta} (\hat{\psi}_{1})$ to be defined in \eqref{bias}, by CS inequality. Hence we have 
\begin{equation}
\label{equiv}
\CSBias_{\theta} (\hat{\psi}_{1}) \asymp \CSBias_{\theta}^{w} (\hat{\psi}_{1}) \equiv \{\E_{\theta} [w_{b} (X) \{\hat{b} (X) - b(X)\}^{2}] \E_{\theta} [w_{p} (X) \{\hat{p} (X) - p (X)\}^{2}]\}^{1 / 2},
\end{equation}
that is, they are equivalent up to a multiplicative positive constant.

\subsection{Standard DML estimators}\leavevmode
\label{sec:dml}

The following algorithm defines the standard DML estimators $\hat{\psi}_{1}$ and $\hat{\psi}_{\mathsf{cf}, 1}$ of a general DR functional $\psi (\theta)$ satisfying the regularity conditions given in Proposition \ref{thm:if1}.

\begin{itemize}
\item[(i)] The $N$ study subjects are randomly split into two parts: an estimation sample of size $n$ and a training (nuisance) sample of size $n_{tr} = N - n$ with $n / N \approx 1 / 2$. Without loss of generality we shall assume that $i = 1, \ldots, n$ corresponds to the estimation sample.

\item[(ii)] Nuisance estimators $\hat{b}$ and $\hat{p}$ are separately constructed from the entire training sample data using machine learning, such as deep neural networks \citep{farrell2021deep, chen2020causal} and define $\hat{\theta} \coloneqq (\hat{b}, \hat{p}, \theta \setminus \{b, p\})$.

\item[(iii)] Denote $\IIFF_{1} \equiv \IIFF_{1} (\theta) \coloneqq \frac{1}{n} \sum_{i = 1}^{n} \IF_{1, \psi} (\theta)$ and $\hat{\IIFF}_{1} \equiv \IIFF_{1} (\hat{\theta}) \coloneqq \frac{1}{n} \sum_{i = 1}^{n} \IF_{1, \psi} (\hat{\theta})$. Then 
\begin{equation*}
\begin{split}
\hat{\psi}_{1} & = \hat{\IIFF}_{1} + \psi (\hat{\theta}) = \frac{1}{n} \sum_{i = 1}^{n} \mathcal{H}_{i} (\hat{b}, \hat{p}) = \frac{1}{n} \sum_{i = 1}^{n} \left( S_{bp, i} \hat{b} (X_{i}) \hat{p} (X_{i}) + m_{1} (O_{i}, \hat{b}) + m_{2} (O_{i}, \hat{p})+ S_{0, i} \right)
\end{split}
\end{equation*}
from $n$ subjects in the estimation sample and $\hat{\psi}_{\cf, 1} = \dfrac{1}{2} \left( \hat{\psi}_{1} + \overline{\hat{\psi}}_{1} \right)$ where $\overline{\hat{\psi}}_{1}$ is $\hat{\psi}_{1}$ but with the training and estimation samples reversed.
\end{itemize}
The next proposition provides asymptotic properties of standard DML estimators $\hat{\psi}_{1}$ (and $\hat{\psi}_{\cf, 1}$) of DR functionals. Its proof is straightforward and can be found in \citet{chernozhukov2018double}, \citet{smucler2019unifying} or \citet[Theorem 3]{farrell2021deep}.
\begin{proposition}\label{thm:drml} 
Under the conditions of Proposition \ref{thm:if1}, conditional on the training sample, if a) the bias of $\hat{\psi}_{1}$ 
\begin{equation} 
\label{bias}
\Bias_{\theta} (\hat{\psi}_{1}) \coloneqq \E_{\theta} [S_{bp} (b (X) - \hat{b} (X)) (p (X) - \hat{p} (X))] \equiv \E_{\theta} [\lambda (X) (b (X) - \hat{b} (X)) (p (X) - \hat{p} (X))]
\end{equation}
is $o (n^{- 1 / 2})$ and b) $\hat{b} (x)$ and $\hat{p} (x)$ converge to $b (x)$ and $p (x)$ in $L_2 (\P_{\theta})$, then:
\begin{enumerate}
\item $\hat{\psi}_{1} - \psi (\theta) = n^{-1} \sum_{i = 1}^n \IF_{1, \psi, i} (\theta) + o (n^{-1/2})$ and $\hat{\psi}_{\cf, 1} - \psi (\theta) = N^{-1} \sum_{i = 1}^N \IF_{1, \psi, i} (\theta) + o_{p} (N^{-1/2})$. Further $n^{1/2} (\hat{\psi}_{1} - \psi(\theta))$ converges conditionally and unconditionally to a normal distribution with mean zero; $\hat{\psi}_{\cf, 1}$ is a regular, asymptotically linear estimator; i.e., $N^{1/2} (\hat{\psi}_{\cf, 1} - \psi (\theta))$ converges unconditionally to a normal distribution with mean zero and variance equal to the semiparametric variance bound $\var_{\theta} \left[ \IF_{1, \psi} (\theta) \right]$.

\item The nominal $(1 - \alpha)$ Wald CIs 
\begin{equation}
\CI_{\alpha} (\hat{\psi}_{1}) \coloneqq \hat{\psi}_{1} \pm z_{\alpha / 2} \hat{\se} [\hat{\psi}_{1}], \ \CI_{\alpha} (\hat{\psi}_{\cf, 1}) \coloneqq \hat{\psi}_{\cf, 1} \pm z_{\alpha / 2} \hat{\se} [\hat{\psi}_{\cf, 1}] \label{eq:ci}
\end{equation}
are asymptotically valid nominal $(1 - \alpha)$ CI for $\psi (\theta)$. Here $\hat{\se} [\hat{\psi}_{1}] = \{\hat{\var} [\hat{\psi}_{1}]\}^{1 / 2}$ with
\begin{equation}
\label{var:dml}
\hat{\var} [\hat{\psi}_{1}] = \frac{1}{n^2} \sum_{i = 1}^{n} \left( \IF_{1, \psi, i} (\hat{\theta}) - \frac{1}{n} \sum_{i = 1}^{n} \IF_{1, \psi, i} (\hat{\theta}) \right)^{2}
\end{equation} 
and $\hat{\se} [\hat{\psi}_{\cf, 1}] = \frac{1}{2} \{\hat{\var} [\hat{\psi}_{1}] + \hat{\var} [\overline{\hat{\psi}}_{1}]\}^{1 / 2}$. The interval $\CI_{\alpha} (\hat{\psi}_{1})$, unlike $\CI_{\alpha} (\hat{\psi}_{\cf, 1})$, is also an asymptotically valid $(1 - alpha)$ CI conditional on the training sample.
\end{enumerate}
\end{proposition}

\allowdisplaybreaks
\subsection{A prelude to our approach}\leavevmode
\label{sec:prelude}

As briefly described in Section \ref{sec:introduction}, our main goal is to develop assumption-lean empirical methods to falsify $\mathsf{NH}_{0, \CS}: \CSBias_{\theta} (\hat{\psi}_{1}) = o (n^{- 1 / 2})$ by falsifying its operationalized pair $\H_{0, \CS} (\delta): \CSBias_{\theta} (\hat{\psi}_{1}) < \se_{\theta} (\hat{\psi}_{1}) \delta$.

However, as we argued in the Introduction, $\CSBias_{\theta} (\hat{\psi}_{1})$ and $\Bias_{\theta} (\hat{\psi}_{1})$ are generally not uniformly consistently estimable without complexity-reducing assumptions on $b$ and $p$ or their estimates $\hat{b}$ and $\hat{p}$. This is evident from the following decomposition of $\Bias_{\theta} (\hat{\psi}_{1})$ because we do not have control over $\Pi_{\theta}^{\perp} [\lambda^{1 / 2} (\hat{b} - b) | \lambda^{1 / 2} \zbar_{k}]$ and $\Pi_{\theta}^{\perp} [\lambda^{1 / 2} (\hat{p} - p) | \lambda^{1 / 2} \zbar_{k}]$ without restrictive complexity-reducing assumptions on $b$ and $p$ or on $\hat{b}$ and $\hat{p}$:
\begin{equation}
\label{decomposition}
\Bias_{\theta} (\hat{\psi}_{1}) = \Bias_{\theta, k} (\hat{\psi}_{1}) + \TB_{\theta, k} (\hat{\psi}_{1})
\end{equation}
where $\TB_{\theta, k} (\hat{\psi}_{1}) \coloneqq \E_{\theta} [\Pi_{\theta}^{\perp} [\lambda^{1 / 2} (\hat{b} - b) | \lambda^{1 / 2} \zbar_{k}] (X) \Pi_{\theta}^{\perp} [\lambda^{1 / 2} (\hat{p} - p) | \lambda^{1 / 2} \zbar_{k}] (X)]$, which was referred to as the truncation bias in \citet{robins2008higher, liu2020nearly} (also see Appendix \ref{app:hoif}), and
\begin{equation}\label{bias_k}
\begin{split}
\Bias_{\theta, k} (\hat{\psi}_{1}) & \coloneqq \E_{\theta} [\Pi_{\theta} [\lambda^{1 / 2} (\hat{b} - b) | \lambda^{1 / 2} \zbar_{k}] (X) \Pi_{\theta} [\lambda^{1 / 2} (\hat{p} - p) | \lambda^{1 / 2} \zbar_{k}] (X)] \\
& \equiv \E_{\theta} [\lambda (X) (\hat{b} (X) - b (X)) \zbar_{k} (X)]^{\top} \Sigma_{k}^{-1} \E_{\theta} [\zbar_{k} (X) \lambda (X) (\hat{p} (X) - p (X))].
\end{split}
\end{equation}
Here $\zbar_{k}$ is a vector of $k$-dimensional vector of dictionary chosen by the analyst satisfying mild regularity conditions (see Condition \ref{cond:sw} in Section \ref{sec:main}) and $\Sigma_{k} \coloneqq \E_{\theta} [\lambda (X) \zbar_{k} (X) \zbar_{k} (X)^{\top}] \equiv \E_{\theta} [S_{bp} \zbar_{k} (X) \zbar_{k} (X)^{\top}]$\footnote{To relate to the discussion in Section \ref{sec:introduction}, $S_{bp} = A$ for the functional $\psi (\theta) = - \E_{\theta} [Y (1)]$.} is the population Gram matrix of $\lambda^{1 / 2} \zbar_{k}$.

Since we generally know neither the sign nor the magnitude of $\TB_{\theta, k} (\hat{\psi}_{1})$, tests of $\H_{0} (\delta)$ based on $\Bias_{\theta, k} (\hat{\psi}_{1})$ fail to protect the nominal level uniformly. But fortunately, as can be shown in the same fashion as in \eqref{cs-heuristic}, the quantity $\Bias_{\theta, k} (\hat{\psi}_{1})$ is a lower bound of $\CSBias_{\theta} (\hat{\psi}_{1})$, making it possible to construct nominal-level tests of $\H_{0, \CS} (\delta)$. The test statistic of $\H_{0, \CS} (\delta)$, introduced in Section \ref{sec:main}, is based on estimators of $\Bias_{\theta, k} (\hat{\psi}_{1})$, derived using the theory of higher-order influence functions (HOIFs) (which are higher-order $U$-statistics) developed in a series of papers by some of the authors \citep{robins2008higher, robins2017minimax, liu2017semiparametric}. Due to space limitation, we refer the interested readers to \citet{robins2008higher} or \citet{van2014higher} for a more comprehensive review. {\it En route} to constructing these HOIF estimators and associated tests, we require access to only the study data and the functions $\hat{b}$ and $\hat{p}$ obtained by analysts. However, our tests are constructed without: i) refitting, modifying, or even having knowledge of the machine learning algorithms that have been employed to compute $\hat{b}, \hat{p}$ from the training sample, and ii) requiring any assumptions at all (aside from a few standard, quite weak assumptions given later) -- in particular, without making any assumptions about the smoothness or sparsity of the nuisance functions $b$ or $p$. The key to achieve i) is by conditioning on the training sample data. For a related discussion, see Remark \ref{rem:finite}.


Formally, our proposed approach begins by noticing that $\Bias_{\theta, k} (\hat{\psi}_{1})$ can be unbiasedly estimated by the following infeasible second order $U$-statistic
\begin{equation}\label{if22}
\begin{split}
\hat{\IIFF}_{22, k} (\Sigma_{k}^{-1}) \coloneqq \frac{1}{n (n - 1)} \sum_{1 \leq i_{1} \neq i_{2} \leq n} \[ \calE_{\hat{b}, m_{2}} (\zbar_{k}) (O) \]_{i_{1}}^{\top} \Sigma_{k}^{-1} \[ \calE_{\hat{p}, m_{1}} (\zbar_{k}) (O) \]_{i_{2}}, \text{ where} \\
\calE_{\hat{b}, m_{2}} (\zbar_{k}) (O) \coloneqq S_{bp} \hat{b} (X) \zbar_{k} (X) + m_{2} (O, \zbar_{k}), \calE_{\hat{p}, m_{1}} (\zbar_{k}) (O) \coloneqq S_{bp} \hat{p} (X) \zbar_{k} (X) + m_{1} (O, \zbar_{k})
\end{split}
\end{equation}
with standard error of order $O \left( \frac{\sqrt{k}}{n} \vee \frac{1}{\sqrt{n}} \right)$ (see Theorem \ref{thm:hoif_stats} for more details). In fact, $\hat{\IIFF}_{22, k} (\Sigma_{k}^{-1})$ is the second-order influence function of $\Bias_{\theta, k} (\hat{\psi}_{1})$\footnote{This is why we adopt the notation $\hat{\IIFF}_{22, k} (\Sigma_{k}^{-1})$ following the convention in \citet{robins2008higher}.}; see Appendix \ref{app:hoif} for more details. $\hat{\IIFF}_{22, k} (\Sigma_{k}^{-1})$ is infeasible because in general $\Sigma_{k}^{-1}$ is unknown. The unbiasedness follows directly from equation \eqref{eq:mean_zero} in Proposition \ref{thm:if1}; the proof of the bound on the standard error of $\hat{\IIFF}_{22, k} (\Sigma_{k}^{-1})$ is deferred to Appendix \ref{app:oracle}, and can also be found in \citet{liu2020nearly}, which mostly focused on the infeasible case by assuming $\Sigma_{k}^{-1}$ to be known. Since $\Sigma_{k}^{-1}$ is generally unknown, we propose to estimate $\Sigma_{k}^{-1}$ by $\hat{\Sigma}_{k}^{-1}$ from the training sample data, where $\hat{\Sigma}_{k} \coloneqq \P_{n_{\tr}} [S_{b p} \zbar_{k} (X) \zbar_{k} (X)^{\top}]$ is simply the empirical Gram matrix estimator. In particular, throughout this paper, we choose $1 \ll k \ll n$. The reason for this choice is discussed in detail in the following remark.

\begin{remark}[On the choice of $k$]\leavevmode
\label{rem:k}
We shall always choose $k$ to be much less than the sample size $n = N / 2$ of the estimation sample, for the standard error of $\hat{\IIFF}_{22, k} (\Sigma_{k}^{-1})$ to be smaller than or equal to the order $n^{-1/2}$ of the standard error of $\hat{\psi}_{1}$, thereby creating the possibility of detecting, for any given $\delta > 0$, that $|\Bias_{\theta, k} (\hat{\psi}_{1})| / \se_{\theta} (\hat{\psi}_{1}) \geq \delta$, when the sample size $n$ is sufficiently large. 

Moreover, we also need $k \ll n_{\tr} = n$, the sample size of the training sample, to ensure $\hat{\Sigma}_{k}^{-1}$ to be a operator-norm consistent estimator of $\Sigma_{k}^{-1}$, without imposing, possibly incorrect, additional smoothness assumption on the distribution of $(S_{bp}, X)$ or sparsity assumptions on $\Sigma_{k}$ or $\Sigma_{k}^{-1}$ required for accurate estimation when $k \geq n$. 

Finally, we take $k \gg 1$ to ensure the asymptotic normality of the $U$-statistic $\hat{\IIFF}_{22, k} (\hat{\Sigma}_{k}^{-1})$ (after appropriate scaling) \citep{bhattacharya1992class} to determine the rejection region of our falsification test to be studied in the next section.

One might view the choice of $k \ll n$ a disadvantage. On the surface, it seems to exclude high-dimensional statistical models. However, this is not the case: (1) The nuisance functions are allowed to be estimated by any procedure, be it high-dimensional or not, from the training sample; (2) $k \ll n$ ensures that our testing procedure can be a valid test without any complexity-reducing assumptions on the nuisance functions, including the distribution of $X$, that is required to evaluate $\Sigma_{k}^{-1}$. We cannot let $k \asymp n$ because, as we will see, our test needs the bias due to estimating $\Sigma_{k}$ by $\hat{\Sigma}_{k}$ to be sufficiently small and recent developments along this direction \citep{cattaneo2018inference, kline2020leave, yadlowsky2022explaining, jiang2022new, jochmans2022heteroscedasticity} cannot be used because of the more stringent assumptions on $\zbar_{k}$, such as sub-Gaussian tail conditions. The jackknife or bootstrap bias correction methods developed by \citet{cattaneo2019two, cattaneo2018kernel} could potentially be applied to our problem but since they only allow $k = O (\sqrt{n})$, the chance of rejecting $\H_{0, \CS} (\delta)$ could be diminished. It will be an interesting problem to study if their approach can be extended to allow $\sqrt{n} \ll k \ll n$, so as to be applied to our problem. We list this as an open problem in Section \ref{sec:conclusion}.
\end{remark}

\section{Main results: A valid falsification test of null hypothesis $\H_{0, \CS} (\delta)$}
\label{sec:main}

Having introduced $\hat{\IIFF}_{22, k} (\hat{\Sigma}_{k}^{-1})$ as an estimator of $\Bias_{\theta, k} (\hat{\psi}_{1})$, we are now prepared to construct a valid nominal level-$\alpha^{\dag}$ falsification test of the non-asymptotic null hypothesis paired with $\NH_{0, \CS}$:
\begin{align*}
\H_{0, \CS} (\delta): \frac{\CSBias_{\theta} (\hat{\psi}_{1})}{\se_{\theta} (\hat{\psi}_{1})} \leq \delta.
\end{align*}

\subsection{What constitutes a valid test of $\H_{0, \CS} (\delta)$?}\leavevmode
\label{sec:what}

Before proceeding to the main result, we note that the infeasible falsification test of $\H_{0, \CS} (\delta)$ proposed in \citet{liu2020nearly} can be generalized to the class of DR functionals as follows\footnote{We also provide relevant results in Appendix \ref{app:oracle_test}.}:
\begin{equation}
\hat{\chi}_{2, k} (\Sigma_{k}^{-1}; \varsigma_{k}, \delta) = \mathbbm{1} \left\{ \frac{|\hat{\IIFF}_{22, k} (\Sigma_{k}^{-1})|}{\hat{\se} [\hat{\psi}_{1}]} - \varsigma_{k} \frac{\hat{\se} [\hat{\IIFF}_{22, k} (\Sigma_{k}^{-1})]}{\hat{\se} [\hat{\psi}_{1}]} \geq \delta \right\}
\end{equation}
where $\hat{\se} [\hat{\psi}_{1}]$ (see \eqref{var:dml}) and $\hat{\se} [\hat{\IIFF}_{22, k} (\Sigma_{k}^{-1})]$ (see \eqref{eq:if22_bvar} in Appendix \ref{app:bootstrap}) are estimators of the standard errors $\se_{\theta} [\hat{\psi}_{1}]$ and $\se_{\theta} [\hat{\IIFF}_{22, k} (\Sigma_{k}^{-1})]$. Here one choose the cut-off $\varsigma_{k} = z_{\alpha^{\dag} / 2}$ by normal approximation. The asymptotic validity of $\hat{\chi}_{2, k} (\Sigma_{k}^{-1}; z_{\alpha^{\dag} / 2}, \delta)$ relies on the unbiasedness of $\hat{\IIFF}_{22, k} (\Sigma_{k}^{-1})$, but the unbiasedness can be relaxed as in the proposition below.

\begin{proposition}
\label{prop:master}
Given any estimator $\hat{\Bias}_{\theta, k} (\hat{\psi}_{1})$ of $\Bias_{\theta, k} (\hat{\psi}_{1})$ that, conditioning on the training sample data, satisfies: \\
(1) Under $\H_{0, \CS} (\delta)$: $\E_{\theta} [\hat{\Bias}_{\theta, k} (\hat{\psi}_{1}) - \Bias_{\theta, k} (\hat{\psi}_{1})] = o \( \frac{\sqrt{k}}{n} \)$; \\
(2) $\se_{\theta} [\hat{\Bias}_{\theta, k} (\hat{\psi}_{1})] \asymp \se_{\theta} [\hat{\IIFF}_{22, k} (\Sigma_{k}^{-1})]$; \\
(3) Under $\H_{0, \CS} (\delta)$: $\frac{\hat{\Bias}_{\theta, k} (\hat{\psi}_{1}) - \Bias_{\theta, k} (\hat{\psi}_{1})}{\se_{\theta} [\hat{\Bias}_{\theta, k} (\hat{\psi}_{1})]} \overset{d}{\rightarrow} N (0, 1)$,
then (again conditioning on the training sample data) the test
\begin{equation}
\hat{\chi}_{k} (\varsigma_{k}, \delta) = \mathbbm{1} \left\{ \frac{|\hat{\Bias}_{\theta, k} (\hat{\psi}_{1})|}{\hat{\se} [\hat{\psi}_{1}]} - \varsigma_{k} \frac{\hat{\se} [\hat{\Bias}_{\theta, k} (\hat{\psi}_{1})]}{\hat{\se} [\hat{\psi}_{1}]} \geq \delta \right\},
\end{equation}
where $\hat{\se} [\hat{\Bias}_{\theta, k} (\hat{\psi}_{1})]$ is a consistent estimator of $\se_{\theta} [\hat{\Bias}_{\theta, k} (\hat{\psi}_{1})]$, rejects $\H_{0, \CS} (\delta)$ with probability less than $\alpha^{\dag}$ as $n \rightarrow \infty$ with $\varsigma_{k} = z_{\alpha^{\dag} / 2}$ under $\H_{0, \CS} (\delta)$. That is, we say that $\hat{\chi}_{k} (z_{\alpha^{\dag} / 2}, \delta)$ is an asymptotically valid test of $\H_{0, \CS} (\delta)$.

The above statement also holds if one replaces $\H_{0, \CS} (\delta)$ by the $k$-projected null hypothesis $\H_{0, k} (\delta)$.
\end{proposition}

We did not give a formal proof as the argument for Proposition \ref{prop:master} is quite simple. Since $\se_{\theta} [\hat{\Bias}_{\theta, k} (\hat{\psi}_{1})]$ can be as small as of order $\frac{\sqrt{k}}{n}$, one needs the bias $\E_{\theta} [\hat{\Bias}_{\theta, k} (\hat{\psi}_{1}) - \Bias_{\theta, k} (\hat{\psi}_{1})]$ to be dominated by this order under the null $\H_{0, \CS} (\delta)$ or the null $\H_{0, k} (\delta)$ to protect the level of the test. The final clause of Proposition \ref{prop:master} will be most relevant to Section \ref{sec:nonstandard}. We also refer readers to Appendix \ref{app:sigma_test} for more details.

\subsection{Bias-reduced estimator of $\Bias_{\theta, k} (\hat{\psi}_{1})$}\leavevmode
\label{sec:soif}

The most natural estimator of $\hat{\Bias}_{\theta, k} (\hat{\psi}_{1})$ is $\hat{\IIFF}_{22, k} (\hat{\Sigma}_{k}^{-1})$ and the corresponding test is
\begin{equation}
\hat{\chi}_{2, k} (\hat{\Sigma}_{k}^{-1}; z_{\alpha^{\dag} / 2}, \delta) = \mathbbm{1} \left\{ \frac{|\hat{\IIFF}_{22, k} (\hat{\Sigma}_{k}^{-1})|}{\hat{\se} [\hat{\psi}_{1}]} - z_{\alpha^{\dag} / 2} \frac{\hat{\se} [\hat{\IIFF}_{22, k} (\hat{\Sigma}_{k}^{-1})]}{\hat{\se} [\hat{\psi}_{1}]} \geq \delta \right\}.
\end{equation}
However, we {\it cannot} prove that the above test is asymptotically valid for $\H_{0, \CS} (\delta)$, because Theorem \ref{thm:soif_toif} below will show the bias of $\hat{\IIFF}_{22, k} (\hat{\Sigma}_{k}^{-1})$ for estimating $\Bias_{\theta, k} (\hat{\psi}_{1})$ is
\begin{equation}
\label{eb2:upper_null}
\EB_{\theta, k, 2} \coloneqq \E_{\theta} [\hat{\IIFF}_{22, k} (\hat{\Sigma}_{k}^{-1}) - \Bias_{\theta, k} (\hat{\psi}_{1})] = O \( \frac{\sqrt{k}}{n} \sqrt{\log k} \)
\end{equation}
under $\H_{0, \CS} (\delta)$. This obtained upper bound $O \left( \frac{\sqrt{k \log k}}{n} \right)$ of the bias due to estimating $\Sigma_{k}^{-1}$ exceeds that which is needed to protect the level of the test $\hat{\chi}_{2, k} (\hat{\Sigma}_{k}^{-1}; z_{\alpha^{\dag} / 2}, \delta)$, as given in Condition (1) of Proposition \ref{prop:master}.

Fortunately, Theorem \ref{thm:soif_toif} will also show that the following third-order $U$-statistic estimator of $\Bias_{\theta, k} (\hat{\psi}_{1})$
\begin{equation}
\label{if2233}
\begin{split}
& \ \hat{\IIFF}_{22 \rightarrow 33, k} (\hat{\Sigma}_{k}^{-1}) \coloneqq \hat{\IIFF}_{22, k} (\hat{\Sigma}_{k}^{-1}) + \hat{\IIFF}_{33, k} (\hat{\Sigma}_{k}^{-1}), \text{ where} \\
\hat{\IIFF}_{33, k} (\hat{\Sigma}_{k}^{-1}) \coloneqq & \ \frac{(n - 2)!}{n!} \sum_{1 \leq i_{1} \neq i_{2} \neq i_{3} \leq n} \[ \calE_{\hat{b}, m_{2}} (\zbar_{k}) (O) \]_{i_{1}}^{\top} \hat{\Sigma}_{k}^{-1} \[ S_{bp} \zbar_{k} (X) \zbar_{k} (X)^{\top} - \hat{\Sigma}_{k} \]_{i_{3}} \hat{\Sigma}_{k}^{-1} \[ \calE_{\hat{p}, m_{1}} (\zbar_{k}) (O) \]_{i_{2}}
\end{split}
\end{equation} 
can reduce the upper bound of the bias due to estimating $\Sigma_{k}^{-1}$ from $O \left( \frac{\sqrt{k \log k}}{n} \right)$ in \eqref{eb2:upper_null} to 
\begin{equation}
\label{eb3:upper_null}
\EB_{\theta, k, 3} \coloneqq \E_{\theta} [\hat{\IIFF}_{22 \rightarrow 33, k} (\hat{\Sigma}_{k}^{-1}) - \Bias_{\theta, k} (\hat{\psi}_{1})] = O \( \frac{\sqrt{k}}{n} \sqrt{\frac{k \log k}{n}} \) = o \( \frac{\sqrt{k}}{n} \)
\end{equation}
as long as we choose $k$ such that $k \log k = o (n)$.

\begin{remark}
$\hat{\IIFF}_{22 \rightarrow 33, k} (\hat{\Sigma}_{k}^{-1})$ can be viewed as a de-biased version of $\hat{\IIFF}_{22, k} (\hat{\Sigma}_{k}^{-1})$, and the bias is partially corrected by $\hat{\IIFF}_{33, k} (\hat{\Sigma}_{k}^{-1})$. As will be shown in Appendix \ref{app:hoif}, $\hat{\IIFF}_{22 \rightarrow 33, k} (\Sigma_{k}^{-1})$ is the third-order influence function of $\Bias_{\theta, k} (\hat{\psi}_{1})$ and $\hat{\IIFF}_{22 \rightarrow 33, k} (\hat{\Sigma}_{k}^{-1})$ is its estimated version.
\end{remark}

To avoid clutter in the remainder of this paper, we introduce the following additional notation for various $L_{q}$-type norms of functions or their weighted $L_{2} (\P_{\theta})$-projections, for $q = 2, 4$: 
\begin{align*}
\BL_{\theta, 2, \hat{b}} & \coloneqq \left\{ \E_{\theta} [\lambda (X) (\hat{b} (X) - b (X))^{2}] \right\}^{1 / 2}, \BL_{\theta, q, \hat{b}, k} \coloneqq \left\{ \E_{\theta} [\{\Pi_{\theta} [\lambda^{1 / 2} (\hat{b} - b) \vert \lambda^{1 / 2} \zbar_{k}] (X)\}^{q}] \right\}^{1 / q}, \\
\BL_{\theta, 2, \hat{p}} & \coloneqq \left\{ \E_{\theta} [\lambda (X) (\hat{p} (X) - p (X))^{2}] \right\}^{1 / 2}, \BL_{\theta, q, \hat{p}, k} \coloneqq \left\{ \E_{\theta} [\{\Pi_{\theta} [\lambda^{1 / 2} (\hat{p} - p) \vert \lambda^{1 / 2} \zbar_{k}] (X)\}^{q}] \right\}^{1 / q}, \\
\BL_{\theta, \infty, \hat{b}, k} & \coloneqq \Vert \Pi_{\theta} [\lambda^{1 / 2} (\hat{b} - b) \vert \lambda^{1 / 2} \zbar_{k}] \Vert_{\infty}, \BL_{\theta, \infty, \hat{p}, k} \coloneqq \Vert \Pi_{\theta} [\lambda^{1 / 2} (\hat{p} - p) \vert \lambda^{1 / 2} \zbar_{k}] \Vert_{\infty}, \BL_{\theta, 2, \hat{\Sigma}, k} \coloneqq \Vert \hat{\Sigma}_{k} - \Sigma_{k} \Vert,
\end{align*}
which will also be useful for the later development of this paper. Note that we define $\CSBias_{\theta, k} (\hat{\psi}_{1}) \equiv \BL_{\theta, 2, \hat{b}, k} \BL_{\theta, 2, \hat{p}, k}$, generalizing the definition of $\CSBias_{\theta, k} (\hat{\psi}_{1})$ given in \eqref{csbias} to the whole class of DR functionals.

We also need to further impose the following weak regularity conditions (Condition \ref{cond:sw}).

\begin{customthm}{W}
\leavevmode\label{cond:sw}

\begin{enumerate}
\item All the eigenvalues of $\Sigma_{k}$ are bounded away from 0 and $\infty$.

\item The observed data $O = (W, X)$, the true nuisance functions $b(X)$ and $p(X)$, the estimated nuisance functions $\hat{b}(X)$ and $\hat{p}(X)$ are bounded with $\P_{\theta}$-probability 1, and $\lambda (X)$ are bounded away from 0 and $\infty$ with $\P_{\theta}$-probability 1.

\item $\Vert \zbar_{k}^{\top} \zbar_{k} \Vert_{\infty} \leq B k$ for some constant $B > 0$.
\end{enumerate}
\end{customthm}

Note that Condition \ref{cond:sw} should also be compared to the following slightly stronger Condition \ref{cond:w} in \citet{liu2020nearly}, which is the same as Condition \ref{cond:sw} except (3) shall be replaced by the following (3'):

\begin{customthm}{S}
\leavevmode\label{cond:w}

\begin{enumerate} [label=(\arabic*')] \setcounter{enumi}{2}

\item $\Vert \zbar_{k}^{\top} \zbar_{k} \Vert_{\infty} \leq B k$ for some constant $B > 0$; in addition, $\BL_{\theta, \infty, \hat{b}, k} < \infty$ and $\BL_{\theta, \infty, \hat{p}, k} < \infty$.
\end{enumerate}
\end{customthm}

\begin{remark}
\label{rem:w} 
Condition \ref{cond:w}(3') is stronger than Condition \ref{cond:sw}(3) and it holds for Cohen-Vial-Daubechies wavelets, local polynomial partition and B-spline series \citep{newey1997convergence, belloni2015some}. But it does not hold in general for Fourier or Legendre polynomial series of $X$ when $X$ is compactly supported \citep{belloni2015some}. We also refer interested readers to \citet[Section 6]{kennedy2020discussion} and \citet[Section 2.1]{liu2020rejoinder} for further motivation on why we decide to relax Condition \ref{cond:w} by Condition \ref{cond:sw}. However, as we will see, when $\Sigma_{k}$ is unknown and need to be estimated from the training sample, violation of Condition \ref{cond:w}(3') but not Condition \ref{cond:sw}(3), will incur a loss in the power of the test (the magnitude depending on the particular series used; see Appendix \ref{app:var}), but the validity of the test (i.e. level) is still guaranteed. Nevertheless, such loss in power happens (or equivalently, Condition \ref{cond:sw} holds but Condition \ref{cond:w} fails to hold), only if the $L_{\infty}$-norms of the projections $\Pi_{\theta} [\lambda^{1 / 2} (\hat{b} - b) \vert \lambda^{1 / 2} \zbar_{k}]$ and $\Pi_{\theta} [\lambda^{1 / 2} (\hat{p} - p) \vert \lambda^{1 / 2} \zbar_{k}]$ are not bounded even though those of $\lambda^{1 / 2} (\hat{b} - b)$ and $\lambda^{1 / 2} (\hat{p} - p)$ are bounded.
\end{remark}

Finally, we summarize the statistical properties of $\hat{\IIFF}_{22, k} (\hat{\Sigma}_{k}^{-1})$ and $\hat{\IIFF}_{22 \rightarrow 33, k} (\hat{\Sigma}_{k}^{-1})$ in Theorem \ref{thm:soif_toif} below:

\begin{theorem}\label{thm:soif_toif}
Under the conditions of Proposition \ref{thm:if1} and Condition \ref{cond:sw}, with $k, n \rightarrow \infty$ but $k \log k = o (n)$, conditioning on the training sample, the following hold on the event that $\hat{\Sigma}_{k}$ is invertible:

\begin{enumerate}
\item The bias and variance of $\hat{\IIFF}_{22, k} (\hat{\Sigma}_{k}^{-1})$ as an estimator of $\Bias_{\theta, k} (\hat{\psi}_{1})$ are of the following order: 
\begin{align*}
\left\vert \EB_{\theta, 2, k} \right\vert & \lesssim \BL_{\theta, 2, \hat{b}, k} \BL_{\theta, 2, \hat{p}, k} \BL_{\theta, 2, \hat{\Sigma}, k} \lesssim \BL_{\theta, 2, \hat{b}, k} \BL_{\theta, 2, \hat{p}, k} \sqrt{\frac{k \log (k)}{n}}, \\
\var_{\theta} \left[\hat{\IIFF}_{22, k} (\hat{\Sigma}_{k}^{-1}) \right] & \lesssim \frac{1}{n} \left\{ \frac{k}{n} + \BL_{\theta, 2, \hat{b}, k}^{2} + \BL_{\theta, 2, \hat{p}, k}^{2} \right\}
\end{align*}
where $\EB_{\theta, 2, k}$ is defined in equation \eqref{eb2:upper_null}.

\item The bias and variance of $\hat{\IIFF}_{22 \rightarrow 33, k} (\hat{\Sigma}_{k}^{-1}) \coloneqq \hat{\IIFF}_{22, k} (\hat{\Sigma}_{k}^{-1}) + \hat{\IIFF}_{33, k} (\hat{\Sigma}_{k}^{-1})$ are of the following order: 
\begin{align*}
\left\vert \EB_{\theta, 3, k} \right\vert & \lesssim \BL_{\theta, 2, \hat{b}, k} \BL_{\theta, 2, \hat{p}, k} \BL_{\theta, 2, \hat{\Sigma}, k}^{2} \lesssim \BL_{\theta, 2, \hat{b}, k} \BL_{\theta, 2, \hat{p}, k} \frac{k \log (k)}{n}, \\
\var_{\theta} \left[ \hat{\IIFF}_{22 \rightarrow 33, k} (\hat{\Sigma}_{k}^{-1}) \right] & \lesssim \frac{1}{n} \left\{ \frac{k}{n} + \BL_{\theta, 2, \hat{b}, k}^{2} + \BL_{\theta, 2, \hat{p}, k}^{2} + k \BL_{\theta, 2, \hat{b}, k}^{2} \BL_{\theta, 2, \hat{p}, k}^{2} \right\}
\end{align*}
where $\EB_{\theta, 3, k}$ is defined in equation \eqref{eb2:upper_null}.

If, however, Condition \ref{cond:sw} is strengthened to Condition \ref{cond:w} 
\begin{equation*}
\var_{\theta} \left[ \hat{\IIFF}_{22 \rightarrow 33, k} (\hat{\Sigma}_{k}^{-1}) \right] \lesssim \frac{1}{n} \left\{ \frac{k}{n} + \BL_{\theta, 2, \hat{b}, k}^{2} + \BL_{\theta, 2, \hat{p}, k}^{2} \right\}.
\end{equation*}

\item Under $\H_{0, \CS} (\delta)$:
\begin{equation*}
\frac{\hat{\IIFF}_{22, k} (\hat{\Sigma}_{k}^{-1}) - \E_{\theta} \left[ \hat{\IIFF}_{22, k} (\hat{\Sigma}_{k}^{-1}) \right]}{\se_{\theta} [\hat{\IIFF}_{22, k} (\hat{\Sigma}_{k}^{-1})]} \text{ and } \frac{\hat{\IIFF}_{22 \rightarrow 33, k} (\hat{\Sigma}_{k}^{-1}) - \E_{\theta}\left[\hat{\IIFF}_{22 \rightarrow 33, k} (\hat{\Sigma}_{k}^{-1})\right]}{\se_{\theta} [\hat{\IIFF}_{22 \rightarrow 33, k} (\hat{\Sigma}_{k}^{-1})]} \overset{d}{\rightarrow} N (0, 1).
\end{equation*}
When we estimate $\se_{\theta} [\hat{\IIFF}_{22, k} (\hat{\Sigma}_{k}^{-1})$ and $\se_{\theta} [\hat{\IIFF}_{22 \rightarrow 33, k} (\hat{\Sigma}_{k}^{-1})$ respectively by their consistent bootstrap estimators $\hat{\se} [\hat{\IIFF}_{22, k} (\hat{\Sigma}_{k}^{-1})$ and $\hat{\se} [\hat{\IIFF}_{22 \rightarrow 33, k} (\hat{\Sigma}_{k}^{-1})$ defined in \eqref{eq:if22_bvar} and \eqref{bvar:if2233}, with $\Sigma_{k}$ replaced by $\hat{\Sigma}_{k}$, in Appendix \ref{app:bootstrap}, we also have 
\begin{equation*}
\frac{\hat{\IIFF}_{22, k} (\hat{\Sigma}_{k}^{-1}) - \E_{\theta} \left[ \hat{\IIFF}_{22, k} (\hat{\Sigma}_{k}^{-1}) \right]}{\hat{\se} [\hat{\IIFF}_{22, k} (\hat{\Sigma}_{k}^{-1})]} \text{ and } \frac{\hat{\IIFF}_{22 \rightarrow 33, k} (\hat{\Sigma}_{k}^{-1}) - \E_{\theta}\left[\hat{\IIFF}_{22 \rightarrow 33, k} (\hat{\Sigma}_{k}^{-1})\right]}{\hat{\se} [\hat{\IIFF}_{22 \rightarrow 33, k} (\hat{\Sigma}_{k}^{-1})]} \overset{d}{\rightarrow} N (0, 1).
\end{equation*}
\end{enumerate}
\end{theorem}

\begin{remark}\leavevmode
\begin{itemize}
\item The bias and variance bounds under Condition \ref{cond:w} have been proved in \citet{liu2017semiparametric}. The change in the proof under Condition \ref{cond:sw} is given in Lemma \ref{lem:hoif_var} in Appendix \ref{app:var}. The normal approximation under $\H_{0, \CS} (\delta)$ follows from Slutsky theorem and asymptotic normality of $U$-statistics with diverging (with $n$) kernels given in Theorem 1 of \citet{bhattacharya1992class}.
\item The generic bias bounds in Theorem \ref{thm:soif_toif} imply those stated in \eqref{eb2:upper_null} and \eqref{eb3:upper_null} under $\H_{0, \CS} (\delta)$, as $$\BL_{\theta, 2, \hat{b}, k} \BL_{\theta, 2, \hat{p}, k} \leq \BL_{\theta, 2, \hat{b}} \BL_{\theta, 2, \hat{p}} = \CSBias_{\theta} (\hat{\psi}_{1}) \lesssim n^{- 1 / 2}$$ when $\H_{0, \CS} (\delta)$ is true.
\end{itemize}
\end{remark}

\subsection{The proposed assumption-lean falsification test of $\H_{0, \CS} (\delta)$}\leavevmode

All the previous discussions in this section culminate in (1) the following test of $\H_{0, \CS} (\delta)$:
\begin{equation}
\label{the_test}
\hat{\chi}_{3, k} (\hat{\Sigma}_{k}^{-1}; z_{\alpha^{\dag} / 2}, \delta) \coloneqq \mathbbm{1} \left\{ \frac{|\hat{\IIFF}_{22 \rightarrow 33, k} (\hat{\Sigma}_{k}^{-1})|}{\hat{\se} [\hat{\psi}_{1}]} - z_{\alpha^{\dag} / 2} \frac{\hat{\se} [\hat{\IIFF}_{22 \rightarrow 33, k} (\hat{\Sigma}_{k}^{-1})]}{\hat{\se} [\hat{\psi}_{1}]} \geq \delta \right\},
\end{equation}
and (2) the following theorem showing its asymptotically validity:
\begin{theorem}
\label{thm:cs}
Assume all the conditions in Theorem \ref{thm:soif_toif}. Under $\H_{0, \CS} (\delta): \frac{\CSBias_{\theta} (\hat{\psi}_{1})}{\se_{\theta} (\hat{\psi}_{1})} \leq \delta$,
\begin{equation}
\lim_{n \rightarrow \infty} \BP_{\theta} \( \hat{\chi}_{3, k} (\hat{\Sigma}_{k}^{-1}; z_{\alpha^{\dag} / 2}, \delta) = 1 \) \leq \alpha^{\dag}.
\end{equation}
That is, $\hat{\chi}_{3, k} (\hat{\Sigma}_{k}^{-1}; z_{\alpha^{\dag} / 2}, \delta)$ is an asymptotically level-$\alpha^{\dag}$ test of $\H_{0, \CS} (\delta)$.
\end{theorem}

The proof of Theorem \ref{thm:cs} is a direct consequence of Theorem \ref{thm:soif_toif} and is deferred to Appendix \ref{app:sigma_test}. Since $\hat{\chi}_{3, k} (\hat{\Sigma}_{k}^{-1}; z_{\alpha^{\dag} / 2}, \delta)$ is not a consistent test, we say $\hat{\chi}_{3, k} (\hat{\Sigma}_{k}^{-1}; z_{\alpha^{\dag} / 2}, \delta)$ can only falsify the null hypothesis $\H_{0, \CS} (\delta)$.

\begin{figure}[H]
\centering
\includegraphics[width=0.8\textwidth]{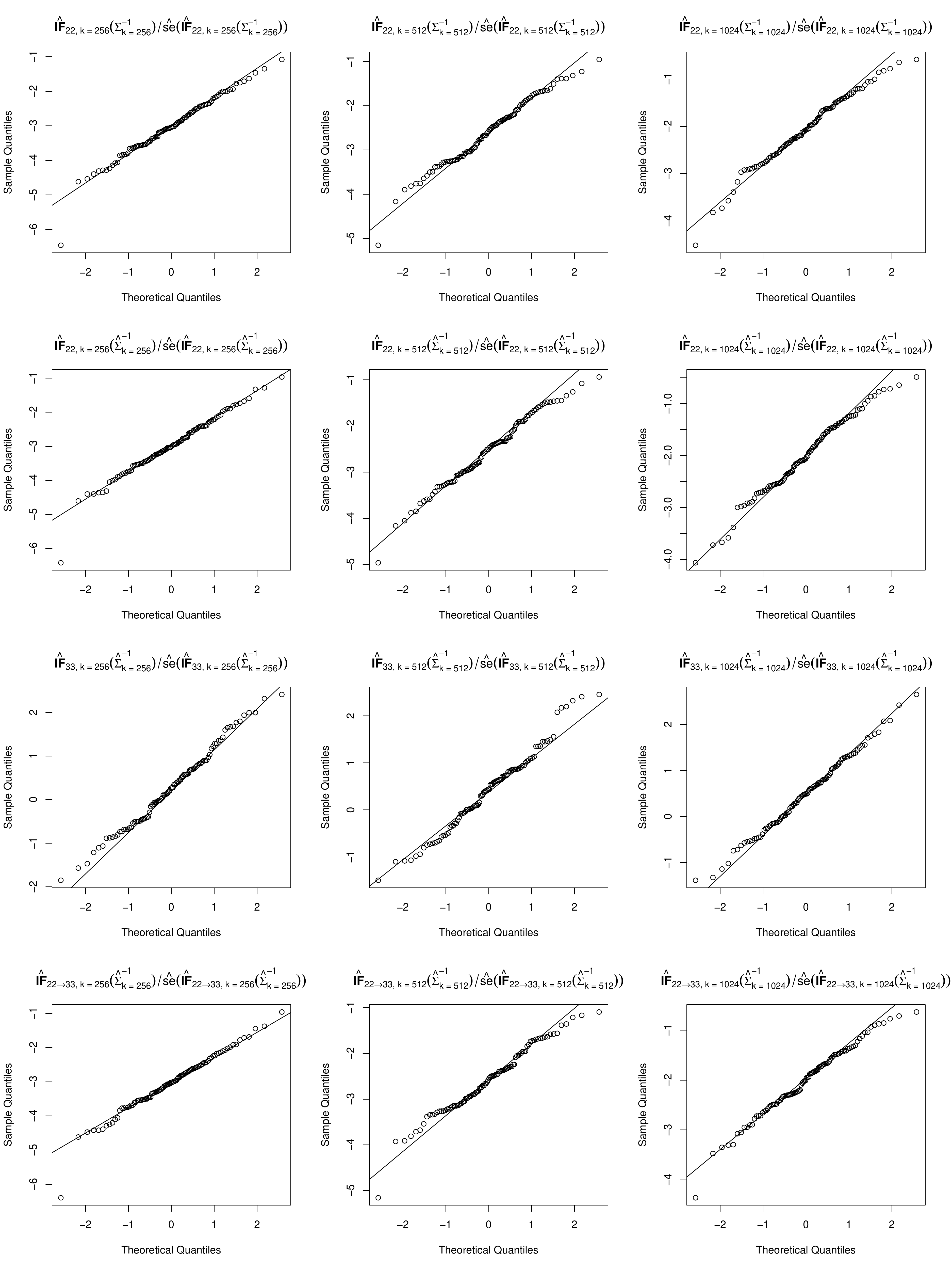}
\caption{For data generating mechanism in simulation setup II: normal qq-plots for $\hat{\IIFF}_{22, k} (\Sigma_{k}^{-1}) / \hat{\se} [\hat{\IIFF}_{22, k} (\Sigma_{k}^{-1})]$ (the first row), $\hat{\IIFF}_{22, k} (\hat{\Sigma}_{k}^{-1}) / \hat{\se} [\hat{\IIFF}_{22, k} (\hat{\Sigma}_{k}^{-1})]$ (the second row), $\hat{\IIFF}_{33, k} (\hat{\Sigma}_{k}^{-1}) / \hat{\se} [\hat{\IIFF}_{33, k} (\hat{\Sigma}_{k}^{-1})]$ (the third row) and $\hat{\IIFF}_{22 \rightarrow 33, k} (\hat{\Sigma}_{k}^{-1}) / \hat{\se} [\hat{\IIFF}_{22 \rightarrow 33, k} (\hat{\Sigma}_{k}^{-1})]$ (the fourth row) for $k = 256$ (left panels), $k = 512$ (middle panels) and $k = 1024$ (right panels).}
\label{fig:qq_h0}
\end{figure}

\begin{remark}
\label{rem:finite}
Because our results are conditioning on the training sample, the test $\hat{\chi}_{3, k} (\hat{\Sigma}_{k}^{-1}; z_{\alpha^{\dag} / 2}, \delta)$ only relies on `asymptopia' to ensure that its rejection probability can be closely approximated by its Gaussian limiting distribution. If Berry-Esseen bound or tail inequalities with constants estimable from the data were available for $\hat{\IIFF}_{22 \rightarrow 33, k} (\hat{\Sigma}_{k}^{-1})$, we could in principle eliminate our dependence on asymptopia. At the sample size used in our simulation (see Section \ref{sec:simulation}), the normal qqplots of $\hat{\IIFF}_{22 \rightarrow 33, k} (\hat{\Sigma}_{k}^{-1})$ in Figure \ref{fig:qq_h0} suggest that normal approximation is reasonably close to the true distribution between $(-2, 2)$ under $\H_{0, \CS} (\delta)$.
\end{remark}

\begin{remark}
\label{rem:temper}
Now we return to the issue raised in footnote 5 on page 7 in the Introduction. Another natural choice for the $k$-projected null hypothesis is
\begin{align*}
\H_{0, k, \CS} (\delta): \CSBias_{\theta, k} (\hat{\psi}_{1})^{2} \equiv \E_{\theta} \[ \Pi_{\theta} [\lambda^{1 / 2} (\hat{b} - b) | \lambda^{1 / 2} \zbar_{k}] (X)^{2} \] \E_{\theta} \[ \Pi_{\theta} [\lambda^{1 / 2} (\hat{p} - p) | \lambda^{1 / 2} \zbar_{k}] (X)^{2} \] \leq \delta^{2} \var_{\theta} (\hat{\psi}_{1})
\end{align*}
because rejection of $\H_{0, k, \CS} (\delta)$ implies rejection of $\H_{0, \CS} (\delta)$. $\CSBias_{\theta, k} (\hat{\psi}_{1})^{2}$ can be unbiasedly estimated by the following infeasible fourth-order $U$-statistic:
\begin{equation*}
\hat{\IIFF}_{44, \CS, k} (\Sigma_{k}^{-1}) \coloneqq \frac{(n - 4)!}{n!} \sum_{1 \leq i_{1} \neq i_{2} \neq i_{3} \neq i_{4} \leq n} [\calE_{\hat{b}, m_{2}} (\zbar_{k}) (O)]^{\top}_{i_{1}} \Sigma_{k}^{-1} [\calE_{\hat{b}, m_{2}} (\zbar_{k}) (O)]_{i_{2}} [\calE_{\hat{p}, m_{1}} (\zbar_{k}) (O)]^{\top}_{i_{3}} \Sigma_{k}^{-1} [\calE_{\hat{p}, m_{1}} (\zbar_{k}) (O)]_{i_{4}}.
\end{equation*}
Using $\hat{\IIFF}_{44, \CS, k} (\Sigma_{k}^{-1})$ as a starting point, one can indeed follow the previous development in Section \ref{sec:main} to construct a feasible test of $\H_{0, k, \CS} (\delta)$. We decide not to further pursue this alternative strategy because it is much more complex to analyze and we leave it to future investigation.
\end{remark}

\section{What if nonstandard DML estimators are used in practice?}
\label{sec:nonstandard}

\subsection{Motivation for testing $\H_{0, k} (\delta)$}\leavevmode

Several recent works \citep{newey2018cross, bradic2019minimax, kline2020leave, kennedy2020towards, mcgrath2022undersmoothing} have exhibited nonstandard DML estimators, denoted also as $\hat{\psi}_{1}$ in this section (to avoid introducing new notation at this point), that improve upon standard DML estimators, with $\Bias_{\theta} (\hat{\psi}_{1}) = o (n^{-1/2})$, even though $\CSBias_{\theta} (\hat{\psi}_{1}) \gg n^{-1/2}$, i.e. $\kappa_{b} + \kappa_{p} \leq 1 / 2$ [recall that $n^{-\kappa_{b}}$ and $n^{-\kappa_{p}}$ are rates of convergence of $\hat{b}$ and $\hat{p}$ to $b$ and $p$ in (weighted) $L_{2} (\P_{\theta})$ norm]. However, to obtain these results, they all (1) use very special nuisance function estimators with $\hat{b}$ and $\hat{p}$ computed from separate non-overlapping subsamples of the training sample, and (2) assume very specific complexity-reducing assumptions such as \Holder{} smoothness \citep{newey2018cross, mcgrath2022undersmoothing} or (approximate) sparsity \citep{bradic2019minimax}. The problem for our methodology when applied to such estimators is that, if (1) and (2) above were true, our test $\hat{\chi}_{3, k} (\hat{\Sigma}_{k}^{-1}; z_{\alpha^{\dag} / 2}, \delta)$ might still reject because $\kappa_{b} + \kappa_{p} \leq 1 / 2$, even though $\Bias_{\theta} (\hat{\psi}_{1}) = o (n^{-1/2})$ holds. 

In this section, we study one possible approach to extend our methodology to apply to nonstandard DML estimators. We will consider an approach in which we simply wish to test the null hypothesis $\H_{0, k} (\delta)$. This can be viewed as testing if the bias of $\hat{\psi}_{1}$ is large along the {\it direction} of the basis/dictionary $\zbar_{k}$ chosen by the analyst. If the analyst has a good grasp of the direction along which $\Bias_{\theta} (\hat{\psi}_{1})$ may be large, say when $b, p$ belong to \Holder{} balls with certain smoothness, then she can test $\H_{0, k} (\delta)$ with $\zbar_{k}$ chosen to be wavelet or B-spline series. If otherwise, we then suggest the analyst tests $\H_{0, k} (\delta)$ with a variety of choices of $\zbar_{k}$, as a form of sensitivity analyses \citep{robins2003general}. If $\H_{0, k} (\delta)$ is rejected with some $\zbar_{k}$ (with a very small p-value), then there is strong evidence that $\Bias_{\theta, k} (\hat{\psi}_{1}) \gtrsim n^{-1/2}$ under the operationalized pairing between $\H_{0, k} (\delta)$ and $\NH_{0, k}: \Bias_{\theta, k} (\hat{\psi}_{1}) = o (n^{-1/2})$. Recall from Section \ref{sec:prelude} that $\Bias_{\theta} (\hat{\psi}_{1})$ can be decomposed as follows:
\begin{equation*}
\Bias_{\theta} (\hat{\psi}_{1}) \equiv \Bias_{\theta, k} (\hat{\psi}_{1}) + \TB_{\theta, k} (\hat{\psi}_{1}).
\end{equation*}
Furthermore, $\TB_{\theta, k} (\hat{\psi}_{1})$ cannot be uniformly consistently estimated under the assumption-lean model being considered here. Even so, $\Bias_{\theta} (\hat{\psi}_{1})$ will not be $o (n^{-1/2})$, and therefore its associated Wald CI centered at $\hat{\psi}_{1}$ will not be valid, unless $\TB_{\theta, k} (\hat{\psi}_{1})$ and $\Bias_{\theta, k} (\hat{\psi}_{1})$ happen to have leading order terms of the same magnitude but opposite signs. As it seems quite fortuitous for such a cancellation to occur, an analyst might agree to retract her claim that the Wald CI is valid. Hence in this section we focus on $\H_{0, k} (\delta)$ and investigate how to test this particular null hypothesis. 

\subsection{Issues with $\hat{\chi}_{3, k} (\hat{\Sigma}_{k}^{-1}; z_{\alpha^{\dag} / 2}, \delta)$ as a test of $\H_{0, k} (\delta)$ and a rescue by HOIFs}\leavevmode

In this section, we begin by showing that $\hat{\chi}_{3, k} (\hat{\Sigma}_{k}^{-1}; z_{\alpha^{\dag} / 2}, \delta)$ is not necessarily an $\alpha^{\dag}$-level test of $\H_{0, k} (\delta)$ for nonstandard DML estimators (neither for standard DML estimators, which does not contradict the claim about the validity of $\hat{\chi}_{3, k} (\hat{\Sigma}_{k}^{-1}; z_{\alpha^{\dag} / 2}, \delta)$ in Theorem \ref{thm:cs} because it is shown to be valid under $\H_{0, \CS} (\delta)$, a different null hypothesis from $\H_{0, k} (\delta)$). Recall from Theorem \ref{thm:soif_toif} that the bias $\hat{\IIFF}_{22 \rightarrow 33, k} (\hat{\Sigma}_{k}^{-1})$ for $\Bias_{\theta, k} (\hat{\psi}_{1})$ was upper bounded by:
\begin{align*}
\EB_{3, \theta, k} = \E_{\theta} [\hat{\IIFF}_{22 \rightarrow 33, k} (\hat{\Sigma}_{k}^{-1}) - \Bias_{\theta, k} (\hat{\psi}_{1})] \lesssim \BL_{\theta, 2, \hat{b}, k} \BL_{\theta, 2, \hat{p}, k} \frac{k \log k}{n}.
\end{align*}
Under $\H_{0, k} (\theta)$, $\Bias_{\theta, k} (\hat{\psi}_{1}) \lesssim n^{- 1 / 2}$, but because $\BL_{\theta, 2, \hat{b}, k} \BL_{\theta, 2, \hat{p}, k} \equiv \CSBias_{\theta, k} (\hat{\psi}_{1}) \geq \Bias_{\theta, k} (\hat{\psi}_{1})$ by CS inequality, we cannot ensure $\BL_{\theta, 2, \hat{b}, k} \BL_{\theta, 2, \hat{p}, k} \lesssim n^{-1/2}$. Hence one cannot guarantee $\EB_{3, \theta, k} = o \( \frac{\sqrt{k}}{n} \)$, as required in Proposition \ref{prop:master} to ensure the validity of $\hat{\chi}_{3, k} (\hat{\Sigma}_{k}^{-1}; z_{\alpha^{\dag} / 2}, \delta)$ as a test of $\H_{0, k} (\delta)$, without making further unverifiable assumptions on $b, p, \hat{b}, \hat{p}$ or $\zbar_{k}$. A natural solution would be to further reduce the bias of $\hat{\IIFF}_{22 \rightarrow 33, k} (\hat{\Sigma}_{k}^{-1})$ due to estimating $\Sigma_{k}^{-1}$ by $\hat{\Sigma}_{k}^{-1}$, which motivates the following $m$-th order $U$-statistic estimator of $\Bias_{\theta, k} (\hat{\psi}_{1})$:
\begin{equation*}
\begin{split}
& \hat{\IIFF}_{22 \rightarrow mm, k} (\hat{\Sigma}_{k}^{-1}) \coloneqq \sum_{j = 2}^{m} \hat{\IIFF}_{jj, k} (\hat{\Sigma}_{k}^{-1}), \text{ where} \\
\small\hat{\IIFF}_{jj, k} (\hat{\Sigma}_{k}^{-1}) & \coloneqq (-1)^{j} \frac{(n - j)!}{n!} \[ \calE_{\hat{b}, m_{2}} (\zbar_{k}) (O) \]^{\top}_{i_{1}} \left\{ \prod_{s = 3}^{j} \hat{\Sigma}_{k}^{-1} \( \[ S_{bp} \zbar_{k} (X) \zbar_{k} (X)^{\top} \]_{i_{s}} - \hat{\Sigma}_{k} \) \right\} \hat{\Sigma}_{k}^{-1} \[ \calE_{\hat{p}, m_{1}} (\zbar_{k}) (O) \]_{i_{2}}.
\end{split}
\end{equation*}
$\hat{\IIFF}_{22 \rightarrow mm, k} (\hat{\Sigma}_{k}^{-1})$ can further reduce the bias [and $\hat{\IIFF}_{22 \rightarrow mm, k} (\Sigma_{k}^{-1})$ is the $m$-th order influence function of $\Bias_{\theta, k} (\hat{\psi}_{1})$ (see HOIF-related theory in Appendix \ref{app:hoif})]. In particular, we have the following theorem on the statistical properties of $\hat{\IIFF}_{22 \rightarrow mm, k} (\hat{\Sigma}_{k}^{-1})$, similar to Theorem \ref{thm:soif_toif}. Again, the bias and variance bounds under Condition \ref{cond:w} are proved in \citet{liu2017semiparametric}. Under Condition \ref{cond:sw}, see the proof of Lemma \ref{lem:hoif_var} in Appendix \ref{app:var}. Recall from Section \ref{sec:main} that Condition \ref{cond:sw} is weaker than Condition \ref{cond:w}.

\begin{theorem}\label{thm:hoif_stats}
Under the conditions as in Theorem \ref{thm:soif_toif} and $k \log k \vee (k m^{2}) = o (n)$: 
\begin{enumerate}[label = (\arabic*)]
\item The bias and variance of $\hat{\IIFF}_{22 \rightarrow mm, k} (\hat{\Sigma}_{k}^{-1})$ as an estimator of $\Bias_{\theta, k} (\hat{\psi}_{1})$ are of the following order:
\begin{align*}
\left\vert \EB_{\theta, m, k} \right\vert & \coloneqq \E_{\theta} \[ \hat{\IIFF}_{22 \rightarrow mm, k} (\hat{\Sigma}_{k}^{-1}) - \Bias_{\theta, k} (\hat{\psi}_{1}) \] \\
& \lesssim \BL_{\theta, 2, \hat{b}, k} \BL_{\theta, 2, \hat{p}, k} \BL_{\theta, 2, \hat{\Sigma}, k}^{\frac{m - 1}{2}} \lesssim \BL_{\theta, 2, \hat{b}, k} \BL_{\theta, 2, \hat{p}, k} \( \frac{k \log k}{n} \)^{\frac{m - 1}{2}}, \\
\var_{\theta} \[ \hat{\IIFF}_{22 \rightarrow mm, k} (\hat{\Sigma}_{k}^{-1}) \] & \lesssim \frac{1}{n} \left\{ \frac{k}{n} + \BL_{\theta, 2, \hat{b}, k} + \BL_{\theta, 2, \hat{p}, k} + k \BL_{\theta, 2, \hat{b}, k}^{2} \BL_{\theta, 2, \hat{p}, k}^{2} \right\}.
\end{align*}
If, however, Condition \ref{cond:sw} is strengthened to Condition \ref{cond:w},
\begin{align*}
\var_{\theta} \[ \hat{\IIFF}_{22 \rightarrow mm, k} (\hat{\Sigma}_{k}^{-1}) \] \lesssim \frac{1}{n} \left\{ \frac{k}{n} + \BL_{\theta, 2, \hat{b}, k} + \BL_{\theta, 2, \hat{p}, k} \right\}.
\end{align*}
\item Under $\H_{0, k} (\delta)$ and Condition \ref{cond:w},
\begin{equation*}
\frac{\hat{\IIFF}_{22 \rightarrow mm, k} (\hat{\Sigma}_{k}^{-1}) - \E_{\theta}\left[\hat{\IIFF}_{22 \rightarrow mm, k} (\hat{\Sigma}_{k}^{-1})\right]}{\se_{\theta} [\hat{\IIFF}_{22 \rightarrow mm, k} (\hat{\Sigma}_{k}^{-1})]} \overset{d}{\rightarrow} N (0, 1).
\end{equation*}
When we estimate $\se_{\theta} [\hat{\IIFF}_{22 \rightarrow mm, k} (\hat{\Sigma}_{k}^{-1})$ by its consistent bootstrap estimator $\hat{\se} [\hat{\IIFF}_{22 \rightarrow mm, k} (\hat{\Sigma}_{k}^{-1})$ (e.g. by \eqref{bvar:if2233} in Appendix \ref{app:bootstrap} with $\Sigma_{k}$ replaced by $\hat{\Sigma}_{k}$), we also have 
\begin{equation*}
\frac{\hat{\IIFF}_{22 \rightarrow mm, k} (\hat{\Sigma}_{k}^{-1}) - \E_{\theta}\left[\hat{\IIFF}_{22 \rightarrow mm, k} (\hat{\Sigma}_{k}^{-1})\right]}{\hat{\se} [\hat{\IIFF}_{22 \rightarrow mm, k} (\hat{\Sigma}_{k}^{-1})]} \overset{d}{\rightarrow} N (0, 1).
\end{equation*}
\end{enumerate}
\end{theorem}

\begin{remark}
\label{rem:comp}
First, note that under $\H_{0, k} (\delta)$ instead of $\H_{0, \CS} (\delta)$ as in the previous section, we only have asymptotic normality of $\hat{\IIFF}_{22 \rightarrow mm, k} (\hat{\Sigma}_{k}^{-1})$ under Condition \ref{cond:w}. Under Condition \ref{cond:sw}, if we were able to show $\BL_{\theta, 4, \hat{b}, k}^{2} \BL_{\theta, 4, \hat{p}, k}^{2} = O (1)$ when $\BL_{\theta, 2, \hat{b}}, \BL_{\theta, 2, \hat{p}}, \BL_{\theta, \infty, \hat{b}}, \BL_{\theta, \infty, \hat{p}}$ are all $O (1)$, then asymptotic normality would also hold. This remains an open question. For more detailed discussion, see Appendix \ref{app:var}.

Further note the following trade-off between the asymptotic statistical properties and computational cost: 
\begin{enumerate}
\item As $m$ grows, the bias of $\hat{\IIFF}_{22 \rightarrow mm, k} (\hat{\Sigma}_{k}^{-1})$ decays to 0 at a rate monotonically increasing with $m$.
\item But larger $m$ incurs higher computational cost to compute $\hat{\IIFF}_{22 \rightarrow mm, k} (\hat{\Sigma}_{k}^{-1})$.
\end{enumerate}
We conjecture that such a trade-off is unavoidable but a rigorous proof is beyond the scope of this paper.

Finally, the extra condition $k m^{2} = o (n)$ imposes restrictions on both $k$ and $m$ and is a result of the variance bound of $\hat{\IIFF}_{mm, k} (\hat{\Sigma}_{k}^{-1})$ shown in Lemma \ref{lem:hoif_var}. For example, if one chooses $k = n / (\log n)^{2}$, then $m$ is at most $o (\log n)$.
\end{remark}

\subsection{The proposed assumption-lean test of $\H_{0, k} (\delta)$ and early-stopping}\leavevmode
\label{sec:hierarchy}

Based on the above discussion, we propose the following nominal $\alpha^{\dag}$-level test of $\H_{0, k} (\delta)$: for any fixed $m \geq 3$,
\begin{equation} 
\label{eq:higher_test}
\hat{\chi}_{m, k} (\hat{\Sigma}_{k}^{-1}; z_{\alpha^{\dag} / 2}, \delta) \coloneqq \mathbbm{1} \left\{ \frac{\vert \hat{\IIFF}_{22 \rightarrow mm, k} (\hat{\Sigma}_{k}^{-1}) \vert}{\hat{\se} [\hat{\psi}_{1}]} - z_{\alpha^{\dag} / 2} \frac{\hat{\se} [\hat{\IIFF}_{22 \rightarrow mm, k} (\hat{\Sigma}_{k}^{-1})]}{\hat{\se} [\hat{\psi}_{1}]} > \delta \right\}.
\end{equation}

Theorem \ref{thm:hoif_stats} immediately implies the following:

\begin{theorem}\leavevmode
\label{prop:hoif_test}
Under the conditions in Theorem \ref{thm:hoif_stats} with Condition \ref{cond:w}, $k \mathsf{log}(k) \vee (k m^{2}) = o(n)$ together with the additional restriction 
\begin{equation}\label{eq:add}
\Bias_{\theta, k} (\hat{\psi}_{1}) \neq o \left( \BL_{\theta, 2, \hat{b}, k} \BL_{\theta, 2, \hat{p}, k} \left( \frac{k \log (k)}{n} \right)^{\frac{m - 1}{2}} \right)
\end{equation}
for any given $\delta > 0$, suppose that $\frac{\vert \Bias_{\theta, k} (\hat{\psi}_{1}) \vert}{\se_{\theta} [\hat{\psi}_{1}]} = \gamma$ for some (sequence) $\gamma = \gamma (n)$ (where $\gamma (n)$ can diverge with $n$), then the rejection probability of $\hat{\chi}_{m, k} (\hat{\Sigma}_{k}^{-1}; z_{\alpha^{\dag} / 2}, \delta)$ converges to 
\begin{equation} 
\label{rejection:3}
2 - \Phi \left( z_{\alpha^{\dag} / 2} - \lim_{n \rightarrow \infty} (\gamma - \delta) \frac{\se_{\theta} [\hat{\psi}_{1}]}{\se_{\theta} [\hat{\IIFF}_{22 \rightarrow mm, k} (\hat{\Sigma}_{k}^{-1})]} \right) - \Phi \left( z_{\alpha^{\dag} / 2} + \lim_{n \rightarrow \infty} (\gamma + \delta) \frac{\se_{\theta} [\hat{\psi}_{1}]}{\se_{\theta} [\hat{\IIFF}_{22 \rightarrow mm, k} (\hat{\Sigma}_{k}^{-1})]} \right)
\end{equation}
as $n \rightarrow \infty$. In particular,

\begin{enumerate}[label=(\arabic*)]
\item under $\H_{0, k} (\delta): \gamma \leq \delta$, $\hat{\chi}_{m, k} (\hat{\Sigma}_{k}^{-1}; z_{\alpha^{\dag} / 2}, \delta)$ rejects the null with probability less than or equal to $\alpha^{\dag}$, as $n \rightarrow \infty$;

\item under the following alternative to $\H_{0, k} (\delta)$: $\gamma = \delta + c$, for any diverging sequence $c = c(n) \rightarrow \infty$, $\hat{\chi}_{m, k} (\hat{\Sigma}_{k}^{-1}; z_{\alpha^{\dag} / 2}, \delta)$ rejects the null with probability converging to 1, as $n \rightarrow \infty$.
\end{enumerate}

\begin{enumerate}[label=(\arabic*')] 
\setcounter{enumi}{2}
\item[(2')] If $\BL_{\theta, 2, \hat{b}, k}$ and $\BL_{\theta, 2, \hat{p}, k}$ converge to 0, under the following alternative to $\H_{0, k} (\delta)$: $\gamma = \delta + c$, for any fixed $c > 0$ or any diverging sequence $c = c(n) \rightarrow \infty$, $\hat{\chi}_{m, k} (\hat{\Sigma}_{k}^{-1}; z_{\alpha^{\dag} / 2}, \delta)$ has rejection probability converging to 1, as $n \rightarrow \infty$.
\end{enumerate}
\end{theorem}

\begin{remark}
Note that the greater $m$ is, the less restrictive \eqref{eq:add} becomes. In fact we can let $m \equiv m (n) \rightarrow \infty$ as $n \rightarrow \infty$, by which \eqref{eq:add} converges to a null requirement.

However, as stated in Remark \ref{rem:comp}, although increasing $m$ relaxes the restriction \eqref{eq:add}, it incurs higher computational cost. In practice, one might want $m$ to be not too large. In view of such computational concern in practice, we propose the following ``early-stopping'' test that terminates at the smallest $m$ at which it fails to reject $\H_{0, k} (\delta)$ and reports ``failure to reject $\H_{0, k} (\delta)$''.

Finally, notice that we only state Theorem \ref{prop:hoif_test} under the stronger Condition \ref{cond:w}. If weakened to Condition \ref{cond:sw}, even under $\H_{0, k} (\delta)$, $k \BL_{\theta, 2, \hat{b}, k}^{2} \BL_{\theta, 2, \hat{p}, k}^{2}$ may still diverge, and the asymptotic normality of $\hat{\IIFF}_{22 \rightarrow mm, k} (\hat{\Sigma}_{k}^{-1})$ cannot be justified unless we were able to show $\BL_{\theta, 4, \hat{b}, k}^{2} \BL_{\theta, 4, \hat{p}, k}^{2}$ to be bounded, as discussed in Remark \ref{rem:comp} and Appendix \ref{app:var}. Thus we need to change the cutoff from $z_{\alpha^{\dag} / 2}$ in \eqref{eq:higher_test} to $(\alpha^{\dag})^{- 1 / 2}$ and justify the level of the test using Chebyshev inequality instead. 


\end{remark}

Formally, for a fixed integer $M > 0$ that is determined by the analyst's computational budget, define 
\begin{equation}
\label{eq:early_test}
\hat{\chi}_{M, k}^{es} (\hat{\Sigma}_{k}^{-1}; z_{\alpha^{\dag} / 2}, \delta) \coloneqq \mathbbm{1} \left\{ \frac{\vert \hat{\IIFF}_{22 \rightarrow mm, k} (\hat{\Sigma}_{k}^{-1}) \vert}{\hat{\se} [\hat{\psi}_{1}]} - z_{\alpha^{\dag} / 2} \frac{\hat{\se} [\hat{\IIFF}_{22 \rightarrow mm, k} (\hat{\Sigma}_{k}^{-1})]}{\hat{\se} [\hat{\psi}_{1}]} > \delta: \forall \ m = 2, \ldots, M \right\}.
\end{equation}
If $\hat{\chi}_{M, k}^{es} (\hat{\Sigma}_{k}^{-1}; z_{\alpha^{\dag} / 2}, \delta)$ fails to reject $\H_{0, k} (\delta)$ at some $m \leq M$, we stop the test at $m$ and claim that we fail to reject $\H_{0, k} (\delta)$. This ``early stopping'' procedure is again an asymptotically valid $\alpha^{\dag}$-level test, under slightly stronger assumptions than those in Theorem \ref{prop:hoif_test}:

\begin{proposition}
\label{prop:early} 
Assume all the conditions of Theorem \ref{prop:hoif_test}, but with \eqref{eq:add} replaced by 
\begin{equation*}
\Bias_{\theta, k} (\hat{\psi}_{1}) \neq o \left( \BL_{\theta, 2, \hat{b}, k} \BL_{\theta, 2, \hat{p}, k} \left( \frac{k \log (k)}{n} \right)^{\frac{M - 1}{2}} \right).
\end{equation*}
Under $\H_{0, k} (\delta)$: 
\begin{align*}
\P_{\theta} \( \hat{\chi}_{M, k}^{es} (\hat{\Sigma}_{k}^{-1}; z_{\alpha^{\dag} / 2}, \delta) = 1 \) \leq \alpha^{\dag}
\end{align*}
that is, $\hat{\chi}_{M, k}^{es} (\hat{\Sigma}_{k}^{-1}; z_{\alpha^{\dag} / 2}, \delta)$ is an asymptotically valid level-$\alpha^{\dag}$ test of $\H_{0, k} (\delta)$.
\end{proposition}

The power of the above ``early-stopping'' procedure is more challenging to characterize, which we leave as future work.

\section{Monte Carlo experiments}
\label{sec:simulation} 
In the Monte Carlo (MC) experiments, we focus on the parameter (up to a minus sign) that was extensively discussed in the Introduction, $\psi (\theta) \equiv \E_{\theta} [Y (a = 1)] \equiv \E_{\theta} [b (X)]$ under ignorability. Recall that $p(X) = 1 / \E_{\theta} [A | X]$ is the inverse propensity score and $b (X) = \E_{\theta} [Y | A = 1, X]$ is the conditional mean of the outcome in the treatment group. For this parameter, $\Sigma_{k} = \E_{\theta} [A \zbar_{k}(X) \zbar_{k}(X)^{\top}]$ and $\hat{\Sigma}_{k} = n^{-1} \sum_{i \in \tr} A_{i} \zbar_{k} (X_{i}) \zbar_{k} (X_{i})^{\top}$. We choose $\psi (\theta) \equiv 0$.

We consider two simulation setups. In simulation setup I, we draw $N=100,000$ i.i.d. $X_{j}$ for $j = 1, \ldots, 4$ (so $d = 4$). The marginal density $f_{j}$ of each $X_{j}$ is supported on $[0, 1]$ with $f_{j} \in \Holder (0.1 + c)$ for some small $c > 0$, as defined in Appendix \ref{app:simulations}. The correlation between each pair of $X_{j}$ and $X_{k}$, with $j \neq k$, is introduced based on the algorithm described in Appendix \ref{app:multiX}. We then simulate $Y$ and $A$ according to the following data generating mechanism: 
\begin{equation*}
Y \sim b (X) + N (0, 1) = \sum_{j = 1}^{4} \tau_{b, j} h_{b} (X_{j}; 0.25) + N (0, 1)
\end{equation*}
and 
\begin{equation*}
A \sim \mathsf{Bernoulli} \left( 1 / p (X) \equiv \mathsf{expit} \left\{ \sum_{j = 1}^{4} \tau_{p, j} h_{p} (X_{j}; 0.25) \right\} \right)
\end{equation*}
where $h_{b} (\cdot; 0.25)$ and $h_{p} (\cdot; 0.25)$ have the forms as defined in Appendix \ref{app:simulations} and hence both belong to $\Holder (0.25 + c)$ for some very small $c > 0$. The numerical values for $\left( \tau_{b, j}, \tau_{p, j} \right)_{j = 1}^{4}$ are provided in Table \ref{tab:s1}. We fix half of the $N = 100,000$ samples as the training sample so $n_{\tr} = n = 50,000$ and only consider the randomness from the estimation sample in the simulation. In simulation setup II, we consider the same data generating mechanism as in setup I except that we choose $b (X) = \sum_{j = 1}^{4} \tau_{b, j} h_{b} (X_{j}; 0.6)$ and $1 / p (X) \equiv \mathsf{expit} \left\{ \sum_{j = 1}^{4} \tau_{p, j} h_{p} (X_{j}; 0.6) \right\}$ where $h_{b} (\cdot; 0.6)$ and $h_{p} (\cdot; 0.6)$ have the forms as defined in Appendix \ref{app:simulations} and hence both belong to $\Holder (0.6 + c)$ for some very small $c > 0$.

We choose D12 (or equivalently db6) Daubechies wavelets at resolutions $\ell \in (6, 7, 8)$ to form the dictionary 
\begin{equation*}
\zbar_{k} (X) = (\zbar_{k'} (X_{1})^{\top}, \zbar_{k'} (X_{2})^{\top}, \zbar_{k'} (X_{3})^{\top}, \zbar_{k'} (X_{4})^{\top})^{\top},
\end{equation*}
with the corresponding $k' \in \{2^{6} = 64, 2^{7} = 128, 2^{8} = 256\}$ and $k \in \{64 \cdot 4 = 256, 128 \cdot 4 = 512, 256 \cdot 4 = 1024\}$. To compute the oracle statistics and tests, we evaluate $\Sigma _{k}$ through MC integration by simulating $L = 10^{7}$ independent $(A, X)$ from the true data generating law. To investigate the finite sample performance of the statistical procedures developed in this article, all the summary statistics of the MC experiments are calculated based on 100 replicates. We estimate the nuisance functions $1 / p (x)$ and $b (x)$ using generalized additive models (GAMs). In particular, the smoothing parameters were selected by generalized cross validation, the default setup in $\mathsf{gam}$ function from $\mathsf{R}$ package $\mathsf{mgcv}$. We choose db6 father wavelets to construct HOIF estimators when analyzing the simulated data.

\subsection{Finite sample performance of tests for $\H_{0, k} (\delta)$}
\label{sec:sim_bias} 

In this section, we consider testing the null hypothesis $\H_{0, k} (\delta)$. Henceforth we investigate the finite sample performance of the oracle statistics and tests $\hat{\IIFF}_{22, k} (\Sigma_{k}^{-1})$, $\hat{\psi}_{2, k} (\Sigma_{k}^{-1})$, and $\hat{\chi}_{2, k} (\Sigma_{k}^{-1})$ where $\hat{\psi}_{2, k} (\Sigma_{k}^{-1}) = \hat{\psi}_{1} - \hat{\IIFF}_{22, k} (\Sigma_{k}^{-1})$, together with the statistics and tests relying on $\hat{\Sigma}_{k}^{-1}$: $\hat{\IIFF}_{22, k} (\hat{\Sigma}_{k}^{-1})$, $\hat{\IIFF}_{33, k} (\hat{\Sigma}_{k}^{-1})$, $\hat{\IIFF}_{22 \rightarrow 33, k} (\hat{\Sigma}_{k}^{-1})$, $\hat{\chi}_{3, k} (\hat{\Sigma}_{k}^{-1})$, $\hat{\psi}_{2, k} (\hat{\Sigma}_{k}^{-1})$, $\hat{\psi}_{3, k} (\hat{\Sigma}_{k}^{-1})$, and $\hat{\chi}_{2, k} (\hat{\Sigma}_{k}^{-1})$, where $\hat{\psi}_{2, k} (\hat{\Sigma}_{k}^{-1}) = \hat{\psi}_{1} - \hat{\IIFF}_{22, k} (\hat{\Sigma}_{k}^{-1})$ and $\hat{\psi}_{2, k} (\hat{\Sigma}_{k}^{-1}) = \hat{\psi}_{1} - \hat{\IIFF}_{22 \rightarrow 33, k} (\hat{\Sigma}_{k}^{-1})$

\begin{figure}[H]
\centering
\includegraphics[width=0.8\textwidth]{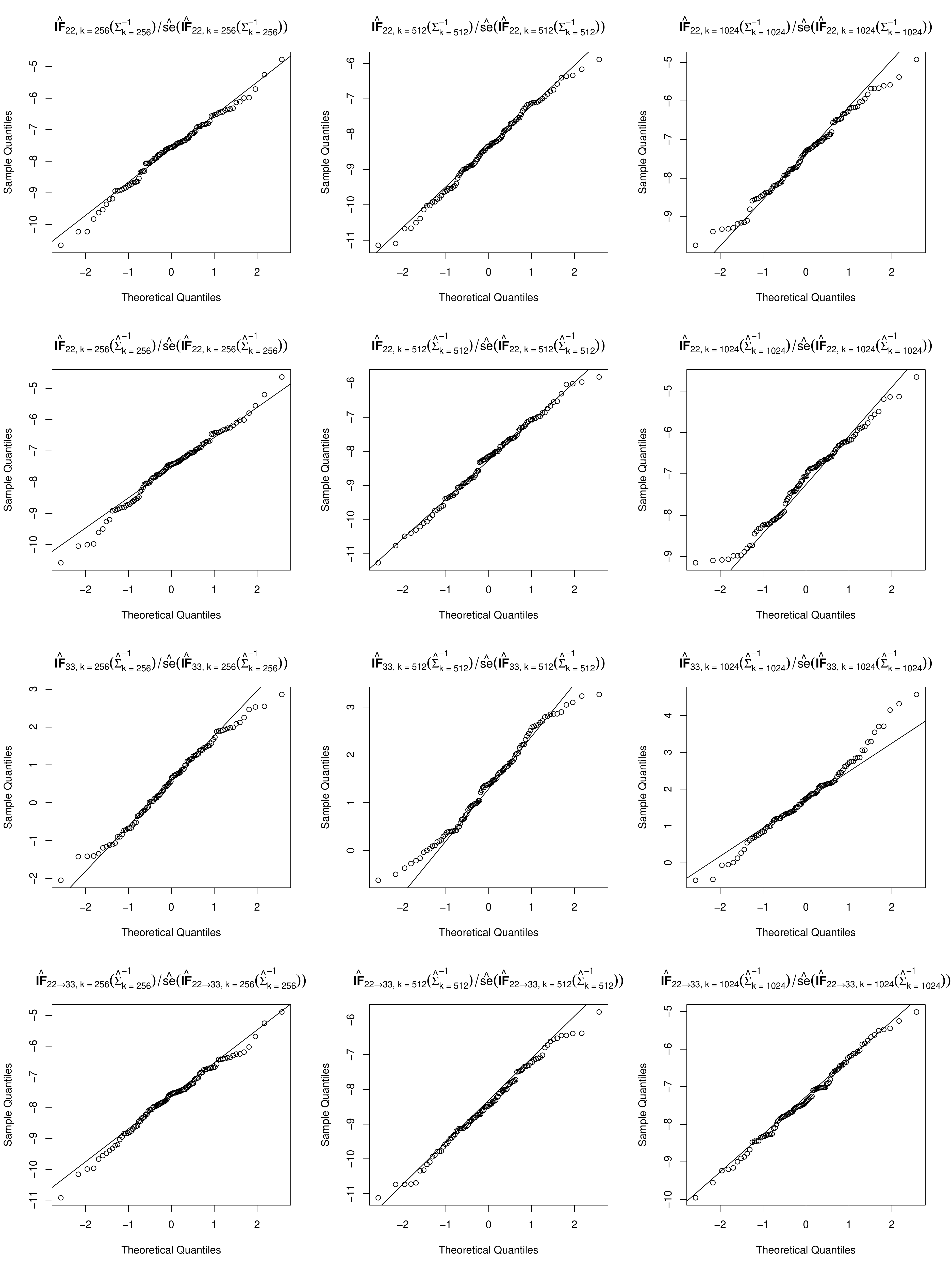}
\caption{For data generating mechanism in simulation setup I: normal qq-plots for $\hat{\IIFF}_{22, k} (\Sigma_{k}^{-1}) / \hat{\se} [\hat{\IIFF}_{22, k} (\Sigma_{k}^{-1})]$ (the first row), $\hat{\IIFF}_{22, k} (\hat{\Sigma}_{k}^{-1}) / \hat{\se} [\hat{\IIFF}_{22, k} (\hat{\Sigma}_{k}^{-1})]$ (the second row), $\hat{\IIFF}_{33, k} (\hat{\Sigma}_{k}^{-1}) / \hat{\se} [\hat{\IIFF}_{33, k} (\hat{\Sigma}_{k}^{-1})]$ (the third row) and $\hat{\IIFF}_{22 \rightarrow 33, k} (\hat{\Sigma}_{k}^{-1}) / \hat{\se} [\hat{\IIFF}_{22 \rightarrow 33, k} (\hat{\Sigma}_{k}^{-1})]$ (the fourth row) for $k = 256$ (left panels), $k = 512$ (middle panels) and $k = 1024$ (right panels).}
\label{fig:qq_ha}
\end{figure}

First, we check the asymptotic normalities of $\frac{\hat{\IIFF}_{22, k} (\Sigma_{k}^{-1})}{\hat{\se} [\hat{\IIFF}_{22, k} (\Sigma_{k}^{-1})]}$, $\frac{\hat{\IIFF}_{22, k} (\hat{\Sigma}_{k}^{-1})}{\hat{\se} [\hat{\IIFF}_{22, k} (\hat{\Sigma}_{k}^{-1})]}$, $\frac{\hat{\IIFF}_{33, k} (\hat{\Sigma}_{k}^{-1})}{\hat{\se} [\hat{\IIFF}_{33, k} (\hat{\Sigma}_{k}^{-1})]}$, and $\frac{\hat{\IIFF}_{22 \rightarrow 33, k} (\hat{\Sigma}_{k}^{-1})}{\hat{\se} [\hat{\IIFF}_{22 \rightarrow 33, k} (\hat{\Sigma}_{k}^{-1})]}$ through normal qq-plots displayed in Figures \ref{fig:qq_ha} (simulation setup I) and \ref{fig:qq_h0} (simulation setup II). We observe that the distributions of most of these statistics are close to normal at different $k$'s ($k = 256$: left panels; $k = 512$: middle panels; $k = 1024$: right panels).

In the simulation, we use nonparametric bootstrap to estimate the standard errors of $\hat{\IIFF}_{22, k} (\Sigma_{k}^{-1})$, $\hat{\IIFF}_{22, k} (\hat{\Sigma}_{k}^{-1})$, $\hat{\IIFF}_{33, k} (\hat{\Sigma}_{k}^{-1})$ and $\hat{\IIFF}_{22 \rightarrow 33, k} (\hat{\Sigma}_{k}^{-1})$, as described in Appendix \ref{app:bootstrap}. We study if the estimated standard errors of $\hat{\IIFF}_{22, k} (\Sigma_{k}^{-1}) $, $\hat{\IIFF}_{22, k} (\hat{\Sigma}_{k}^{-1})$, $\hat{\IIFF}_{33, k} (\hat{\Sigma}_{k}^{-1})$, and $\hat{\IIFF}_{22 \rightarrow 33, k} (\hat{\Sigma}_{k}^{-1})$ by nonparametric bootstrap are close to their true standard errors (calibrated by the MC standard deviations from 100 replicates in the simulation). We use $B = 100$ bootstrap samples to compute the bootstrapped standard errors as the estimated standard errors for all four statistics. In Tables \ref{tab:var_ha} (simulation setup I) and \ref{tab:var_h0} (simulation setup II), we display the MC standard deviations (the upper numerical values in each cell), accompanied with the MC averages of the estimated standard errors (the lower numerical values outside the parenthesis in each cell) and \textit{MC standard deviations} of the estimated standard errors (the lower numerical values inside the parenthesis in each cell) of all three statistics at $k = 256$ (left panel), $k = 512$ (middle panel) and $k = 1024$ (right panel). From Table \ref{tab:var_ha} and Table \ref{tab:var_h0}, we observe that the estimated standard errors only slightly differ from the MC standard deviations.

Then we investigate (1) the finite sample performance of $\hat{\IIFF}_{22, k} (\Sigma_{k}^{-1})$, $\hat{\IIFF}_{22, k} (\hat{\Sigma}_{k}^{-1})$, and $\hat{\IIFF}_{22 \rightarrow 33, k} (\hat{\Sigma}_{k}^{-1})$ and evaluate how close they are to $\Bias_{\theta} (\hat{\psi}_{1})$, which is evaluated based on the MC bias of 100 replicates, and (2) the rejection rate of the tests $\hat{\chi}_{2, k} (\Sigma_{k}^{-1}; z_{0.10 / 2}, \delta)$, $\hat{\chi}_{2, k} (\hat{\Sigma}_{k}^{-1}; z_{0.10 / 2}, \delta)$ and $\hat{\chi}_{3, k} (\Sigma_{k}^{-1}; z_{0.10 / 2}, \delta)$ for the null hypothesis $\H_{0, k} (\delta)$. The numerical results are shown in Tables \ref{tab:main_ha} (simulation setup I) and \ref{tab:main_h0} (simulation setup II).

\begin{itemize}
\item In the upper panel of Table \ref{tab:main_ha} (simulation setup I), the first row and the third column, we display the MC bias of the DML estimator $\hat{\psi}_{1}$ and the MC average of $\hat{\se}[\hat{\psi}_{1}]$, which are $-34.26 \times 10^{-3}$ and $8.77 \times 10^{-3}$. Since the ratio between the bias and standard error is around 4, we expect the associated 90\% Wald CI $\hat{\psi}_{1} \pm z_{0.05} \hat{\se}(\hat{\psi}_{1})$ does not have the nominal coverage. This is indeed the case by reading from the first row, the second column of the upper panel of Table \ref{tab:main_ha}, showing the MC coverage probability for the 90\% Wald CI of $\hat{\psi}_{1}$ is 0\%.

Similarly, in the upper panel of Table \ref{tab:main_h0} (simulation setup II), the first row and the third column, we display the MC bias of the DML estimator $\hat{\psi}_{1}$ and the MC average of $\hat{\se}[\hat{\psi}_{1}]$, which are $-8.88 \times 10^{-3}$ and $8.10 \times 10^{-3}$. This is as expected because the true nuisance functions $b$ and $p$ belong to $\Holder (0.6 + c)$ for some $c > 0$ and DML estimator (with $b$ and $p$ estimated in optimal rate in $L_{2}$ norm) is expected to have bias of $o (n^{-1/2})$ when the average smoothness between $b$ and $p$ is above 0.5 \citep{robins2009semiparametric}. Here the ratio between the bias and standard error is around 1 and we expect the associated 90\% Wald CI $\hat{\psi}_{1} \pm z_{0.05} \hat{\se}(\hat{\psi}_{1})$ is slightly undercovered. This is indeed the case by reading from the first row, the second column of the upper panel of Table \ref{tab:main_h0}, showing the MC coverage probability for the 90\% Wald CI of $\hat{\psi}_{1}$ is 83\% (note that $\hat{\psi}_{2, k} = \hat{\psi}_{1}$ when $k = 0$).

\item In the second column of the upper panel of Table \ref{tab:main_ha}, we display the MC averages of $\hat{\IIFF}_{22, k} (\Sigma_{k}^{-1})$ and the MC averages of its estimated standard errors; in the middle column, we display those of $\hat{\IIFF}_{22, k} (\hat{\Sigma}_{k}^{-1})$; and in the lower panel, we display those of $\hat{\IIFF}_{22 \rightarrow 33, k} (\hat{\Sigma}_{k}^{-1})$ after third-order bias correction. In the upper panel, we observe that increasing $k$ from 256 to 512 does improve the amount of bias recovered by $\hat{\IIFF}_{22, k} (\Sigma_{k}^{-1})$ (from $-18.76 \times 10^{-3}$ to $- 25.34 \times 10^{-3}$), but there is no obvious difference in the MC averages of $\hat{\IIFF}_{22, k} (\Sigma_{k}^{-1})$ between $k = 512$ and $k = 1024$ ($- 25.34 \times 10^{-3}$ and $- 25.43 \times 10^{-3}$, about 75\% of the total bias). The MC average of the estimated standard error increases with $k$, from $2.45 \times 10^{-3}$ at $k = 256$ to $3.43 \times 10^{-3}$ at $k = 1024$. This is consistent with the theoretical prediction that the variability of $\hat{\IIFF}_{22, k} (\Sigma_{k}^{-1})$ should grow linearly with $k$. In the middle, we observe very similar numerical results between $\hat{\IIFF}_{22, k} (\Sigma_{k}^{-1})$ and $\hat{\IIFF}_{22, k} (\hat{\Sigma}_{k}^{-1})$ for all the statistics that we are interested in. Interestingly, we did not see much improvement of using $\hat{\IIFF}_{22 \rightarrow 33, k} (\hat{\Sigma}_{k}^{-1})$ instead of $\hat{\IIFF}_{22, k} (\hat{\Sigma}_{k}^{-1})$, when compared to $\hat{\IIFF}_{22, k} (\Sigma_{k}^{-1})$, at least in this example.

In the second column of the upper panel of Table \ref{tab:main_h0}, in the upper panel, we also observe that increasing $k$ from 256 to 512 also slightly improve the amount of bias recovered by $\hat{\IIFF}_{22, k} (\Sigma_{k}^{-1})$ (from $-4.54 \times 10^{-3}$ to $- 4.94 \times 10^{-3}$). The MC average of the estimated standard error increases with $k$, from $1.51 \times 10^{-3}$ at $k = 256$ to $2.43 \times 10^{-3}$ at $k = 1024$. In the middle panel, we observe very similar numerical results between $\hat{\IIFF}_{22, k} (\Sigma_{k}^{-1})$ and $\hat{\IIFF}_{22, k} (\hat{\Sigma}_{k}^{-1})$ for all the statistics that we are interested in. Interestingly, we again did not see much improvement of using $\hat{\IIFF}_{22 \rightarrow 33, k} (\hat{\Sigma}_{k}^{-1})$ instead of $\hat{\IIFF}_{22, k} (\hat{\Sigma}_{k}^{-1})$, when compared to $\hat{\IIFF}_{22, k} (\Sigma_{k}^{-1})$, at least in this example.

\item In the third columns of Table \ref{tab:main_ha}, we display the MC coverage probabilities of the 90\% two-sided CIs of $\hat{\psi}_{2, k} (\Sigma_{k}^{-1})$ (upper panel) and $\hat{\psi}_{3, k} (\hat{\Sigma}_{k}^{-1})$ (lower panel). Comparing the oracle bias-corrected estimator $\hat{\psi}_{2, k} (\Sigma_{k}^{-1})$ vs. $\hat{\psi}_{1}$, the coverage probability improves from 0\% to 33\% (82\%) at $k = 256$ (at $k = 1024$). Similar observations can be made for $\hat{\psi}_{2, k} (\hat{\Sigma}_{k}^{-1})$ and $\hat{\psi}_{3, k} (\hat{\Sigma}_{k}^{-1})$ as well when $\Sigma_{k}$ is unknown.

In the third columns of Table \ref{tab:main_h0}, we display the MC coverage probabilities of the 90\% two-sided CIs of $\hat{\psi}_{2, k} (\Sigma_{k}^{-1})$ (upper panel) and $\hat{\psi}_{3, k} (\hat{\Sigma}_{k}^{-1})$ (lower panel). Comparing the oracle bias-corrected estimator $\hat{\psi}_{2, k} (\Sigma_{k}^{-1})$ vs. $\hat{\psi}_{1}$, the coverage probability improves from 83\% to 100\% at $k = 256$ and $k = 1024$. Similar observations can be made for $\hat{\psi}_{2, k} (\hat{\Sigma}_{k}^{-1})$ and $\hat{\psi}_{3, k} (\hat{\Sigma}_{k}^{-1})$ as well when $\Sigma_{k}$ is unknown.

\item In the fourth columns of Tables \ref{tab:main_ha} and \ref{tab:main_h0}, we display the MC biases of $\hat{\psi}_{2, k} (\Sigma_{k}^{-1})$ (upper panel) and $\hat{\psi}_{3, k} (\hat{\Sigma}_{k}^{-1})$ (lower panel), together with their MC standard deviations (in the parentheses). As expected, the standard deviations of the estimators after bias correction are very similar to those of $\hat{\psi}_{1}$. This indeed confirms that we are able to correct bias without drastically inflating the variance.

\item In the fifth column of Table \ref{tab:main_ha}, we display the MC rejection rates of the test statistic: the upper panel shows the MC rejection rates of the oracle test $\hat{\chi}_{2, k} (\Sigma_{k}^{-1}; z_{0.10 / 2}, \delta = 3 / 4)$ and the lower panel shows the MC rejection rates of the test $\hat{\chi}_{3, k} (\hat{\Sigma}_{k}^{-1}; z_{0.10 / 2}, \delta = 3 / 4)$. These simulation results demonstrate that when a ``good'' dictionary (db6 father wavelets) is used, our proposed test does have power to reject the null hypothesis of actual interest $\H_{0} (\delta)$ that the ratio between the bias of $\hat{\psi}_{1}$ and its standard error is lower than $\delta$. All the rejection rates in Table \ref{tab:main_ha} are 100\% when $\delta = 3 / 4$, regardless of whether we are using the oracle test or not. When $\delta = 2$, we obtained nontrivial rejection rates (data not shown). This is as expected because the ratio between $\hat{\IIFF}_{22, k}$'s and $\hat{\se} (\hat{\psi}_{1})$ is around 2 to 3. In the fifth column of Table \ref{tab:main_h0}, all the rejection rates are 0\% when $\delta = 3 / 4$. This is also as expected because the ratio between $\hat{\IIFF}_{22, k}$'s and $\hat{\se} (\hat{\psi}_{1}) $ is close to $1 / 2$.
\end{itemize}

\begin{remark}
Here we choose $k \lesssim n / (\log n)^{2}$, which is on the order of 1000 when $n = 50,000$. Larger $k$ could easily lead to numerical instability due to inverting a large-dimensional sample Gram matrix (also see \citet{liu2020nearly}). In a technical report by one of the authors \citep{liu2023hoif}, a new class of empirical HOIF estimators is proposed that overcomes the above issues by replacing $\hat{\Sigma}_{k}$ computed from the training sample by that computed from the estimation sample. The theoretical results are somewhat more complicated to state than the estimators used in this paper. We will thus only refer interested readers to \citet{liu2023hoif}. In a separate manuscript \citep{wanis2023falsification}, we report the finite-sample and real-world data performance of these new estimators and the corresponding test statistics, together with a data-driven method for choosing $k$ in practice. We also refer readers to~\citet{breunig2020adaptive} and \citet{liu2021adaptive} for data-driven methods of choosing $k$ in a theoretically-oriented manner.
\end{remark}

\begin{table}[tbp]
\centering
\begin{tabular}{c|c|c|c}
\hline
$k$ & 256 & 512 & 1024 \\ 
\hline
$\hat{\IIFF}_{22, k} (\Sigma_{k}^{-1})$ & \begin{tabular}{@{}c}
2.437 \\ 
2.268 (0.218)
\end{tabular} & \begin{tabular}{@{}c}
3.011 \\ 
2.716 (0.268)
\end{tabular} & \begin{tabular}{@{}c}
3.247 \\ 
2.841 (0.322)
\end{tabular} \\ 
\hline
$\hat{\IIFF}_{22, k} (\hat{\Sigma}_{k}^{-1})$ & \begin{tabular}{@{}c}
2.456 \\ 
2.271 (0.218)
\end{tabular} & \begin{tabular}{@{}c}
3.075 \\ 
2.778 (0.278)
\end{tabular} & \begin{tabular}{@{}c}
3.546 \\ 
3.032 (0.378)
\end{tabular} \\ 
\hline
$\hat{\IIFF}_{33, k} (\hat{\Sigma}_{k}^{-1})$ & \begin{tabular}{@{}c}
0.495 \\ 
0.603 (0.0598)
\end{tabular} & \begin{tabular}{@{}c}
0.814 \\ 
1.126 (0.126)
\end{tabular} & \begin{tabular}{@{}c}
1.518 \\ 
1.998 (0.285)
\end{tabular} \\ 
\hline
$\hat{\IIFF}_{22 \rightarrow 33, k} (\hat{\Sigma}_{k}^{-1})$ & \begin{tabular}{@{}c}
2.282 \\ 
2.251 (0.227)
\end{tabular} & \begin{tabular}{@{}c}
2.745 \\ 
2.738 (0.290)
\end{tabular} & \begin{tabular}{@{}c}
2.743 \\ 
3.009 (0.427)
\end{tabular} \\ 
\hline
\end{tabular}
\caption{For data generating mechanism in simulation setup I: We reported the MC standard deviations $\times 10^{-3}$ (upper values in each cell), the MC averages of the estimated standard errors $\times 10^{-3}$ (lower values in each cell outside the parenthesis) and the MC standard deviations of the estimated standard errors $\times 10^{-3}$ (lower values in each cell inside the parenthesis) through nonparametric bootstrap resampling for $\hat{\IIFF}_{22, k} (\Sigma_{k}^{-1})$, $\hat{\IIFF}_{22, k} (\hat{\Sigma}_{k}^{-1})$, $\hat{\IIFF}_{33, k} (\hat{\Sigma}_{k}^{-1})$ and $\hat{\IIFF}_{22 \rightarrow 33, k} (\hat{\Sigma}_{k}^{-1})$.}
\label{tab:var_ha}
\end{table}

\begin{table}[tbp]
\resizebox{\columnwidth}{!}{
\begin{tabular}{c|c|c|c|c}
\hline
k & $\hat{\IIFF}_{22, k} (\Sigma_{k}^{-1}) \times 10^{-3}$ & \shortstack{MC Coverage \\ ($\hat{\psi}_{2, k} (\Sigma_{k}^{-1})$ 90\% Wald CI)} & $\Bias (\hat{\psi}_{2, k} (\Sigma_{k}^{-1})) \times 10^{-3}$ & \shortstack{$\hat{\chi}_{2, k} (\Sigma_{k}^{-1}; z_{0.10 / 2}, \delta = 3 / 4)$} \\
\hline
$0$ & $\mathsf{NA}$ ($\mathsf{NA}$) & 0\% & -34.26 (8.77) & $\mathsf{NA}$ \\
$256$ & -18.76 (2.27) & 31\% & -15.50 (8.60) & 100\% \\
$512$ & -25.34 (2.72) & 83\% & -8.92 (8.61) & 100\% \\
$1024$ & -25.43 (2.84) & 82\% & -8.83 (8.79) & 100\% \\
\hline
\hline
k & $\hat{\IIFF}_{22 \rightarrow 33, k} (\hat{\Sigma}_{k}^{-1}) \times 10^{-3}$ & \shortstack{MC Coverage \\ ($\hat{\psi}_{3, k} (\hat{\Sigma}_{k}^{-1})$ 90\% Wald CI)} & $\Bias (\hat{\psi}_{3, k} (\hat{\Sigma}_{k}^{-1})) \times 10^{-3}$ & \shortstack{$\hat{\chi}_{3, k} (\hat{\Sigma}_{k}^{-1}; z_{0.10 / 2}, \delta = 3 / 4)$} \\
\hline
$256$ & -18.54 (2.25) & 30\% & -15.72 (8.60) & 100\% \\
$512$ & -24.65 (2.74) & 81\% & -9.61 (8.60) & 100\% \\
$1024$ & -23.82 (3.01) & 76\% & -10.44 (8.76) & 100\% \\ 
\hline
\end{tabular}}
\caption{For data generating mechanism in simulation setup I: We reported the MC averages of point estimates and standard errors (the first column in each panel) of $\hat{\IIFF}_{22, k} (\Sigma_{k}^{-1})$ (upper panel) and $\hat{\IIFF}_{22 \rightarrow 33, k} (\hat{\Sigma}_{k}^{-1})$ (lower panel), together with the coverage probabilities of two-sided 90\% Wald CIs (the second column in each panel) of $\hat{\psi}_{2, k} (\Sigma_{k}^{-1})$ (upper panel) and $\hat{\psi}_{3, k} (\hat{\Sigma}_{k}^{-1})$ (lower panel), the MC biases and MC standard deviations (the third column in each panel) of $\hat{\psi}_{2, k} (\Sigma_{k}^{-1})$ (upper panel) and $\hat{\psi}_{3, k} (\hat{\Sigma}_{k}^{-1})$ (lower panel) and the empirical rejection rates (the fourth column in each panel) of $\hat{\chi}_{2, k}^{(1)} (\Sigma_{k}^{-1}; z_{0.10 / 2}, \delta = 3 / 4)$ (upper panel) and $\hat{\chi}_{3, k}^{(1)} (\hat{\Sigma}_{k}^{-1}; z_{0.10 / 2}, \delta = 3 / 4)$ (lower panel). In the upper panel, for $k = 0$, $\hat{\psi}_{2, k = 0} \equiv \hat{\psi}_{1}$.}
\label{tab:main_ha}
\end{table}

\begin{table}[tbp]
\centering
\begin{tabular}{c|c|c|c}
\hline
$k$ & 256 & 512 & 1024 \\ 
\hline
$\hat{\IIFF}_{22, k} (\Sigma_{k}^{-1})$ & \begin{tabular}{@{}c}
1.166 \\ 
1.209 (0.156)
\end{tabular} & \begin{tabular}{@{}c}
1.367 \\ 
1.408 (0.191)
\end{tabular} & \begin{tabular}{@{}c}
1.673 \\ 
1.584 (0.267)
\end{tabular} \\ 
\hline
$\hat{\IIFF}_{22, k} (\hat{\Sigma}_{k}^{-1})$ & \begin{tabular}{@{}c}
1.187 \\ 
1.219 (0.159)
\end{tabular} & \begin{tabular}{@{}c}
1.433 \\ 
1.439 (0.198)
\end{tabular} & \begin{tabular}{@{}c}
1.828 \\ 
1.697 (0.285)
\end{tabular} \\ 
\hline
$\hat{\IIFF}_{33, k} (\hat{\Sigma}_{k}^{-1})$ & \begin{tabular}{@{}c}
0.251 \\ 
0.331 (0.0451)
\end{tabular} & \begin{tabular}{@{}c}
0.417 \\ 
0.629 (0.0997)
\end{tabular} & \begin{tabular}{@{}c}
0.854 \\ 
1.251 (0.247)
\end{tabular} \\ 
\hline
$\hat{\IIFF}_{22 \rightarrow 33, k} (\hat{\Sigma}_{k}^{-1})$ & \begin{tabular}{@{}c}
1.109 \\ 
1.186 (0.156)
\end{tabular} & \begin{tabular}{@{}c}
1.256 \\ 
1.346 (0.191)
\end{tabular} & \begin{tabular}{@{}c}
1.488 \\ 
1.560 (0.383)
\end{tabular} \\ 
\hline
\end{tabular}
\caption{For data generating mechanism in simulation setup II: We reported the MC standard deviations $\times 10^{-3}$ (upper values in each cell), the MC averages of the estimated standard errors $\times 10^{-3}$ (lower values in each cell outside the parenthesis) and the MC standard deviations of the estimated standard errors $\times 10^{-3}$ (lower values in each cell inside the parenthesis) through nonparametric bootstrap resampling for $\hat{\IIFF}_{22, k} (\Sigma_{k}^{-1})$, $\hat{\IIFF}_{22, k} (\hat{\Sigma}_{k}^{-1})$, $\hat{\IIFF}_{33, k} (\hat{\Sigma}_{k}^{-1})$ and $\hat{\IIFF}_{22 \rightarrow 33, k} (\hat{\Sigma}_{k}^{-1})$.}
\label{tab:var_h0}
\end{table}

\begin{table}[tbp]
\resizebox{\columnwidth}{!}{
\begin{tabular}{c|c|c|c|c}
\hline
k & $\hat{\IIFF}_{22, k} (\Sigma_{k}^{-1}) \times 10^{-3}$ & \shortstack{MC Coverage \\ ($\hat{\psi}_{2, k} (\Sigma_{k}^{-1})$ 90\% Wald CI)} & $\Bias (\hat{\psi}_{2, k} (\Sigma_{k}^{-1})) \times 10^{-3}$ & \shortstack{$\hat{\chi}_{2, k} (\Sigma_{k}^{-1}; z_{0.10 / 2}, \delta = 3 / 4)$} \\
\hline
$0$ & $\mathsf{NA}$ ($\mathsf{NA}$) & 83\% & -8.88 (8.10) & $\mathsf{NA}$ \\
$256$ & -4.54 (1.21) & 99\% & -4.34 (8.13) & 0\% \\
$512$ & -4.89 (1.41) & 99\% & -3.99 (8.21) & 0\% \\
$1024$ & -4.94 (1.58) & 100\% & -3.94 (8.35) & 0\% \\
\hline
\hline
k & $\hat{\IIFF}_{22 \rightarrow 33, k} (\hat{\Sigma}_{k}^{-1}) \times 10^{-3}$ & \shortstack{MC Coverage \\ ($\hat{\psi}_{3, k} (\hat{\Sigma}_{k}^{-1})$ 90\% Wald CI)} & $\Bias (\hat{\psi}_{3, k} (\hat{\Sigma}_{k}^{-1})) \times 10^{-3}$ & \shortstack{$\hat{\chi}_{3, k} (\hat{\Sigma}_{k}^{-1}; z_{0.10 / 2}, \delta = 3 / 4)$} \\
\hline
$256$ & -4.49 (1.19) & 99\% & -4.39 (8.13) & 0\% \\
$512$ & -4.76 (1.35) & 99\% & -4.12 (8.20) & 0\% \\
$1024$ & -4.62 (1.56) & 100\% & -4.26 (8.32) & 0\% \\ 
\hline
\end{tabular}}
\caption{For data generating mechanism in simulation setup II: We reported the MC averages of point estimates and standard errors (the first column in each panel) of $\hat{\IIFF}_{22, k} (\Sigma_{k}^{-1})$ (upper panel) and $\hat{\IIFF}_{22 \rightarrow 33, k} (\hat{\Sigma}_{k}^{-1})$ (lower panel), together with the coverage probabilities of two-sided 90\% Wald CIs (the second column in each panel) of $\hat{\psi}_{2, k} (\Sigma_{k}^{-1})$ (upper panel) and $\hat{\psi}_{3, k} (\hat{\Sigma}_{k}^{-1})$ (lower panel), the MC biases and MC standard deviations (the third column in each panel) of $\hat{\psi}_{2, k} (\Sigma_{k}^{-1})$ (upper panel) and $\hat{\psi}_{3, k} (\hat{\Sigma}_{k}^{-1})$ (lower panel) and the empirical rejection rates (the fourth column in each panel) of $\hat{\chi}_{2, k}^{(1)} (\Sigma_{k}^{-1}; z_{0.10 / 2}, \delta = 3 / 4)$ (upper panel) and $\hat{\chi}_{3, k}^{(1)} (\hat{\Sigma}_{k}^{-1}; z_{0.10 / 2}, \delta = 3 / 4)$ (lower panel). In the upper panel, for $k = 0$, $\hat{\psi}_{2, k = 0} \equiv \hat{\psi}_{1}$.}
\label{tab:main_h0}
\end{table}

\section{Concluding remarks}
\label{sec:conclusion} 
In this paper, we developed a valid assumption-lean test, based on third-order $U$-statistics, that can empirically falsify the justification for a Wald CI centered at a standard DML estimator $\hat{\psi}_{1}$ covering the underlying DR functional $\psi (\theta)$ at the nominal rate (see Section \ref{sec:main}). When nonstandard DML estimators are used, we also develop a test that is based on higher-order $U$-statistics, which are in fact HOIFs of $\Bias_{\theta, k} (\hat{\psi}_{1})$. We mention a few interesting future directions to end our manuscript. First, it is important to extend the proposed approach to allow endogeneity \citep{angrist1996identification, ai2003efficient, chen2005measurement, newey2003instrumental, ai2007estimation, chen2013optimal, breunig2019simple, chen2018optimal, chen2016methods}, when some auxiliary variables are available for point-identifying the causal effects. Machine learning, including deep learning, has also been applied to these problems in recent years \citep{chen2023efficient, kompa2022deep}. In Appendix \ref{app:prox_hoif}, we document the HOIFs for the running example $\psi (\theta) = - \E_{\theta} [Y (a = 1)]$ in Section \ref{sec:introduction} under the so-called ``proximal causal learning'' framework \citep{tchetgen2020introduction}. This framework has been shown to be closely related to other approaches dealing with endogeneity, e.g. synthetic controls \citep{abadie2010synthetic, shi2021theory} or quadratic functionals of nonparametric instrumental variable regression \citep{breunig2019simple}. Based on the form of these HOIFs, it is straightforward to generalize our method to scenarios under endogeneity. Second, as mentioned earlier in \textbf{Literature Overview}, parameters implicitly defined via (conditional) moment restrictions involving nonparametric nuisance functions include, as special cases, DR functionals and certain parameters related to instrumental variables and proximal causal inference \citep{ai2003efficient, ai2007estimation, ai2012semiparametric}. Developing the theory of higher-order influence functions for these parameters will be a natural next step. Third, extending our framework to heterogeneous treatment effect \citep{chernozhukov2017generic, kennedy2022minimax} could also be of practical interest, given its significance in personalized decision making. Fourth, another interesting direction is to explore if it is possible to use other bias correction strategies to construct the assumption-lean falsification test, such as the (iterative) bootstrap approach investigated in \citet{cattaneo2018kernel, cattaneo2019two}, and \citet{koltchinskii2020estimation}. Finally, we are also considering to directly learn a data-driven representation $\tilde{\zbar}_{k}$ via the penultimate layer \citep{ansuini2019intrinsic, damian2022neural} or distillation \citep{ha2021adaptive} of a deep neural network trained to predict the residuals of the nuisance function estimators, hoping to increase the chance of rejection when the bias or the CS bias indeed exceeds $n^{-1/2}$ in real-world settings.

\printbibliography

\clearpage
\newpage \newgeometry{margin = 0.5in, paperwidth=11in, paperheight=12in} \vfill\eject \pdfpagewidth=11in \pdfpageheight=12in \oddsidemargin +0.2in \evensidemargin +0.0in \topmargin 5pt \linespread{1.5}\parskip .05in 

\allowdisplaybreaks
\appendix
\newrefsection

\section{Derivations and proofs}
\allowdisplaybreaks

\subsection{More detailed review of doubly-robust functionals (DR functionals)}
\label{app:regular}

All the theoretical results in this paper rely on the following regularity condition (Condition 1 in \citet{rotnitzky2021characterization}):
\begin{condition}\label{cond:r}
There exists a dense set $\mathcal{H}_{b}$ of $L_{2} (\P_{F})$ such that $\mathcal{F}_{b} \cap \mathcal{B} \neq \emptyset$, and for each $\theta \in \Theta$ and $h \in \mathcal{H}_{b}$, there exists $\varepsilon (\theta, h) > 0$ such that $b + t h \in \mathcal{B}$ if $|t| < \varepsilon (\theta, h)$. The same holds replacing $b$ with $p$ and $\mathcal{B}$ with $\mathcal{P}$. Furthermore $\E_{\theta} [|S_{bp} p (X) h (X)|] < \infty$ for any $h \in \mathcal{H}_{b}$ and $\E_{\theta} [|S_{bp} b (X) h (X)|] < \infty$ for any $h \in \mathcal{H}_{p}$. Moreover for all $\theta \in \Theta$, $\E_{\theta} [|S_{bp} b' (X) p' (X)|] < \infty$ for any $b' \in \mathcal{B}$ and $p' \in \mathcal{P}$.
\end{condition}

\subsection{More details on training strategies for nuisance functions}
\label{app:loss}
In this section, we discuss training strategies for nuisance functions in further details. First, we prove Lemma \ref{lem:loss}.
\begin{proof}[Proof of Lemma \ref{lem:loss}]
The following algebra makes this statement explicit: 
\begin{align*}
\frac{1}{2} \E_{\theta} [\lambda (X) (b(X) - h(X))^{2}] & = \frac{1}{2} \E_{\theta} [S_{bp} (b(X) - h(X))^{2}] \\
& = \E_{\theta} [S_{bp} \frac{h(X)^{2}}{2} - S_{bp} b(X) h(X)] + \frac{1}{2} \E_{\theta} [S_{bp} b(X)^{2}] \\
& = \E_{\theta} [S_{bp} \frac{h(X)^{2}}{2} + m_{2}(O, h)] + \frac{1}{2} \E_{\theta} [S_{bp} b(X)^{2}]
\end{align*}
where in the third line we use equation \eqref{eq:mean_zero} and the extra term $\frac{1}{2} \E_{\theta} [S_{bp} b(X)^{2}]$ is unrelated to minimizing $\frac{1}{2} \E_{\theta} [\lambda (X) (b(X) - h(X))^{2}]$. By symmetry, 
\begin{equation*}
\frac{1}{2} \E_{\theta} [\lambda(X) (p(X) - h(X))^{2}] = \E_{\theta} [S_{bp} \frac{h(X)^{2}}{2} + m_{1} (O, h)] + \frac{1}{2} \E_{\theta} [S_{bp} p(X)^{2}].
\end{equation*}
\end{proof}

For cross-entropy loss commonly employed in machine learning (in particular deep learning) when the response is binary, it is also approximately solving an $L_{2} (\P_{\theta})$ loss minimization: here we denote $\pi = p^{-1}$ and $\tilde{\pi}$ is the one being fitted in the minimization problem, by using Taylor expansion
\begin{align*}
& \ \E_{\theta} \left[ A \log \tilde{\pi} (X) + (1 - A) \log (1 - \pi (X)) \right] \\
= & \ \E_{\theta} \left[ \pi (X) \log \tilde{\pi} (X) + (1 - \pi (X)) \log (1 - \tilde{\pi} (X)) \right] \\
= & \ \E_{\theta} \left[ \pi (X) \left\{ \log \pi (X) + \frac{\tilde{\pi} (X) - \pi (X)}{\pi (X)} - \frac{(\tilde{\pi} (X) - \pi (X))^{2}}{\pi (X)^{2}} \right\} + (1 - \pi (X)) \left\{ \log (1 - \pi (X)) - \frac{\tilde{\pi} (X) - \pi (X)}{1 - \pi (X)} + \frac{(\tilde{\pi} (X) - \pi (X))^{2}}{(1 - \pi (X))^{2}} \right\} \right] \\
& + \text{remainders} \\
= & \ \text{constant} + \E_{\theta} \left[ (\tilde{\pi} (X) - \pi (X))^{2} \left( \frac{1}{\pi (X)} + \frac{1}{1 - \pi (X)} \right) \right].
\end{align*}

We also want to point out the connection to recent works on using adversarial loss \citep{chernozhukov2020adversarial, ghassami2022minimax} to train the nuisance estimators. Adversarial loss is essentially trying to leverage the moment conditions in equation \eqref{eq:mean_zero}. As shown in Theorem 1 of \citet{ghassami2022minimax}, the solution to the population adversarial loss with $L_{2}$-regularization is equivalent to the solution to the loss functions in equation \eqref{loss} and they establish the convergence rates exactly in the $\lambda$-weighted $L_{2}$-loss. Therefore, at least in principle, our testing framework is also applicable when nuisance functions are trained using adversarial loss according to \citet{ghassami2022minimax} and \citet{chernozhukov2020adversarial} with an additional $L_{2}$-regularization term.

\subsection{The infeasible procedure when $\Sigma_{k}^{-1} \coloneqq \{\E_{\theta} [\lambda (X) \zbar_{k} (X) \zbar_{k} (X)^{\top}]\}^{-1}$ is known}
\label{app:oracle} 
It is pedagogically useful to first consider an infeasible procedure that would only be implementable if $\Sigma_{k}^{-1}$ were known. We will focus on such situation in this section.

\subsubsection{Truncated parameters and their unbiased estimators}
\label{sec:hoif_zkpsi} 

Though there does not exist uniformly (in $\theta$) consistent test of the null hypothesis that the bias $\Bias_{\theta} (\hat{\psi}_{1})$ of $\hat{\psi}_{1}$ for estimating $\psi (\theta)$ is smaller than a fraction $\delta$ of its standard error $\se_{\theta} (\hat{\psi}_{1})$, this is not the case for the bias of $\hat{\psi}_{1}$ as an estimator of the truncated parameter $\tilde{\psi}_{k} (\theta)$ of $\psi (\theta)$ (\citet[Section 3.2]{robins2008higher}), defined next. Recall from Section \ref{sec:introduction} that the bias of $\hat{\psi}_{1}$ for $\tilde{\psi}_{k} (\theta)$ is the estimable part of the bias of $\hat{\psi}_{1}$ for $\psi (\theta)$.

\begin{definition}[The truncated parameter $\tilde{\psi}_{k} (\theta)$, truncation bias $\TB_{\theta, k}$, and the projected bias $\Bias_{\theta, k} (\hat{\psi}_{1})$] \label{def:truncated_parameter}\leavevmode
Given estimators $\hat{b} \in \mathcal{B}$ and $\hat{p} \in \mathcal{P}$ of $b$ and $p$ computed from the training sample and some $k$ dimensional basis functions $\zbar_{k} (x)$ in $L_{2} (\P_{F})$:

\begin{itemize}
\item Define the (conditional) truncated parameter of a DR functional $\psi (\theta)$ as 
\begin{equation*}
\tilde{\psi}_{k} (\theta) \coloneqq \E_{\theta} \left[ \mathcal{H} \left( \tilde{b}_{k, \theta}, \tilde{p}_{k, \theta} \right) \right],
\end{equation*}
where $\tilde{b}_{k, \theta}$ and $\tilde{p}_{k, \theta}$ are defined as follows. Consider the working linear models 
\begin{equation*}
\left\{ \begin{array}{c}
b_{k}^{\ast} (X; \bar{\zeta}_{b, k}) = \hat{b}(X) + \bar{\zeta}_{b, k}^{\top} \zbar_{k} (X), \\ 
p_{k}^{\ast} (X; \bar{\zeta}_{p, k}) = \hat{p}(X) + \bar{\zeta}_{p, k}^{\top} \zbar_{k} (X).
\end{array}
\right.
\end{equation*}
Define $\tilde{\bar{\zeta}}_{b} (\theta)$ and $\tilde{\bar{\zeta}}_{p} (\theta)$ as the solutions to the following system of equations 
\begin{equation*}
\left\{ \begin{array}{l}
0 = \E_{\theta} \left[ \dfrac{\partial \mathcal{H} \left( b_{k}^{\ast} (X; \bar{\zeta}_{b, k}), p^{\ast} (X; \bar{\zeta}_{p, k}) \right)}{\partial \bar{\zeta}_{b, k}} \right] = \E_{\theta} \left[ S_{bp} p_{k}^{\ast} (X; \bar{\zeta}_{p, k}) \zbar_{k} (X) + m_{1} (O, \zbar_{k}) \right] \\ 
0 = \E_{\theta} \left[ \dfrac{\partial \mathcal{H} \left( b_{k}^{\ast} (X; \bar{\zeta}_{b, k}), p^{\ast} (X; \bar{\zeta}_{p, k}) \right)}{\partial \bar{\zeta}_{p, k}} \right] = \E_{\theta} \left[ S_{bp} b_{k}^{\ast} (X; \bar{\zeta}_{b, k}) \zbar_{k} (X) + m_{2} (O, \zbar_{k}) \right].
\end{array} \right.
\end{equation*}
where, for $j = 1, 2$, $m_{j} (O, \zbar_{k}) = \left( m_{j} (O, z_{1}), \ldots, m_{j} (O, z_{k}) \right)$. Hence 
\begin{align} \label{eq:zeta}
\left\{ \begin{array}{c}
\tilde{\bar{\zeta}}_{b, k} (\theta) = - \Sigma_{k}^{-1} \E_{\theta} \left[ \left( S_{bp} \hat{b}(X) \zbar_{k}(X) + m_{2} (O, \zbar_{k}) \right) \right], \\ 
\tilde{\bar{\zeta}}_{p, k} (\theta) = - \Sigma_{k}^{-1} \E_{\theta} \left[ \left( S_{bp} \hat{p}(X) \zbar_{k}(X) + m_{1} (O, \zbar_{k}) \right) \right]
\end{array} \right.
\end{align}
where 
\begin{equation*}
\Sigma_{k} \coloneqq \E_{\theta} \left[ S_{bp} \zbar_{k} (X) \zbar_{k} (X)^{\top} \right] \equiv \E_{\theta} \left[ \lambda(X) \zbar_{k} (X) \zbar_{k} (X)^{\top} \right].
\end{equation*}

Now define 
\begin{equation*}
\left\{ \begin{array}{c}
\tilde{b}_{k, \theta} (X) \coloneqq b_{k}^{\ast} (X; \tilde{\bar{\zeta}}_{b, k} (\theta)) \equiv \hat{b}(X) + \tilde{\bar{\zeta}}_{b, k}(\theta)^{\top} \zbar_{k} (X), \\ 
\tilde{p}_{k, \theta} (X) \coloneqq p_{k}^{\ast} (X; \tilde{\bar{\zeta}}_{p, k} (\theta)) \equiv \hat{p}(X) + \tilde{\bar{\zeta}}_{p, k}(\theta)^{\top} \zbar_{k} (X).
\end{array} \right.
\end{equation*}

\item Define the difference between $\tilde{\psi}_{k} (\theta)$ and $\psi (\theta)$ as the truncation bias $\TB_{\theta, k}$: 
\begin{equation}
\TB_{\theta, k} \coloneqq \tilde{\psi}_{k} (\theta) - \psi(\theta) = \E_{\theta} \left[ S_{bp} \left( \tilde{b}_{k, \theta} (X) - b (X) \right) \left( \tilde{p}_{k, \theta} (X) - p (X) \right) \right]. \label{eq:tb1}
\end{equation}

\item Define the bias of $\hat{\psi}_{1}$ as an estimator of $\tilde{\psi}_{k} (\theta)$ as 
\begin{align}
\Bias_{\theta, k} (\hat{\psi}_{1}) \coloneqq \E_{\theta} \left[ \hat{\psi}_{1} - \tilde{\psi}_{k} (\theta) \right].  \label{eq:biask}
\end{align}
\end{itemize}
\end{definition}

By definition, $\Bias_{\theta, k} (\hat{\psi}_{1}) = \Bias_{\theta} (\hat{\psi}_{1}) - \TB_{\theta, k}$. Combining equation \eqref{eq:tb1}, we have $\tilde{\psi}_{k} (\theta) - \psi (\theta) \equiv \Bias_{\theta} (\hat{\psi}_{1}) - \Bias_{\theta, k}(\hat{\psi}_{1})$. As discussed further below, $\TB_{\theta, k}$ is not consistently estimable without imposing smoothness/sparsity assumptions on the nuisance function spaces $\mathcal{B}$ and $\mathcal{P}$. But $\Bias_{\theta, k} (\hat{\psi}_{1})$ can be unbiasedly estimated by the following oracle (i.e. $\Sigma _{k}^{-1}$ known) second-order U-statistic: 
\begin{equation*}
\hat{\IIFF}_{22, k} (\Sigma_{k}^{-1}) \coloneqq \frac{1}{n (n - 1)} \sum_{1 \leq i_{1} \neq i_{2} \leq n} \hat{\IF}_{22, k, \bar{\imath}_{2}} (\Sigma_{k}^{-1})
\end{equation*}
where 
\begin{equation*}
\hat{\IF}_{22, k, \bar{\imath}_{2}} (\Sigma_{k}^{-1}) \coloneqq \left[ \calE_{\hat{b}, m_{2}} (\zbar_{k}) (O) \right]_{i_{1}}^{\top} \Sigma_{k}^{-1} \left[ \calE_{\hat{p}, m_{1}} (\zbar_{k}) (O) \right]_{i_{2}}.
\end{equation*}
Here $\calE_{1, b'} (h)$ and $\calE_{2, p'} (h)$ are linear functionals defined as: given any $h \in L_{2} (\P_{F})$, $b' \in \mathcal{B}$ and $p' \in \mathcal{P}$ 
\begin{equation}
\calE_{b', m_{2}} (h) (O) \coloneqq S_{bp} b' (X) h (X) + m_{2} (O, h) \label{calE:b}
\end{equation}
and 
\begin{equation}
\calE_{p^{\prime}, m_{1}} (h) (O) \coloneqq S_{bp} p^{\prime} (X) h(X) + m_{1} (O, h). \label{calE:p}
\end{equation}
When the range of $h$ is multidimensional (e.g. $h = \zbar_{k}$), the linear functionals $\calE_{1, b'}$ and $\calE_{2, p'}$ act on $h$ component-wise.

For the examples covered in Example \ref{eg:riesz}, $\calE_{\hat{b}, m_{2}} (\zbar_{k}) (O)$ and $\calE_{\hat{p}, m_{1}} (\zbar_{k}) (O)$ are

\begin{enumerate}
\item Example \ref{eg:riesz}(1): 
\begin{align*}
\left\{ \begin{array}{l}
\calE_{\hat{b}, m_{2}} \left( \zbar_{k} \right) (O) = A \hat{b}(X) \zbar_{k} (X) - A Y \zbar_{k} (X), \\ 
\calE_{\hat{p}, m_{1}} \left( \zbar_{k} \right) (O) = A \hat{p}(X) \zbar_{k} (X) - \zbar_{k} (X).
\end{array} \right.
\end{align*}

\item Example \ref{eg:riesz}(2): 
\begin{equation*}
\left\{ \begin{array}{l}
\calE_{\hat{b},m_{2}} \left( \zbar_{k} \right) (O) = \hat{b}(X) \zbar_{k}(X) - Y \zbar_{k} (X), \\ 
\calE_{\hat{p},m_{1}} \left( \zbar_{k} \right) (O) = \hat{p}(X) \zbar_{k}(X) - A \zbar_{k} (X).
\end{array} \right.
\end{equation*}
\end{enumerate}

The following lemma proves that (i) $\hat{\IIFF}_{22, k} (\Sigma_{k}^{-1})$ is an unbiased estimator of $\Bias_{\theta, k} (\hat{\psi}_{1})$ and (ii) $\Bias_{\theta, k} (\hat{\psi}_{1})$ is the expected product of the $L_{2} (\P_{\theta})$ projections of the weighted residuals 
\begin{align*}
\Delta_{1, \hat{b}} & \coloneqq \lambda(X)^{1 / 2} \left( \hat{b}(X) - b(X) \right), \\
\Delta_{2, \hat{p}} & \coloneqq \lambda(X)^{1 / 2} \left( \hat{p}(X) - p(X) \right)
\end{align*}
onto the subspace spanned by the $\lambda^{1 / 2}$-weighted basis functions $\bar{v}_{k} \coloneqq \lambda^{1 / 2} \zbar_{k}$:

\begin{lemma}
\label{lem:projection} 
Define $\beta_{\hat{b}, k} \coloneqq \E_{\theta} \left[ \calE_{\hat{b}, m_{2}} (\zbar_{k}) (O) \right]^{\top} \Sigma _{k}^{-1}$ and $\beta_{\hat{p}, k} \coloneqq \E_{\theta} \left[ \calE_{\hat{p}, m_{1}} (\zbar_{k}) (O) \right]^{\top} \Sigma_{k}^{-1}$. Under the conditions of Proposition \ref{thm:if1}, we have 
\begin{align}
& \ \Bias_{\theta, k} (\hat{\psi}_{1}) = \E_{\theta} \left[ \hat{\IIFF}_{22, k} (\Sigma_{k}^{-1}) \right] \nonumber \\
= & \ \E_{\theta} \left[ \Pi_{\theta} \left[ \Delta_{1, \hat{b}} \vert \bar{v}_{k} \right] (X) \Pi_{\theta} \left[ \Delta_{2, \hat{p}} \vert \bar{v}_{k} \right] (X) \right] = \beta_{\hat{b}, k}^{\top} \Sigma_{k} \beta_{\hat{p}, k} \label{eq:proj_bias}
\end{align}
and 
\begin{align}
\TB_{\theta, k} & \coloneqq \tilde{\psi}_{k} (\theta) - \psi(\theta) = \Bias_{\theta} (\hat{\psi}_{1}) - \Bias_{\theta, k} (\hat{\psi}_{1}) = \E_{\theta} \left[ \Pi_{\theta}^{\perp} \left[ \Delta_{1, \hat{b}} \vert \bar{v}_{k} \right] (X) \Pi_{\theta}^{\perp} \left[ \Delta_{2, \hat{p}} \vert \bar{v}_{k} \right] (X) \right]. \label{eq:tb2}
\end{align}
where $\Pi_{\theta}^{\perp} [\cdot | \bar{v}_{k}]$ is the projection operator onto the ortho-complement of the linear span of $\bar{v}_{k}$.

Furthermore, define the bias corrected estimator $\hat{\psi}_{2, k} (\Sigma_{k}^{-1}) \coloneqq \hat{\psi}_{1} - \hat{\IIFF}_{22, k} (\Sigma_{k}^{-1})$. Then 
\begin{equation}
\begin{split}
\Bias_{\theta} (\hat{\psi}_{2, k} (\Sigma_{k}^{-1})) & \equiv \E_{\theta} \left[ \hat{\psi}_{2, k} (\Sigma_{k}^{-1}) - \psi (\theta) \right] = \TB_{\theta, k}, \\
\text{and thus } \E_{\theta} \left[ \hat{\psi}_{2, k} (\Sigma_{k}^{-1}) \right] & = \psi (\theta) + \TB_{\theta, k} = \tilde{\psi}_{k} (\theta).
\end{split}
\label{eq:psi2}
\end{equation}
\end{lemma}

\begin{remark}
\label{rem:regression} 
The proof of the above lemma is deferred to Appendix \ref{proof:projection}. By $\Bias_{\theta, k} (\hat{\psi}_{1}) = \E_{\theta} \left[ \hat{\IIFF}_{22, k} (\Sigma_{k}^{-1}) \right] = \beta_{\hat{b}, k}^{\top} \Sigma_{k}^{-1} \beta_{\hat{p}, k}$, we can view $\hat{\IIFF}_{22, k} (\Sigma_{k}^{-1})$ as an unbiased estimator of the bilinear functional $\beta_{\hat{b}, k}^{\top} \Sigma_{k}^{-1} \beta_{\hat{p}, k}$, where $\beta_{\hat{b}, k}$ and $\beta_{\hat{p}, k}$ are the population coefficients of the following linear regressions: 
\begin{align*}
\text{regress } \lambda^{1/2} (\hat{b} - b) \text{ against } \lambda^{1/2} \zbar_{k}, \\
\text{regress } \lambda^{1/2} (\hat{p} - p) \text{ against } \lambda^{1/2} \zbar_{k}.
\end{align*}
\end{remark}

\subsection{Projection, truncation bias, and the motivation of higher-order influence functions}
\label{app:projection}
\subsubsection{Proof of Lemma \ref{lem:projection}} \label{proof:projection}
\begin{proof}
First, we show the first part of equation \eqref{eq:proj_bias}. 
\begin{align*}
\mathsf{TB}_{k} (\theta) = & \; \E_{\theta} \left[ S_{bp} \left( \tilde{b}_{k, \theta} (X) - b(X) \right) \left( \tilde{p}_{k, \theta} (X) - p(X) \right) \right] \\
= & \; \E_{\theta} \left[ \begin{array}{c}
S_{bp} \left\{ \hat{b} (X) - b (X) - \E_{\theta} \left[ (S_{bp} \hat{b} (X) \zbar_{k} (X) + m_{2} (O, \zbar_{k})) \right]^{\top} \Sigma_{k}^{-1} \zbar_{k} (X) \right\} \\ 
\times \left\{ \hat{p} (X) - p (X) - \E_{\theta} \left[ (S_{bp} \hat{p} (X) \zbar_{k} (X) + m_{1} (O, \zbar_{k})) \right]^{\top} \Sigma_{k}^{-1} \zbar_{k} (X) \right\}
\end{array} \right] \\
= & \; \E_{\theta} \left[ S_{bp} (\hat{b} (X) - b (X)) (\hat{p} (X) - p (X)) \right] \\
& - \E_{\theta} \left[ S_{bp} (\hat{b} (X) - b (X)) \zbar_{k} (X) \right]^{\top} \Sigma_{k}^{-1} \E_{\theta} \left[ S_{bp} (\hat{p} (X) - p (X)) \zbar_{k} (X) \right] \\
& - \E_{\theta} \left[ S_{bp} (\hat{p}(X) - p(X)) \zbar_{k} (X) \right]^{\top} \Sigma_{k}^{-1} \E_{\theta} \left[ S_{bp} (\hat{b} (X) - b (X)) \zbar_{k} (X) \right] \\
& + \E_{\theta} \left[ S_{bp} (\hat{p} (X) - p (X)) \zbar_{k} (X) \right]^{\top} \Sigma_{k}^{-1} \E_{\theta} \left[ S_{bp} (\hat{b} (X) - b (X)) \zbar_{k}(X) \right] \\
= & \; \underbrace{\E_{\theta} \left[ S_{bp} (\hat{b} (X) - b (X)) (\hat{p} (X) - p (X)) \right]}_{\equiv \ \Bias_{\theta} (\hat{\psi}_{1})} \\
& - \E_{\theta} \left[ S_{bp} (\hat{b} (X) - b (X)) \zbar_{k} (X) \right]^{\top} \Sigma_{k}^{-1} \E_{\theta} \left[ S_{bp} (\hat{p} (X) - p (X)) \zbar_{k} (X) \right]
\end{align*}
where in the third equality we apply Proposition \ref{thm:if1}(3). It follows that 
\begin{align*}
\Bias_{\theta, k} (\hat{\psi}_{1}) = \E_{\theta} \left[ S_{bp} (\hat{b} (X) - b (X)) \zbar_{k} (X) \right]^{\top} \Sigma_{k}^{-1} \E_{\theta} \left[ S_{bp} (\hat{p} (X) - p (X)) \zbar_{k} (X) \right].
\end{align*}

Again by Proposition \ref{thm:if1}(3), 
\begin{align*}
\E_{\theta} \left[ \calE_{\hat{b}, m_{2}} (\zbar_{k}) (O) \right] & = \E_{\theta} \left[ S_{bp} \hat{b} (X) \zbar_{k} (X) + m_{2} (O, \zbar_{k}) \right] = \E_{\theta} \left[ S_{bp} \left( \hat{b} (X) - b (X) \right) \zbar_{k} (X) \right] \\
\E_{\theta} \left[ \calE_{\hat{p}, m_{1}} (\zbar_{k}) (O) \right] & = \E_{\theta} \left[ S_{bp} \hat{p} (X) \zbar_{k} (X) + m_{1} (O, \zbar_{k}) \right] = \E_{\theta} \left[ S_{bp} \left( \hat{p} (X) - p (X) \right) \zbar_{k} (X) \right]
\end{align*}
and thus 
\begin{align*}
\E_{\theta} \left[ \hat{\IIFF}_{22, k} (\Sigma_{k}^{-1}) \right] = \E_{\theta} \left[ \calE_{\hat{b}, m_{2}} (\zbar_{k}) (O) \right]^{\top} \Sigma_{k}^{-1} \E_{\theta} \left[ \calE_{\hat{p}, m_{1}} (\zbar_{k}) (O) \right] \equiv \Bias_{\theta, k} (\hat{\psi}_{1}).
\end{align*}

In terms of the second part of equation \eqref{eq:proj_bias}: 
\begin{align*}
\Bias_{\theta, k} (\hat{\psi}_{1}) = & \ \E_{\theta} \left[ S_{bp} (\hat{b} (X) - b (X)) \zbar_{k} (X) \right]^{\top}
\Sigma_{k}^{-1} \E_{\theta} \left[ S_{bp} (\hat{p} (X) - p (X)) \zbar_{k} (X) \right] \\
= & \ \E_{\theta} \left[ \lambda(X) (\hat{b} (X) - b (X)) \zbar_{k} (X) \right]^{\top} \Sigma_{k}^{-1} \E_{\theta} \left[ \lambda (X) (\hat{p} (X) - p (X)) \zbar_{k} (X) \right] \\
= & \ \E_{\theta} \left[ \lambda(X)^{1 / 2} (\hat{b} (X) - b (X)) \bar{v}_{k} (X) \right]^{\top} \Sigma_{k}^{-1} \E_{\theta} \left[ \lambda (X)^{1 / 2} (\hat{p} (X) - p (X)) \bar{v}_{k} (X) \right] \\
= & \ \E_{\theta} \left[ \Delta_{1, \hat{b}} \bar{v}_{k} (X) \right]^{\top} \Sigma_{k}^{-1} \E_{\theta} \left[ \Delta_{2, \hat{p}} \bar{v}_{k} (X) \right] \\
= & \ \E_{\theta} \left[ \Pi_{\theta} \left[ \Delta_{1, \hat{b}} \vert \bar{v}_{k} \right] (X) \Pi_{\theta} \left[ \Delta_{2, \hat{p}} \vert \bar{v}_{k} \right] (X) \right].
\end{align*}
where in the last line we use the definition of the projection operator $\Pi_{\theta} \left[ \cdot \vert \bar{v}_{k} \right]$ (see equation \eqref{eq:projection}).

Then equation \eqref{eq:tb2} immediately follows: 
\begin{align*}
\TB_{\theta, k} = & \ \Bias_{\theta} (\hat{\psi}_{1}) - \Bias_{\theta, k} (\hat{\psi}_{1}) \\
= & \ \E_{\theta} \left[ \Delta_{1, \hat{b}} \Delta_{2, \hat{p}} \right] - \E_{\theta} \left[ \Pi_{\theta} \left[ \Delta_{1, \hat{b}} \vert \bar{v}_{k} \right] (X)^{\top} \Pi_{\theta} \left[ \Delta_{2, \hat{p}} \vert \bar{v}_{k} \right] (X) \right] \\
= & \ \E_{\theta}\left[ \Pi_{\theta}^{\perp} \left[ \Delta_{1, \hat{b}} \vert \bar{v}_{k} \right] (X) \right]^{\top} \E_{\theta} \left[ \Pi_{\theta}^{\perp} \left[ \Delta_{2, \hat{p}} \vert \bar{v}_{k} \right] (X) \right].
\end{align*}

Finally, we prove equation \eqref{eq:psi2}: 
\begin{align*}
\Bias_{\theta} (\hat{\psi}_{2, k} (\Sigma_{k}^{-1})) & = \E_{\theta} \left[ \hat{\psi}_{2, k} (\Sigma_{k}^{-1}) - \psi (\theta) \right] \\
& = \E_{\theta} \left[ \hat{\psi}_{1} - \psi (\theta) - \hat{\IIFF}_{22, k} (\Sigma_{k}^{-1}) \right] \\
& = \Bias_{\theta} (\hat{\psi}_{1}) - \Bias_{\theta, k} (\hat{\psi}_{1}) = \TB_{\theta, k}
\end{align*}
where the third line follows from the definition of $\Bias_{\theta} (\hat{\psi}_{1})$ and equation \eqref{eq:proj_bias}.
\end{proof}

The statistical properties of the oracle estimator $\hat{\IIFF}_{22, k} (\Sigma_{k}^{-1})$ and the oracle biased-corrected estimator $\hat{\psi}_{2, k} (\Sigma_k^{-1})$ are given by the following proposition.

\begin{proposition}
\label{thm:oracle_soif} 
Under the conditions of Proposition \ref{thm:if1} and Condition \ref{cond:sw}, with $k, n \rightarrow \infty$, and $k = o (n^{2})$, conditional on the training sample, we have

\begin{enumerate}
\item $\hat{\IIFF}_{22, k} (\Sigma_{k}^{-1})$ is unbiased for $\Bias_{\theta, k} (\hat{\psi}_{1})$ with variance of order 
\begin{equation*}
\frac{1}{n} \left( \frac{k}{n} + \BL_{\theta, 2, \hat{b}, k}^{2} + \BL_{\theta, 2, \hat{p}, k}^{2} \right).
\end{equation*}

\item $\frac{\hat{\IIFF}_{22, k} (\Sigma_{k}^{-1}) - \Bias_{\theta, k} (\hat{\psi}_{1})}{\se_{\theta} [\hat{\IIFF}_{22, k} (\Sigma_{k}^{-1})]} \overset{d}{\rightarrow} N (0, 1)$, where $\overset{d}{\rightarrow}$ stands for convergence in distribution. Further, $\se_{\theta} [\hat{\IIFF}_{22, k} (\Sigma_{k}^{-1})] \coloneqq \var_{\theta}^{1 / 2} [\hat{\IIFF}_{22, k} (\Sigma_{k}^{-1})]$ can be estimated by the bootstrap estimator $\hat{\se} [\hat{\IIFF}_{22, k} (\Sigma_{k}^{-1})] \coloneqq \hat{\var}^{1 / 2} [\hat{\IIFF}_{22, k} (\Sigma_{k}^{-1})]$ defined in Appendix \ref{app:bootstrap}.

\item $\hat{\IIFF}_{22, k} (\Sigma_{k}^{-1}) \pm z_{\alpha^{\dag} / 2} \hat{\se} [\hat{\IIFF}_{22, k} (\Sigma_{k}^{-1})]$ is a $(1 - \alpha^{\dag})$ asymptotic two-sided Wald CI for $\Bias_{\theta, k} (\hat{\psi}_{1})$ with length of order 
\begin{equation*}
\frac{1}{\sqrt{n}} \left( \sqrt{\frac{k}{n}} + \BL_{\theta, 2, \hat{b}, k} + \BL_{\theta, 2, \hat{p}, k} \right).
\end{equation*}
\end{enumerate}
\end{proposition}

\begin{proof}
The variance order of $\hat{\IIFF}_{22, k} (\Sigma_{k}^{-1})$ is proved in Appendix \ref{app:var}. When $k = o(n^2)$ and $k \rightarrow \infty$ as $n \rightarrow \infty$, the conditional asymptotic normality of $\frac{\hat{\IIFF}_{22, k} (\Sigma_{k}^{-1}) - \Bias_{\theta, k} (\hat{\psi}_{1})}{\se_{\theta} [\hat{\IIFF}_{22, k} (\Sigma_{k}^{-1})]}$ follows directly from Hoeffding decomposition, with the conditional asymptotic normality of the degenerate second-order U-statistic part implied by \citet[Corollary 1.2]{bhattacharya1992class}. Appendix \ref{app:bootstrap} proves that the bootstrap standard error estimators $\hat{\se} [\hat{\IIFF}_{22, k} (\Sigma_{k}^{-1})]$ satisfy $\frac{\hat{\se} [\hat{\IIFF}_{22, k} (\Sigma_{k}^{-1})]}{\se_{\theta} [\hat{\IIFF}_{22, k} (\Sigma_{k}^{-1})]} = 1 + o_{\P_{\theta}} (1)$.
\end{proof}

\begin{remark}
\label{rem:k1} 
When $k \gtrsim n^{2}$, the Gaussian limit of $\hat{\IIFF}_{22, k} (\Sigma_{k}^{-1})$ does not hold. Further if $k \gg n^{2}$, $\var_{\theta} [\hat{\IIFF}_{22, k} (\Sigma_{k}^{-1})] = O (\frac{k}{n^{2}})$ is of order greater than 1, and therefore $\hat{\IIFF}_{22, k} (\Sigma_{k}^{-1})$ cannot consistently estimate $\Bias_{\theta, k} (\hat{\psi}_{1})$ regardless of its magnitude.
\end{remark}

\subsection{Derivation of the higher order influence functions of $\tilde{\psi}_{k} (\theta)$ in Theorem \ref{thm:hoif}}
\label{app:hoif}

In this section, we give an almost self-contained derivation of higher order influence functions (HOIFs) of $\tilde{\psi}_{k} (\theta)$, except for \citet[Theorem 2.3]{robins2008higher}, which provides the ``calculus'' of deriving $m$-th order influence functions from $(m - 1)$-th order influence functions.

\subsubsection{A review of the theory of higher order influence functions}

As mentioned above, in this section, we present the higher order influence functions (HOIFs) for the DR functionals \citep{rotnitzky2021characterization} under a (locally) nonparametric model of every order with the derivation deferred to Appendix \ref{app:hoif}. First, we need to formally define HOIFs of a generic functional $\psi (\theta)$, not necessarily in the DR functional/DR class. In particular, we use the characterization of HOIFs under the (locally) nonparametric models given in Theorem 2.3 of \citet{robins2008higher}. We note that in the following $\IIFF_{m} (\theta)$ is a $m$-th order U-statistic and $\IIFF_{mm} (\theta)$ is a degenerate $m$-th order U-statistic under $\P_{\theta}$.

\begin{definition}
\label{def:hoif} 
For a generic functional $\psi (\theta)$, under a (locally) nonparametric model for which the first order influence function $\IF_{1} (\theta)$ exists. For $m \geq 2$, if the $(m - 1)$-th order influence function $\IIFF_{m - 1} (\theta)$ exists, then there exists a $m$-th order influence function $\IIFF_{m} (\theta)$ of $\psi (\theta)$, defined recursively via the following rule: 
\begin{align*}
\IIFF_{1} (\theta) & \equiv \IIFF_{11} (\theta) = \frac{1}{n} \sum_{i = 1}^{n} \IF_{1, i} (\theta), \\
\IIFF_{m} (\theta) & \equiv \IIFF_{m - 1} (\theta) + \IIFF_{mm} (\theta)
\end{align*}
where $m \IIFF_{mm} (\theta)$ is the $m$-th order degenerate U-statistic component of the Hoeffding decomposition of 
\begin{equation*}
\frac{(n - m)!}{n!} \sum_{1 \leq i_{1} \neq \cdots \neq i_{m} \leq n} \mathsf{if}_{1, \mathsf{if}_{m - 1, m - 1} (O_{i_{1}}, \cdots, O_{i_{m - 1}}; \cdot)} (O_{i_{m}}; \theta),
\end{equation*}
if, for each $\left( o_{i_{1}}, \cdots, o_{i_{m - 1}} \right)$ in the support of $(O_{i_{1}}, \cdots, O_{i_{m - 1}})$, the first order influence function 
$$
\IF_{1, \mathsf{if}_{m - 1, m - 1} (o_{i_{1}}, \cdots, o_{i_{m - 1}}; \cdot)} (O_{i_{m}}; \theta)
$$ 
exists for the functional   
\begin{equation*}
\mathsf{if}_{m - 1, m - 1} (o_{i_{1}}, \cdots, o_{i_{m - 1}}; \theta))
\end{equation*}
given by the kernel of the $(m - 1)$-th order U-statistic $\IIFF_{m - 1, m - 1} (\theta)$ evaluated at fixed data $o_{i_{1}}, \cdots, o_{i_{m - 1}}$. Otherwise $\IIFF_{m} (\theta)$ does not exist.
\end{definition}

The general use of HOIF estimators is to decrease bias as given in the following Theorem 2.2 of \citet{robins2008higher} .

\begin{theorem}[Theorem 2.2 of \citet{robins2008higher}]
\label{ext} 
Given an $m$-th order influence function $\IIFF_{m} (\theta)$ for $\psi (\theta)$, if both $\E_{\theta} [\IIFF_{m} (\hat{\theta})]$ and $\psi (\hat{\theta}) - \psi (\theta) \equiv \hat{\psi} - \psi (\theta)$ have bounded Fr\'{e}chet derivatives with respect to $\hat{\theta}$ to order $m$ for a norm $\Vert \cdot \Vert$, then the estimator $\psi (\hat{\theta}) + \IIFF_{m} (\hat{\theta})$ has bias of order $\Vert \hat{\theta} - \theta \Vert^{m + 1}$ as an estimator of $\psi (\theta )$ where $\hat{\theta}$ and $\IIFF_{m} (\cdot)$ are computed from independent samples.
\end{theorem}

\subsubsection{Higher order influence functions of the DR functionals}

\label{sec:hoif_dr}

With the above preparation, we are ready to state the following theorem characterizing the HOIFs of the DR functionals:

\begin{theorem}
\label{thm:hoif} 
Under the conditions of Proposition \ref{thm:if1}, in a locally nonparametric model, under $\P_{\hat{\theta}}$, the $m$-th order influence function of $\tilde{\psi}_{k} (\theta)$ is $\hat{\IIFF}_{m, k} (\Sigma_{k}^{-1}) = \hat{\IIFF}_{1} - \hat{\IIFF}_{22 \rightarrow mm, k} (\Sigma_{k}^{-1})$, where $\hat{\IIFF}_{\ell \ell \rightarrow m m, k} (\Sigma_{k}^{-1}) \coloneqq \sum_{j = \ell}^{m}\hat{\IIFF}_{j j, k} (\Sigma_{k}^{-1})$, 
\begin{equation*}
\hat{\IIFF}_{22, k} (\Sigma_{k}^{-1}) \coloneqq \frac{1}{n (n - 1)} \sum_{1 \leq i_{1} \neq i_{2} \leq n} \hat{\IF}_{22, k, \bar{i}_{2}} (\Sigma_{k}^{-1})
\end{equation*}
and for $j > 2$, 
\begin{equation*}
\hat{\IIFF}_{j j, k} (\Sigma_{k}^{-1}) \coloneqq \frac{(n - j)!}{n!} \sum_{1 \leq i_{1} \neq \ldots \neq i_{j} \leq n} \hat{\IF}_{j j, k, \bar{i}_{j}} (\Sigma_{k}^{-1}).
\end{equation*}
Here 
\begin{equation*}
\hat{\IF}_{2 2, k, \bar{i}_{2}} (\Sigma_{k}^{-1}) \equiv \hat{\IF}_{2 2, \tilde{\psi}_{k}, \bar{i}_{2}} (\theta) = \left[ \calE_{1, \hat{b}} (\zbar_{k}) (O) \right]_{i_{1}}^{\top} \Sigma_{k}^{-1} \left[ \calE_{2, \hat{p}} (\zbar_{k}) (O) \right]_{i_{2}},
\end{equation*}
and 
\begin{align}
& \ \hat{\IF}_{j j, k, \bar{i}_{j}} (\Sigma_{k}^{-1}) \equiv \hat{\IF}_{j j, \tilde{\psi}_{k}, \bar{i}_{j}} (\theta) \notag \\
=& \ (-1)^{j} \left[ \calE_{1, \hat{b}} (\zbar_{k}) (O) \right]_{i_{1}}^{\top} \left\{ \prod\limits_{s = 3}^{j} \Sigma_{k}^{-1} \left( \left[ S_{bp} \zbar_{k} (X) \zbar_{k} (X)^{\top} \right]_{i_{s}} - \Sigma_{k} \right) \right\} \Sigma_{k}^{-1} \left[ \calE_{2, \hat{p}} (\zbar_{k}) (O) \right]_{i_{2}}. \label{eq:ifjj}
\end{align}
are the kernels of second order influence function $\hat{\IIFF}_{22, k} (\Sigma_{k}^{-1})$ and $j$-th order influence function $\hat{\IIFF}_{j j, k} (\Sigma_{k}^{-1})$ respectively, where $\bar{i}_{j} \coloneqq \{i_{1}, \ldots, i_{j}\}$.

Furthermore, define 
\begin{align*}
\hat{\IIFF}_{22 \rightarrow mm, k} (\Sigma_{k}^{-1}) \coloneqq \sum_{j = 2}^{m} \hat{\IIFF}_{j j, k} (\Sigma_{k}^{-1}).
\end{align*}
As a consequence, $\hat{\IIFF}_{22 \rightarrow mm, k} (\Sigma_{k}^{-1})$ is the $m$-th order influence function of $\Bias_{\theta, k} (\hat{\psi}_{1})$ under $\P_{\hat{\theta}}$.
\end{theorem}

Note that in the main text, we simply call $\hat{\IIFF}_{22 \rightarrow mm, k} (\Sigma_{k}^{-1})$ the $m$-th order influence function of $\Bias_{\theta, k} (\hat{\psi}_{1})$, without the qualification of ``under the law $\P_{\hat{\theta}}$'' to simplify our exposition. Similarly, we call $\hat{\IIFF}_{22 \rightarrow mm, k} (\hat{\Sigma}_{k}^{-1})$ the estimated $m$-th order influence function of $\Bias_{\theta, k} (\hat{\psi}_{1})$.

\begin{proof}
Without loss of generality, we assume $\P_{\theta} (S_{bp} \geq 0) = 1$. First, let's recall the results from Proposition \ref{thm:if1} (or \citet[Theorems 2 (ii)]{rotnitzky2021characterization}), the influence function of $\psi (\theta)$ has the following form 
\begin{equation*}
\IF_{1, \psi (\theta)} = S_{bp} b (X) p (X) + m_{1} (O, b) + m_{2} (O, p) + S_{0} - \psi (\theta),
\end{equation*}
where the linear functionals $m_{1} (O, h)$ and $m_{2} (O, h)$ have the Riesz representers $\calR_{1} (X)$ and $\calR_{2} (X)$ respectively. By the definition of Riesz representers, they must obey the following identities: 
\begin{align*}
\left\{ \begin{array}{c}
\E_{\theta} \left[ m_{1} (O, h) \right] = \E_{\theta} \left[ h (X) \mathcal{R}_{1} (X) \right] \\ 
\E_{\theta} \left[ m_{2} (O, h) \right] = \E_{\theta} \left[ h (X) \mathcal{R}_{2} (X) \right]
\end{array} \right.
\end{align*}
for any $h \in L_{2} (g)$.

To derive the HOIFs for $\tilde{\psi}_{k} (\theta)$, we need its first-order influence function. Recall from Definition \ref{def:truncated_parameter} that $\tilde{\psi}_{k} (\theta)$ is of the following form: 
\begin{align*}
& \; \tilde{\psi}_{k} (\theta) \\
\equiv & \; \E_{\theta}\left[ \mathcal{H} \left( \tilde{b}_{k, \theta}, \tilde{p}_{k, \theta} \right) \right] \\
= & \; \E_{\theta}\left[ S_{bp} \tilde{b}_{k, \theta}(X) \tilde{p}_{k, \theta}(X) + m_{1} (O, \tilde{b}_{k, \theta}) + m_{2} (O, \tilde{p}_{k, \theta}) + S_{0} \right] \\
= & \; \E_{\theta} \left[ \begin{array}{c}
S_{bp} \left( \hat{b}(X) + \tilde{\bar{\zeta}}_{b, k}^{\top} (\theta) \zbar_{k}(X) \right) \left( \hat{p}(X) + \tilde{\bar{\zeta}}_{p, k}^{\top} (\theta) \zbar_{k}(X)^{\top} \right) \\  
+ \ m_{1} (O, \hat{b} + \tilde{\bar{\zeta}}_{b, k}^{\top} (\theta) \zbar_{k}) + m_{2} (O, \hat{p} + \tilde{\bar{\zeta}}_{p, k}^{\top} (\theta) \zbar_{k}) + S_{0} 
\end{array} \right] \\
= & \; \E_{\theta} \left[ \begin{array}{c}
S_{bp} \left( \hat{b}(X) - \Sigma_{k}^{-1} \E_{\theta} \left[ S_{bp} \hat{b}(X) \zbar_{k}(X)^{\top} + m_{2} (O, \zbar_{k}(X))^{\top} \right] \zbar_{k}(X) \right) \\ 
\times \left( \hat{p}(X) - \Sigma_{k}^{-1} \E_{\theta} \left[ S_{bp} \hat{p}(X) \zbar_{k}(X)^{\top} + m_{1} (O, \zbar_{k}(X))^{\top} \right] \zbar_{k}(X) \right) \\ 
+ \ m_{1} (O, \hat{b} - \Sigma_{k}^{-1} \E_{\theta} \left[ S_{bp} \hat{b}(X) \zbar_{k}(X)^{\top} + m_{2} (O, \zbar_{k}(X))^{\top} \right] \zbar_{k}) \\ 
+ \ m_{2} (O, \hat{p} - \Sigma_{k}^{-1} \E_{\theta} \left[ S_{bp} \hat{p}(X) \zbar_{k}(X)^{\top} + m_{1} (O, \zbar_{k}(X))^{\top} \right] \zbar_{k}) + S_{0}
\end{array} \right].
\end{align*}
Then 
\begin{align*}
\IF_{1, \tilde{\psi}_{k} (\theta), i} (\theta) = & \; \mathcal{H} (\tilde{b}_{k} (X, \theta), \tilde{p}_{k} (X, \theta)) - \tilde{\psi}_{k} (\theta) \\
& + \ \E_{\theta} \left[ \partial \mathcal{H} \left( b_{k}^{\ast} (X, \tilde{\bar{\zeta}}_{b, k} (\theta)), p_{k}^{\ast} (X, \tilde{\bar{\zeta}}_{b, k} (\theta)) \right) / \partial \bar{\zeta}_{b, k}^{\top} \right] \IF_{1, \tilde{\bar{\zeta}}_{b, k}} (\theta) \\
& + \ \E_{\theta} \left[ \partial \mathcal{H} \left( b_{k}^{\ast} (X, \tilde{\bar{\zeta}}_{b, k} (\theta)), p_{k}^{\ast} (X, \tilde{\bar{\zeta}}_{b, k} (\theta)) \right) / \partial \bar{\zeta}_{p, k}^{\top} \right] \IF_{1, \tilde{\bar{\zeta}}_{p, k}} (\theta) \\
= & \; \mathcal{H} (\tilde{b}_{k, \theta}, \tilde{p}_{k, \theta}) - \tilde{\psi}_{k} (\theta) \\
= & \; S_{bp, i} \tilde{b}_{k, \theta} (X_{i}) \tilde{p}_{k, \theta} (X_{i}) + m_{1} (O_{i}, \tilde{b}_{k, \theta}) + m_{2} (O_{i}, \tilde{p}_{k, \theta}) + S_{0, i} - \tilde{\psi}_{k} (\theta).
\end{align*}
where the second equality follows from the definitions of $\tilde{\bar{\zeta}}_{b, k} (\theta)$ and $\tilde{\bar{\zeta}}_{p, k} (\theta)$ in equation \eqref{eq:zeta}. Then the influence function of $- \IF_{1, \tilde{\psi}_{k} (\theta), i} (\theta)$, denoted as $\IF_{1, - \mathsf{if}_{1, \tilde{\psi}_{k} (\theta), i_{1}}, i_{2}} (\theta)$, can be calculated as follows: 
\begin{align*}
\IF_{1, - \mathsf{if}_{1, \tilde{\psi}_{k} (\theta), i_{1}}, i_{2}} (\theta) = & \; - \frac{\partial \mathcal{H}_{i_{1}} \left( \tilde{b}_{k, \theta} (X_{i_{1}}), \tilde{p}_{k, \theta} (X_{i_{1}}) \right)}{\partial \bar{\zeta}_{b, k}^{\top}} \IF_{1, \tilde{\bar{\zeta}}_{b, k}, i_{2}} (\theta) - \frac{\partial \mathcal{H}_{i_{1}} \left( \tilde{b}_{k, \theta} (X_{i_{1}}), \tilde{p}_{k, \theta} (X_{i_{1}}) \right)}{\partial \bar{\zeta}_{p, k}^{\top}} \IF_{1, \tilde{\bar{\zeta}}_{p, k}, i_{2}} (\theta)
\end{align*}
where 
\begin{align*}
\left\{ \begin{array}{c}
\frac{\partial \mathcal{H}_{i_{1}} \left( \tilde{b}_{k, \theta} (X_{i_{1}}), \tilde{p}_{k, \theta} (X_{i_{1}}) \right)}{\partial \bar{\zeta}_{b, k}} = S_{bp, i_{1}} \tilde{p}_{k, \theta} (X_{i_{1}}) \zbar_{k} (X_{i_{1}}) + m_{1} (O_{i_{1}}, \zbar_{k}), \\ 
\frac{\partial \mathcal{H}_{i_{1}} \left( \tilde{b}_{k, \theta} (X_{i_{1}}), \tilde{p}_{k, \theta} (X_{i_{1}}) \right)}{\partial \bar{\zeta}_{p, k}} = S_{bp, i_{1}} \tilde{b}_{k, \theta} (X_{i_{1}}) \zbar_{k} (X_{i_{1}}) + m_{2} (O_{i_{1}}, \zbar_{k})
\end{array} \right.
\end{align*}
and by Lemma \ref{lem:app_zeta} 
\begin{align*}
\left\{ \begin{array}{c}
\IF_{1, \tilde{\bar{\zeta}}_{b, k}, i_{2}} (\theta) = - \Sigma_{k}^{-1} \left( S_{bp, i_{2}} \tilde{b}_{k, \theta} (X_{i_{2}}) \zbar_{k} (X_{i_{2}}) + m_{2} (O_{i_{2}}, \zbar_{k}) \right), \\ 
\IF_{1, \tilde{\bar{\zeta}}_{p, k}, i_{2}} (\theta) = - \Sigma_{k}^{-1} \left( S_{bp, i_{2}} \tilde{p}_{k, \theta} (X_{i_{2}}) \zbar_{k} (X_{i_{2}}) + m_{1} (O_{i_{2}}, \zbar_{k}) \right).
\end{array} \right.
\end{align*}
Then 
\begin{align*}
& \; \IF_{1, - \mathsf{if}_{1, \tilde{\psi}_{k} (\theta), i_{1}}, i_{2}} (\theta) \\
= & \left[ S_{bp} \tilde{p}_{k, \theta} (X) \zbar_{k} (X) + m_{1} (O, \zbar_{k}) \right]_{i_{1}}^{\top} \Sigma_{k}^{-1} \left[ S_{bp} \tilde{b}_{k, \theta} (X) \zbar_{k} (X) + m_{2} (O, \zbar_{k}) \right]_{i_{2}} \\
& + \left[ S_{bp} \tilde{b}_{k, \theta} (X) \zbar_{k} (X) + m_{2} (O, \zbar_{k}) \right]_{i_{1}}^{\top} \Sigma_{k}^{-1} \left[ S_{bp} \tilde{p}_{k, \theta} (X) \zbar_{k} (X) + m_{1} (O, \zbar_{k}) \right]_{i_{2}}.
\end{align*}

Then by \citet[part 5.c of Theorem 2.3]{robins2008higher}, 
\begin{align*}
\IIFF_{22, k} (\Sigma_{k}^{-1}) = \frac{1}{2} \Pi_{\theta} \left[ \mathbb{V}_{2} \left[ \IF_{1, - \mathsf{if}_{1, \tilde{\psi}_{k} (\theta), i_{1}}, i_{2}} (\theta) \right] \vert \mathcal{U}_{1}^{\perp, 2} (\theta) \right].
\end{align*}
Here $\mathcal{U}_{m} (\theta)$ denotes the Hilbert space of all $m$-th order U-statistics with mean zero and finite variance with inner product defined by covariances w.r.t. the $n$-fold product measure $\P_{\theta} (\cdot)^{n}$ and $\mathcal{U}_{m}^{\perp, m + 1} (\theta)$ denotes the orthocomplement of $\mathcal{U}_{m} (\theta)$ in $\mathcal{U}_{m + 1} (\theta)$. The notation $\mathbb{V}_{m} [ U_{i_{1}, \ldots, i_{m}} ]$ maps an $m$-th order U-statistic kernel $U_{i_{1}, \ldots, i_{m}}$ to an $m$-th order U-statistic: 
\begin{align*}
\mathbb{V}_{m} [U_{i_{1}, \ldots, i_{m}}] = \frac{(n - m)!}{n!} \sum_{1 \leq i_{1} \neq \cdots \neq i_{m} \leq n} U_{i_{1}, \ldots, i_{m}}.
\end{align*}
For the 2nd order U-statistic kernel $\IF_{1, \mathsf{if}_{1, \tilde{\psi}_{k} (\theta), i_{1}}, i_{2}} (\theta)$, we thus have 
\begin{align*}
& \; \mathbb{V}_{2} \left[ \IF_{1, - \mathsf{if}_{1, \tilde{\psi}_{k} (\theta), i_{1}}, i_{2}} (\theta) \right] \\
= & \; \mathbb{V}_{2} \left[ \begin{array}{c}
\left[ S_{bp} \tilde{p}_{k, \theta} (X) \zbar_{k} (X) + m_{1} (O, \zbar_{k}) \right]_{i_{1}}^{\top} \Sigma_{k}^{-1} \left[ S_{bp} \tilde{b}_{k, \theta} (X) \zbar_{k} (X) + m_{2} (O, \zbar_{k}) \right]_{i_{2}} \\ 
+ \ \left[ S_{bp} \tilde{b}_{k, \theta} (X) \zbar_{k} (X) + m_{2} (O, \zbar_{k}) \right]_{i_{1}}^{\top} \Sigma_{k}^{-1} \left[ S_{bp} \tilde{p}_{k, \theta} (X) \zbar_{k} (X) + m_{1} (O, \zbar_{k}) \right]_{i_{2}}
\end{array} \right] \\
= & \; \mathbb{V}_{2} \left[ 2 \left[ S_{bp} \tilde{p}_{k, \theta} (X) \zbar_{k} (X) + m_{1} (O, \zbar_{k}) \right]_{i_{1}}^{\top} \Sigma_{k}^{-1} \left[ S_{bp} \tilde{b}_{k, \theta} (X) \zbar_{k} (X) + m_{2} (O, \zbar_{k}) \right]_{i_{2}} \right]
\end{align*}
where the last equality is due to the symmetrization of $\mathbb{V}_{2} [\cdot]$. In addition, we have 
\begin{align*}
\Pi_{\theta} \left[ \mathbb{V}_{2} \left[ \mathsf{IF}_{1, - \mathsf{if}_{1, \tilde{\psi}_{k} (\theta), i_{1}}, i_{2}} (\theta) \right] \vert \mathcal{U}_{1} (\theta) \right] = 0
\end{align*}
since 
\begin{align*}
\left\{ \begin{array}{c}
\E_{\theta} \left[ S_{bp} \tilde{p}_{k, \theta} (X) \zbar_{k} (X) + m_{1} (O, \zbar_{k}) \right] = 0, \\ 
\E_{\theta} \left[ S_{bp} \tilde{b}_{k, \theta} (X) \zbar_{k} (X) + m_{2} (O, \zbar_{k}) \right] = 0.
\end{array} \right.
\end{align*}
Thus 
\begin{align*}
& \IIFF_{22, k} (\Sigma_{k}^{-1}) = \frac{1}{2} \mathbb{V}_{2} \left[ \IF_{1, - \mathsf{if}_{1, \tilde{\psi}_{k} (\theta), i_{1}}, i_{2}} (\theta) \right] \\
& = \frac{1}{n (n - 1)} \sum_{1 \leq i_{1} \neq i_{2} \leq n} \left[ S_{bp} \tilde{p}_{k, \theta} (X) \zbar_{k} (X) + m_{1} (O, \zbar_{k}) \right]_{i_{1}}^{\top} \Sigma_{k}^{-1} \left[ S_{bp} \tilde{b}_{k, \theta} (X) \zbar_{k} (X) + m_{2} (O, \zbar_{k}) \right]_{i_{2}}.
\end{align*}
Hence formally we have 
\begin{align*}
\IF_{22, k, \bar{i}_{2}} (\Sigma_{k}^{-1}) = \left[ S_{bp} \tilde{p}_{k, \theta} (X) \zbar_{k} (X) + m_{1} (O, \zbar_{k}) \right]_{i_{1}}^{\top} \Sigma_{k}^{-1} \left[ S_{bp} \tilde{b}_{k, \theta} (X) \zbar_{k} (X) + m_{2} (O, \zbar_{k}) \right]_{i_{2}}
\end{align*}
and 
\begin{align*}
\hat{\IF}_{22, k, \bar{i}_{2}} (\Sigma_{k}^{-1}) = \left[ S_{bp} \hat{p} (X) \zbar_{k} (X) + m_{1} (O, \zbar_{k}) \right]_{i_{1}}^{\top} \Sigma_{k}^{-1} \left[ S_{bp} \hat{b} (X) \zbar_{k} (X) + m_{2} (O, \zbar_{k}) \right]_{i_{2}}.
\end{align*}

We now complete the proof by induction. We assume that the form for $\hat{\IF}_{jj, k, \bar{i}_{j}} (\Sigma_{k}^{-1})$ is true and thus by the induction hypothesis 
\begin{align*}
& \; \IF_{jj, k, \bar{i}_{j}} (\Sigma_{k}^{-1}) \\
= & \; (-1)^{j} \left[ S_{bp} \tilde{p}_{k, \theta} (X) \zbar_{k} (X) + m_{1} (O, \zbar_{k}) \right]^{\top}_{i_{1}} \prod_{s = 3}^{j} \Sigma_{k}^{-1} \left( \left[ S_{bp} \zbar_{k} (X) \zbar_{k} (X)^{\top} \right]_{i_{s}} - \Sigma_{k} \right) \\
& \times \ \Sigma_{k}^{-1} \left[ S_{bp} \tilde{b}_{k, \theta} (X) \zbar_{k} (X) + m_{2} (O, \zbar_{k}) \right]_{i_{2}}
\end{align*}
and 
\begin{equation*}
\IIFF_{jj, k} (\Sigma_{k}^{-1}) = \mathbb{V}_{j} \left[ \IF_{jj, k, \bar{i}_{j}} (\Sigma_{k}^{-1}) \right].
\end{equation*}
Then 
\begin{align*}
& \; \mathbb{V}_{j + 1} \left[ \mathsf{IF}_{(j + 1), (j + 1), k, \bar{i}_{j + 1}} (\Sigma_{k}^{-1}) \right] \\
= & \; \frac{1}{j + 1} \mathbb{V}_{j + 1} \left[ \Pi_{\theta} \left[ \IF_{1, \mathsf{if}_{jj, k, \bar{i}_{j}}, i_{j + 1}} (\theta) \vert \mathcal{U}_{j}^{\perp, j + 1} (\theta) \right] \right].
\end{align*}
Then the derivatives w.r.t. $\theta$'s in $\tilde{\bar{\zeta}}_{b, k} (\theta)$, $\tilde{\bar{\zeta}}_{p, k} (\theta)$ and $(j - 1)$ terms of $\Sigma_{k}^{-1}$ will be contributing terms to $\mathbb{V}_{j + 1} \left[ \IF_{(j + 1), (j + 1), k, \bar{i}_{j + 1}} (\Sigma_{k}^{-1}) \right]$. But differentiating w.r.t. $(j - 2)$ terms of $\Sigma_{k}$ will not contribute terms to $\mathbb{V}_{j + 1} \left[ \IF_{(j + 1), (j + 1), k, \bar{i}_{j + 1}} (\Sigma_{k}^{-1}) \right]$ as the contribution of these $(j - 2)$ terms to $\IF_{1, \mathsf{if}_{jj, k, \bar{i}_{j}}, i_{j + 1}} (\theta)$ is a function of $j$ units' data and is thus an element of $\mathcal{U}_{j} (\theta)$. Then as in the proof of Lemma \ref{lem:app_zeta}, 
\begin{align*}
\IF_{1, \Sigma_{k}^{-1}} (\theta) = - \Sigma_{k}^{-1} \left( S_{bp} \zbar_{k} (X) \zbar_{k} (X)^{\top} - \Sigma_{k} \right) \Sigma_{k}^{-1}
\end{align*}
and upon permuting the indices, the contribution of each of these $(j - 1)$ terms to $\mathsf{IF}_{1, \mathsf{if}_{jj, k, \bar{i}_{j}}, i_{j + 1}} (\theta)$ is 
\begin{equation}  \label{kernel}
\begin{split}
& - (-1)^{j} \left[ S_{bp} \tilde{p}_{k, \theta} (X) \zbar_{k} (X) + m_{1} (O, \zbar_{k}) \right]^{\top}_{i_{1}} \\
& \times \sum_{s = 3}^{j + 1} \Sigma_{k}^{-1} \left\{ \left( S_{bp} \zbar_{k} (X) \zbar_{k} (X)^{\top} \right)_{i_{s}} - \Sigma_{k} \right\} \\
& \times \Sigma_{k}^{-1} \left[ S_{bp} \tilde{b}_{k, \theta} (X) \zbar_{k} (X) + m_{2} (O, \zbar_{k}) \right]_{i_{2}}
\end{split}
\end{equation}
which is degenerate and thus orthogonal to $\mathcal{U}_{j} (\theta)$. Notice that equation \eqref{kernel} is exactly the kernel of $\IIFF_{(j + 1), (j + 1), k} (\Sigma_{k}^{-1})$. Then differentiating w.r.t. $\theta$'s in $\tilde{p}_{k, \theta}$ and $\tilde{b}_{k, \theta}$, we further obtain terms of the following form contributing to $\IF_{1, \mathsf{if}_{jj, k, \bar{i}_{j}}, i_{j + 1}} (\theta)$: 
\begin{equation}  
\label{re_kernel}
\begin{split}
& (-1)^{j} \IF_{1, \tilde{\bar{\zeta}}_{p, k}, i_{j + 1}} (\theta)^{\top} \left[ S_{bp} \zbar_{k} (X) \zbar_{k} (X)^{\top} \right]_{i_{1}} \left[ \sum_{s = 3}^{j} \Sigma_{k}^{-1} \left\{ \left( S_{bp} \zbar_{k} (X) \zbar_{k} (X)^{\top} \right)_{i_{s}} - \Sigma_{k} \right\} \right] \\
& \times \Sigma_{k}^{-1} \left[ S_{bp} \tilde{b}_{k, \theta} (X) \zbar_{k} (X) + m_{2} (O, \zbar_{k}) \right]_{i_{2}} \\
& + (-1)^{j} \left[ S_{bp} \tilde{p}_{k, \theta} (X) \zbar_{k} (X) + m_{1} (O, \zbar_{k}) \right]_{i_{1}}^{\top} \left[ \sum_{s = 3}^{j} \Sigma_{k}^{-1} \left\{ \left( S_{bp} \zbar_{k} (X) \zbar_{k} (X)^{\top} \right)_{i_{s}} - \Sigma_{k} \right\} \right] \\
& \times \Sigma_{k}^{-1} \left[ S_{bp} \zbar_{k} (X) \zbar_{k} (X)^{\top} \right]_{i_{2}} \IF_{1, \tilde{\bar{\zeta}}_{b, k}, i_{j + 1}} (\theta).
\end{split}
\end{equation}
Applying Lemma \ref{lem:app_zeta} again and projecting the above display onto $\mathcal{U}_{j}^{\perp, j + 1} (\theta)$, we obtain that 
\begin{align*}
\Pi_{\theta} \left[ \text{equation } \eqref{re_kernel} \vert \mathcal{U}_{j}^{\perp, j + 1} (\theta) \right] \equiv 2 \times \text{equation } \eqref{kernel}.
\end{align*}
Recall that there are $(j - 1)$ terms of equation \eqref{kernel} in total contributing to $\IF_{1, \mathsf{if}_{jj, k, \bar{i}_{j}}, i_{j + 1}} (\theta)$, thus eventually there are $(j + 1)$ terms of equation \eqref{kernel} in total contributing to $\mathbb{V}_{j + 1} \left[ \Pi_{\theta} \left[ \IF_{1, \mathsf{if}_{jj, k, \bar{i}_{j}}, i_{j + 1}} (\theta) \vert \mathcal{U}_{j}^{\perp, j + 1} (\theta) \right] \right]$. Further recall that 
\begin{align*}
& \; \mathbb{V}_{j + 1} \left[ \IF_{(j + 1), (j + 1), k, \bar{i}_{j + 1}} (\Sigma_{k}^{-1}) \right] \\
= & \; \frac{1}{j + 1} \mathbb{V}_{j + 1} \left[ \Pi_{\theta} \left[ \IF_{1, \mathsf{if}_{jj, k, \bar{i}_{j}}, i_{j + 1}} (\theta) \vert \mathcal{U}_{j}^{\perp, j + 1} (\theta) \right] \right].
\end{align*}
Therefore we have 
\begin{align*}
\IIFF_{(j + 1), (j + 1), k} (\Sigma_{k}^{-1}) & \equiv \mathbb{V}_{j + 1} \left[ \IF_{(j + 1), (j + 1), k, \bar{i}_{j + 1}} (\Sigma_{k}^{-1}) \right] \\
& = \frac{1}{j + 1} \mathbb{V}_{j + 1} \left[ (j + 1) \times \text{equation } \eqref{kernel} \right] \\
& = \mathbb{V}_{j + 1} \left[ \text{equation } \eqref{kernel} \right].
\end{align*}
We thus prove the results for general $j$ by induction.
\end{proof}

\begin{lemma}
\label{lem:app_zeta} 
As defined in equation \eqref{eq:zeta}, the influence functions of $\tilde{\bar{\zeta}}_{b, k} (\theta)$ and $\tilde{\bar{\zeta}}_{p, k} (\theta)$ are 
\begin{align}  
\label{if1_bp}
\left\{ \begin{array}{c}
\IF_{1, \tilde{\bar{\zeta}}_{b, k}} (\theta) = - \Sigma_{k}^{-1} \left( S_{bp} \tilde{b}_{k, \theta} (X) \zbar_{k} (X) + m_{2} (O, \zbar_{k}) \right), \\ 
\IF_{1, \tilde{\bar{\zeta}}_{p, k}} (\theta) = - \Sigma_{k}^{-1} \left( S_{bp} \tilde{p}_{k, \theta} (X) \zbar_{k} (X) + m_{1} (O, \zbar_{k}) \right).
\end{array} \right.
\end{align}
\end{lemma}

\begin{proof}
We only need to show $\hat{\IF}_{1, \tilde{\bar{\zeta}}_{b, k}} (\theta)$ and $\hat{\IF}_{1, \tilde{\bar{\zeta}}_{p, k}} (\theta)$ will follow by symmetry. Recall that 
\begin{equation*}
\tilde{\bar{\zeta}}_{b, k} (\theta) = - \left\{ \E_{\theta} \left[ S_{bp} \zbar_{k} (X) \zbar_{k} (X)^{\top} \right] \right\}^{-1} \E_{\theta} \left[ S_{bp} \hat{b} (X) \zbar_{k} (X) + m_{2} (O, \zbar_{k}) \right].
\end{equation*}
Then 
\begin{align*}
& \; \IF_{1, \tilde{\bar{\zeta}}_{b, k}} (\theta) \\
= & - \left\{ \E_{\theta}\left[ S_{bp} \zbar_{k} (X) \zbar_{k} (X)^{\top} \right] \right\}^{-1} \mathsf{IF}_{1, \E_{\cdot} \left[ S_{bp} \hat{b} (X) \zbar_{k} (X) + m_{2} (O, \zbar_{k}) \right]} (\theta) \\
& - \IF_{1, \left\{ \E_{\cdot} \left[ S_{bp} \zbar_{k} (X) \zbar_{k} (X)^{\top} \right] \right\}^{-1}} (\theta) \E_{\theta} \left[ S_{bp} \hat{b} (X) \zbar_{k} (X) + m_{2} (O, \zbar_{k}) \right] \\
= & - \Sigma_{k}^{-1} \left( S_{bp} \hat{b} (X) \zbar_{k} (X) + m_{2} (O, \zbar_{k}) - \E_{\theta} \left[ S_{bp} \hat{b} (X) \zbar_{k} (X) + m_{2} (O, \zbar_{k}) \right] \right) \\
& + \Sigma_{k}^{-1} \left( S_{bp} \zbar_{k} (X) \zbar_{k} (X)^{\top} - \Sigma_{k} \right) \Sigma_{k}^{-1} \E_{\theta} \left[ S_{bp} \hat{b} (X) \zbar_{k} (X) + m_{2} (O, \zbar_{k}) \right] \\
= & - \Sigma_{k}^{-1} \left( S_{bp} \hat{b} (X) \zbar_{k} (X) + m_{2} (O, \zbar_{k}) - S_{bp} \zbar_{k} (X)^{\top} \Sigma_{k}^{-1} \E_{\theta} \left[ S_{bp} \hat{b} (X) \zbar_{k} (X) + m_{2} (O, \zbar_{k}) \right] \zbar_{k} (X)
\right) \\
= & - \Sigma_{k}^{-1} \left( S_{bp} \hat{b} (X) \zbar_{k} (X) + m_{2} (O, \zbar_{k}) + S_{bp} \zbar_{k} (X)^{\top} \tilde{\bar{\zeta}}_{b, k} (\theta) \zbar_{k} (X) \right) \\
= & - \Sigma_{k}^{-1} \left\{ S_{bp} \left( \hat{b} (X) + \zbar_{k} (X)^{\top} \tilde{\bar{\zeta}}_{b, k} (\theta) \right) \zbar_{k} (X) + m_{2} (O, \zbar_{k}) \right\} \\
= & - \Sigma_{k}^{-1} \left( S_{bp} \tilde{b}_{k, \theta} (X) \zbar_{k} (X) + m_{2} (O, \zbar_{k}) \right)
\end{align*}
where the fourth equality follows from the definition of $\tilde{\bar{\zeta}}_{b, k} (\theta)$ and the sixth equality follows from the definition of $\tilde{b}_{k, \theta} (X)$.
\end{proof}

Under conditions in Proposition \ref{thm:if1}, $\tilde{b}_{k, \hat{\theta}} = \hat{b}$ and $\tilde{p}_{k, \hat{\theta}} = \hat{p}$ or equivalently $\tilde{\bar{\zeta}}_{b, k} (\hat{\theta}) = \tilde{\bar{\zeta}}_{p, k} (\hat{\theta}) = 0$. This follows from results in Proposition \ref{thm:if1} with $\theta$ evaluated at $\hat{\theta}$: 
\begin{align*}
\E_{\hat{\theta}} \left[ S_{bp} \hat{b} (X) h (X) + m_{2} (O, h) \right] = 0 \text{ for all } h \in \mathcal{B}, \E_{\hat{\theta}} \left[ S_{bp} \hat{p} (X) h (X) + m_{1} (O, h) \right] = 0 \text{ for all } h \in \mathcal{P}.
\end{align*}
Thus under $\hat{\theta}$, the HOIFs of $\tilde{\psi}_{k} (\hat{\theta})$ are: $\hat{\IIFF}_{m, k} (\Sigma_{k}^{-1}) = \hat{\IIFF}_{1} - \hat{\IIFF}_{22 \rightarrow mm, k} (\Sigma_{k}^{-1})$, where $\hat{\IIFF}_{\ell \ell \rightarrow mm, k} (\Sigma_{k}^{-1}) \coloneqq \sum_{j = \ell}^{m} \hat{\IIFF}_{jj, k} (\Sigma_{k}^{-1})$, 
\begin{align}
\hat{\IIFF}_{22, k} (\Sigma_{k}^{-1}) \coloneqq \frac{1}{n (n - 1)} \sum_{1 \leq i_{1} \neq i_{2} \leq n} \hat{\IF}_{22, k, \bar{i}_{2}} (\Sigma_{k}^{-1}) \label{eq:IF22_hat}
\end{align}
and for $j > 2$, 
\begin{align}
\hat{\IIFF}_{jj, k} (\Sigma_{k}^{-1}) \coloneqq \frac{(n - j)!}{n!} \sum_{1 \leq i_{1} \neq \ldots \neq i_{j} \leq n} \hat{\IF}_{jj, k, \bar{i}_{j}} (\Sigma_{k}^{-1}). \label{eq:IFjj_hat}
\end{align}
Here 
\begin{align}
\hat{\IF}_{22, k, \bar{i}_{2}} (\Sigma_{k}^{-1}) \equiv \IF_{22, \tilde{\psi}_{k}, \bar{i}_{2}} (\hat{\theta}) = & \ \left[ \calE_{\hat{b}, m_{2}} \left( \zbar_{k} \right) (O) \right]^{\top}_{i_{1}} \Sigma_{k}^{-1} \left[ \calE_{\hat{p}, m_{1}} \left( \zbar_{k} \right) (O) \right]_{i_{2}}, \label{eq:if22_hat}
\end{align}
and 
\begin{align}
& \; \hat{\IF}_{jj, k, \bar{i}_{j}} (\Sigma_{k}^{-1}) \equiv \IF_{jj, \tilde{\psi}_{k}, \bar{i}_{j}} (\hat{\theta}) \notag \\
= & \; (-1)^{j} \left[ \calE_{\hat{b}, m_{2}} \left( \zbar_{k} \right) (O) \right]^{\top}_{i_{1}} \left\{ \prod\limits_{s = 3}^{j} \Sigma_{k}^{-1} \left( \left[ S_{bp} \zbar_{k} (X) \zbar_{k} (X)^{\top} \right]_{i_{s}} - \Sigma_{k} \right) \right\} \Sigma_{k}^{-1} \left[ \calE_{\hat{p}, m_{1}} \left( \zbar_{k} \right) (O) \right]_{i_{2}}. \label{eq:ifjj_hat}
\end{align}
are the kernels of second order influence function $\hat{\IIFF}_{22, k} (\Sigma_{k}^{-1})$ and $j$-th order influence function $\hat{\IIFF}_{jj, k} (\Sigma_{k}^{-1})$ respectively, where $\calE_{\hat{b}, m_{2}} $ and $\calE_{\hat{p}, m_{1}}$ are defined in equations \eqref{calE:b} and \eqref{calE:p}.

As a consequence of Theorem \ref{thm:hoif}, the $m$-th order influence function for the truncated parameter $\tilde{\psi}_{k} (\theta)$ under $\P_{\hat{\theta}}$ is simply $\hat{\psi}_{m, k} (\Sigma_{k}^{-1}) - \tilde{\psi}_{k} (\theta) \equiv \hat{\IIFF}_{m, k} (\Sigma_{k}^{-1})$ and the $m$-th order influence function for the bias $\Bias_{\theta, k} (\hat{\psi}_{1})$ is $\hat{\IIFF}_{22 \rightarrow mm, k} (\Sigma_{k}^{-1})$, with the $b$ and $p$ in $\IIFF_{m, k} (\Sigma_{k}^{-1})$ and $\IIFF_{22 \rightarrow mm, k} (\Sigma_{k}^{-1})$ replaced by $\hat{b}$ and $\hat{p}$.

\subsection{Statistical properties of $\hat{\chi}_{2, k} (\Sigma_{k}^{-1}; \varsigma_{k}, \delta)$}
\label{app:oracle_test} 
As mentioned above, based on the statistical properties of $\hat{\psi}_{1}$ and $\hat{\IIFF}_{22, k} (\Sigma_{k}^{-1})$ summarized in Propositions \ref{thm:drml} and \ref{thm:oracle_soif}, for a DR functional $\psi (\theta)$, we study the statistical property of the oracle two-sided test $\hat{\chi}_{2, k} (\Sigma_{k}^{-1}; \varsigma_{k}, \delta)$ as a test of the surrogate null hypothesis $\H_{0, k} (\delta)$.

The following theorem characterizes the asymptotic level and power of $\hat{\chi}_{2, k} (\Sigma_{k}^{-1}; \varsigma_{k}, \delta)$ for $\H_{0, k} (\delta)$ and the asymptotic level under $\H_{0, \CS} (\delta)$.

\begin{theorem}
\label{thm:oracle_test} 
For a DR functional $\psi (\theta)$, under the conditions in Proposition \ref{thm:if1}, Proposition \ref{thm:oracle_soif} and Condition \ref{cond:sw}, when $k \rightarrow \infty$ but $k = o(n)$, for any given $\varsigma_{k}, \delta > 0$, suppose that $\frac{\vert \Bias_{\theta, k} (\hat{\psi}_{1}) \vert}{\se_{\theta} [\hat{\psi}_{1}]} = \gamma$ for some (sequence) $\gamma = \gamma(n)$ (where $\gamma(n)$ can diverge with $n$), then the rejection probability of $\hat{\chi}_{2, k} (\Sigma_{k}^{-1}; \varsigma_{k}, \delta)$ converges to 
\begin{equation}  
\label{rejection:2}
2 - \Phi \left( \varsigma_{k} - \lim_{n \rightarrow \infty} (\gamma - \delta) \frac{\se_{\theta} [\hat{\psi}_{1}]}{\se_{\theta} [\hat{\IIFF}_{22, k} (\Sigma_{k}^{-1})]} \right) - \Phi \left( \varsigma_{k} + \lim_{n \rightarrow \infty} (\gamma + \delta) \frac{\se_{\theta} [\hat{\psi}_{1}]}{\se_{\theta} [\hat{\IIFF}_{22, k} (\Sigma_{k}^{-1})]} \right)
\end{equation}
as $n \rightarrow \infty$. In particular,

\begin{enumerate}[label=(\arabic*)]
\item under $\H_{0, k} (\delta): \gamma \leq \delta$, $\hat{\chi}_{2, k} (\Sigma_{k}^{-1}; \varsigma_{k}, \delta)$ rejects the null with probability less than or equal to $2 (1 - \Phi (\varsigma_{k}))$, as $n \rightarrow \infty$;
\item under the following alternative to $\H_{0, k} (\delta)$: $\gamma = \delta + c$, for any diverging sequence $c = c(n) \rightarrow \infty$, $\hat{\chi}_{2, k} (\Sigma_{k}^{-1}; \varsigma_{k}, \delta)$ rejects the null with probability converging to 1, as $n \rightarrow \infty$.

\begin{enumerate}[label=(\arabic*')] \setcounter{enumi}{2}
\item[(2')] If $\BL_{\theta, 2, \hat{b}, k}$ and $\BL_{\theta, 2, \hat{p}, k}$ converge to 0, under the following alternative to $\H_{0, k} (\delta)$: $\gamma = \delta + c$, for any fixed $c > 0$ or any diverging sequence $c = c(n) \rightarrow \infty$, $\hat{\chi}_{2, k} (\Sigma_{k}^{-1}; \varsigma_{k}, \delta)$ has rejection probability converging to 1, as $n \rightarrow \infty$.
\end{enumerate}
\end{enumerate}
As an immediate consequence, under $\H_{0, \CS} (\delta)$, the rejection probability of $\hat{\chi}_{2, k} (\Sigma_{k}^{-1}; z_{\alpha^{\dag} / 2}, \delta)$ is asymptotically no greater than $\alpha^{\dag}$.
\end{theorem}

\begin{remark}
\leavevmode\label{rem:oracle_test} 
In Appendix \ref{app:oracle_test_proof}, we prove equation \eqref{rejection:2}. We now prove that equation \eqref{rejection:2} implies Theorem \ref{thm:oracle_test}(1)-(2) and (2').

\begin{itemize}
\item Regarding (1), under $\H_{0, k} (\delta): \gamma \leq \delta$, 
\begin{equation*}
- (\gamma - \delta) \frac{\se_{\theta} [\hat{\psi}_{1}]}{\se_{\theta} [\hat{\IIFF}_{22, k}]} \geq 0 \text{ and } (\gamma + \delta) \frac{\se_{\theta} [\hat{\psi}_{1}]}{\se_{\theta} [\hat{\IIFF}_{22, k}]} \geq 0,
\end{equation*}
which implies that the rejection probability is less than or equal to $2 - 2 \Phi (\varsigma_{k})$. Choose $\varsigma_{k} = z_{\alpha^{\dag} / 2}$, $2 (1 - \Phi (\varsigma_{k})) = 2 \alpha^{\dag} / 2 = \alpha^{\dag}$ and conclude that the test is a valid level $\alpha^{\dag}$ test of the null.

\item Under the alternative to $\H_{0, k} (\delta)$ with $\gamma = \delta + c$ for some $c > 0$, it follows from Proposition \ref{thm:oracle_soif} and equation \eqref{rejection:2} that the rejection probability of $\hat{\chi}_{2, k} (\Sigma_{k}^{-1}; \varsigma_{k}, \delta)$, as $n \rightarrow \infty$, is no smaller than 
\begin{align*}
\begin{array}{c}
2 - \Phi \left( \varsigma_{k} - c \Theta(b, p, \hat{b}, \hat{p}, f, \zbar_{k}) \left\{ \dfrac{k}{n} + \BL_{\theta, 2, \hat{b}, k} + \BL_{\theta, 2, \hat{p}, k} \right\}^{-1} \right) - \Phi (\infty)
\end{array}
\end{align*}
where $\Theta (b, p, \hat{b}, \hat{p}, f, \zbar_{k})$ is some positive constant depending on the true regression functions $b$ and $p$, the estimated functions $\hat{b}, \hat{p}$ from the training sample, the density $f$ of $X$ and the chosen basis functions $\zbar_{k}$. To have power approaching 1 to reject $\H_{0, k} (\delta)$, we need one of the following:

\begin{itemize}
\item If one of $\BL_{\theta, 2, \hat{b}, k}$ and $\BL_{\theta, 2, \hat{p}, k}$ is $O(1)$, we need $c \rightarrow \infty$ to guarantee that the rejection probability of $\hat{\chi}_{2, k} (\Sigma_{k}^{-1}; \varsigma_{k}, \delta)$ converges to $1 - \Phi(-\infty) = 1$. Hence we have Theorem \ref{thm:oracle_test}(2).

\item If $c$ is fixed, we need both $\BL_{\theta, 2, \hat{b}, k}$ and $\BL_{\theta, 2, \hat{p}, k}$ to be $o(1)$ to guarantee that the rejection probability of $\hat{\chi}_{2, k} (\Sigma_{k}^{-1}; \varsigma_{k}, \delta)$ converges to $1 - \Phi(-\infty) = 1$. Hence we have Theorem \ref{thm:oracle_test}(2').
\end{itemize}
\end{itemize}
\end{remark}

\subsubsection{Proof of Theorem \ref{thm:oracle_test}}
\label{app:oracle_test_proof}
\allowdisplaybreaks
\begin{proof}
The rejection probability of $\hat{\chi}_{2, k} (\Sigma_{k}^{-1}; \varsigma_{k}, \delta)$ is computed as follows: 
\begin{align*}
& \; \lim_{n \rightarrow \infty} \P_{\theta} \left( \frac{\vert \hat{\IIFF}_{22, k} (\Sigma_{k}^{-1}) \vert}{\hat{\se} [\hat{\psi}_{1}]} - \varsigma_{k} \frac{\hat{\se} [\hat{\IIFF}_{22, k} (\Sigma_{k}^{-1})]}{\hat{\se} [\hat{\psi}_{1}]} > \delta \right) \\
= & \; \lim_{n \rightarrow \infty} \left\{ \begin{array}{c}
\P_{\theta} \left( \frac{\hat{\IIFF}_{22, k} (\Sigma_{k}^{-1})}{\hat{\se} [\hat{\IIFF}_{22, k} (\Sigma_{k}^{-1})]} > \varsigma_{k} + \delta \frac{\hat{\se} [\hat{\psi}_{1}]}{\hat{\se} [\hat{\IIFF}_{22, k} (\Sigma_{k}^{-1})]} \right) \\ 
+ \ \P_{\theta} \left( \frac{\hat{\IIFF}_{22, k} (\Sigma_{k}^{-1})}{\hat{\se} [\hat{\IIFF}_{22, k} (\Sigma_{k}^{-1})]} < - \varsigma_{k} - \delta \frac{\hat{\se} [\hat{\psi}_{1}]}{\hat{\se} [\hat{\IIFF}_{22, k} (\Sigma_{k}^{-1})]} \right)
\end{array} \right\} \\
= & \; \lim_{n \rightarrow \infty} \P_{\theta} \left( \frac{\hat{\IIFF}_{22, k} (\Sigma_{k}^{-1}) - \Bias_{\theta, k} (\hat{\psi}_{1})}{\hat{\se} [\hat{\IIFF}_{22, k} (\Sigma_{k}^{-1})]} > \varsigma_{k} - \frac{\Bias_{\theta, k} (\hat{\psi}_{1})}{\hat{\se} [\hat{\IIFF}_{22, k} (\Sigma_{k}^{-1})]} + \delta \frac{\hat{\se} [\hat{\psi}_{1}]}{\hat{\se} [\hat{\IIFF}_{22, k} (\Sigma_{k}^{-1})]} \right) \\
& \; + \lim_{n \rightarrow \infty} \P_{\theta} \left( \frac{\hat{\IIFF}_{22, k} (\Sigma_{k}^{-1}) - \Bias_{\theta, k} (\hat{\psi}_{1})}{\hat{\se} [\hat{\IIFF}_{22, k} (\Sigma_{k}^{-1})]} < - \varsigma_{k} - \frac{\Bias_{\theta, k} (\hat{\psi}_{1})}{\hat{\se} [\hat{\IIFF}_{22, k} (\Sigma_{k}^{-1})]} - \delta \frac{\hat{\se} [\hat{\psi}_{1}]}{\hat{\se} [\hat{\IIFF}_{22, k} (\Sigma_{k}^{-1})]} \right) \\
= & \; \lim_{n \rightarrow \infty} \P_{\theta} \left( \frac{\hat{\IIFF}_{22, k} (\Sigma_{k}^{-1})) - \Bias_{\theta, k} (\hat{\psi}_{1})}{\se_{\theta} [\hat{\IIFF}_{22, k} (\Sigma_{k}^{-1})]} (1 + o_{\P_{\theta}}(1)) > \varsigma_{k} - (\gamma - \delta) \frac{\se_{\theta} [\hat{\psi}_{1}]}{\se_{\theta} [\hat{\IIFF}_{22, k} (\Sigma_{k}^{-1})]} (1 + o_{\P_{\theta}}(1)) \right) \\
& \; + \lim_{n \rightarrow \infty} \P_{\theta} \left( \frac{\hat{\IIFF}_{22, k} (\Sigma_{k}^{-1}) - \Bias_{\theta, k} (\hat{\psi}_{1})}{\se_{\theta} [\hat{\IIFF}_{22, k} (\Sigma_{k}^{-1})} (1 + o_{\P_{\theta}}(1)) < - \varsigma_{k} - (\gamma + \delta) \frac{\se_{\theta} [\hat{\psi}_{1}]}{\se_{\theta} [\hat{\IIFF}_{22, k} (\Sigma_{k}^{-1})]} (1 + o_{\P_{\theta}}(1)) \right) \\
= & \; 1 - \Phi \left( \varsigma_{k} - (\gamma - \delta) \frac{\se_{\theta} [\hat{\psi}_{1}]}{\se_{\theta} [\hat{\IIFF}_{22, k} (\Sigma_{k}^{-1})]} \right) + \Phi \left( - \varsigma_{k} - (\gamma + \delta) \frac{\se_{\theta} [\hat{\psi}_{1}]}{\se_{\theta} [\hat{\IIFF}_{22, k} (\Sigma_{k}^{-1})]} \right) \\
= & \; 2 - \Phi \left( \varsigma_{k} - (\gamma - \delta) \frac{\se_{\theta} [\hat{\psi}_{1}]}{\se_{\theta} [\hat{\IIFF}_{22, k} (\Sigma_{k}^{-1})]} \right) - \Phi \left( \varsigma_{k} + (\gamma + \delta) \frac{\se_{\theta} [\hat{\psi}_{1}]}{\se_{\theta} [\hat{\IIFF}_{22, k} (\Sigma_{k}^{-1})]} \right).
\end{align*}
\end{proof}

\subsection{General formula of the variance order of $\hat{\IIFF}_{mm, k} (\hat{\Sigma}_{k}^{-1})$}
\label{app:var} 

In this section we derive the following formula for the order of $\var_{\theta} \left[ \hat{\IIFF}_{mm, k} (\hat{\Sigma}_{k}^{-1}) \right]$ for general $m$. Then we have

\begin{lemma}
\label{lem:hoif_var} 
When $m \geq 3$, there exists a constant $C > 0$ that depends only on the spectra of $\Sigma$ and $\hat{\Sigma}$, such that
\begin{equation}  
\label{eq:hoif_var}
\begin{split}
& \ \var_{\theta} \left[ \hat{\IIFF}_{mm, k} (\hat{\Sigma}_{k}^{-1}) \right] \\
\lesssim & \ \frac{1}{n} \left\{ \begin{array}{c}
m^{2 m} \left( \dfrac{C k}{n} \right)^{m - 1} + \sum\limits_{j = 2}^{m - 1} m^{2 j} \left( \dfrac{C k}{n} \right)^{j - 1} \BL_{\theta, 2, \hat{\Sigma}, k}^{2 (m - j - 2) \vee 0} \left\{ \begin{array}{c}
\BL_{\theta, 2, \hat{\Sigma}, k}^{4} + \BL_{\theta, 2, \hat{b}, k}^{2} \BL_{\theta, 2, \hat{\Sigma}, k}^{2} + \BL_{\theta, 2, \hat{p}, k}^{2} \BL_{\theta, 2, \hat{\Sigma}, k}^{2} + \BL_{\theta, 4, \hat{b}, k}^{2} \BL_{\theta, 4, \hat{p}, k}^{2}
\end{array} \right\} \\ 
+ \ m^{2} \BL_{\theta, 2, \hat{\Sigma}, k}^{2 (m - 3)} \left\{ \BL_{\theta, 2, \hat{b}, k}^{2} \BL_{\theta, 2, \hat{\Sigma}, k}^{2} + \BL_{\theta, 2, \hat{p}, k}^{2} \BL_{\theta, 2, \hat{\Sigma}, k}^{2} + (\BL_{\theta, 4, \hat{b}, k}^{2} \BL_{\theta, 4, \hat{p}, k}^{2} \wedge \BL_{\theta, 2, \hat{b}, k}^{2} \BL_{\theta, \infty, \hat{p}, k}^{2} \wedge \BL_{\theta, \infty, \hat{b}, k}^{2} \BL_{\theta, 2, \hat{p}, k}^{2})\right\}
\end{array} \right\} \\
\overset{Condition \ \ref{cond:sw}}{\lesssim} & \ \frac{1}{n} \left\{ \begin{array}{c}
m^{2m} \left( \dfrac{C k}{n} \right)^{m - 1} + \sum\limits_{j = 2}^{m - 1} m^{2 j} \left( \dfrac{C k}{n} \right)^{j - 1} \BL_{\theta, 2, \hat{\Sigma}, k}^{2 (m - j - 2) \vee 0} \left\{ \begin{array}{c}
\BL_{\theta, 2, \hat{\Sigma}, k}^{4} + \BL_{\theta, 2, \hat{b}, k}^{2} \BL_{\theta, 2, \hat{\Sigma}, k}^{2} + \BL_{\theta, 2, \hat{p}, k}^{2} \BL_{\theta, 2, \hat{\Sigma}, k}^{2} + k \BL_{\theta, 2, \hat{b}, k}^{2} \BL_{\theta, 2, \hat{p}, k}^{2}
\end{array} \right\} \\ 
+ \ m^{2} \BL_{\theta, 2, \hat{\Sigma}, k}^{2 (m - 3)} \left\{ \BL_{\theta, 2, \hat{b}, k}^{2} \BL_{\theta, 2, \hat{\Sigma}, k}^{2} + \BL_{\theta, 2, \hat{p}, k}^{2} \BL_{\theta, 2, \hat{\Sigma}, k}^{2} + k \BL_{\theta, 2, \hat{b}, k}^{2} \BL_{\theta, 2, \hat{p}, k}^{2} \right\}
\end{array} \right\} \\
\overset{Condition \ \ref{cond:w}}{\lesssim} & \ \frac{1}{n} \left\{ \begin{array}{c}
m^{2 m} \left( \dfrac{C k}{n} \right)^{m - 1} + \sum\limits_{j = 2}^{m - 1} m^{2 j} \left( \dfrac{C k}{n} \right)^{j - 1} \BL_{\theta, 2, \hat{\Sigma}, k}^{2 (m - j - 2) \vee 0} \left\{ \begin{array}{c}
\BL_{\theta, 2, \hat{\Sigma}, k}^{4} + \BL_{\theta, 2, \hat{b}, k}^{2} \BL_{\theta, 2, \hat{\Sigma}, k}^{2} + \BL_{\theta, 2, \hat{p}, k}^{2} \BL_{\theta, 2, \hat{\Sigma}, k}^{2} \\ 
+ \ \left\{ \BL_{\theta, 2, \hat{b}, k}^{2} \BL_{\theta, \infty, \hat{p}, k}^{2} \wedge \BL_{\theta, \infty, \hat{b}, k}^{2} \BL_{\theta, 2, \hat{p}, k}^{2} \right\}
\end{array} \right\} \\ 
+ \ m^{2} \BL_{\theta, 2, \hat{\Sigma}, k}^{2 (m - 3)} \left\{ \BL_{\theta, 2, \hat{b}, k}^{2} \BL_{\theta, 2, \hat{\Sigma}, k}^{2} + \BL_{\theta, 2, \hat{p}, k}^{2} \BL_{\theta, 2, \hat{\Sigma}, k}^{2} + \left\{ \BL_{\theta, 2, \hat{b}, k}^{2} \BL_{\theta, \infty, \hat{p}, k}^{2} \wedge \BL_{\theta, \infty, \hat{b}, k}^{2} \BL_{\theta, 2, \hat{p}, k}^{2} \right\} \right\}
\end{array} \right\}.
\end{split}
\end{equation}
\end{lemma}

\begin{proof}
The detailed calculations can be found in the Appendix of \citet{liu2017semiparametric}. The two different versions of \eqref{eq:hoif_var} are due to the application of \Holder{} inequality with different \Holder{} conjugate pairs (i.e. $(2, 2)$ and $(1, \infty)$). When Condition \ref{cond:w} does not hold but Condition \ref{cond:sw} holds, we have
\begin{align*}
\BL_{\theta, 4, \hat{b}, k}^{2} \BL_{\theta, 4, \hat{p}, k}^{2} \leq k \BL_{\theta, 2, \hat{b}, k}^{2} \BL_{\theta, 2, \hat{p}, k}^{2},
\end{align*}
hence the second line inequality in \eqref{eq:hoif_var} under only Condition \ref{cond:sw}. But this upper bound may not be tight. For example, when $\zbar_{k}$ is Fourier or polynomials series that violates Condition \ref{cond:w} yet satisfies Condition \ref{cond:sw} \citep{belloni2015some},
\begin{align*}
\BL_{\theta, 4, \hat{b}, k}^{2} \BL_{\theta, 4, \hat{p}, k}^{2} \leq \log k \BL_{\theta, 2, \hat{b}, k}^{2} \BL_{\theta, 2, \hat{p}, k}^{2}
\end{align*}
because $\BL_{\theta, \infty, \hat{b}, k}, \BL_{\theta, \infty, \hat{p}, k} \lesssim \log k$ when $\BL_{\theta, \infty, \hat{b}}, \BL_{\theta, \infty, \hat{p}}$ are $O (1)$. But for Legendre polynomial series which also violates Condition \ref{cond:w} yet satisfies Condition \ref{cond:sw} \citep{belloni2015some}, we currently can only have the following bound
\begin{align*}
\BL_{\theta, 4, \hat{b}, k}^{2} \BL_{\theta, 4, \hat{p}, k}^{2} \leq k \BL_{\theta, 2, \hat{b}, k}^{2} \BL_{\theta, 2, \hat{p}, k}^{2}.
\end{align*}
\end{proof}

\subsection{Bootstrapping higher order influence functions}
\label{app:bootstrap} 
In this section, we will use higher order moments of multinomial distributions frequently. Two good references are \citet{mosimann1962compound, newcomer2008computation}. For notational convenience, we suppress the dependence on $\Sigma_{k}^{-1}$ or $\hat{\Sigma}_{k}^{-1}$ in $\hat{\IIFF}_{mm, k} (\Sigma_{k}^{-1})$ and $\hat{\IIFF}_{mm, k} (\hat{\Sigma}_{k}^{-1})$ as the proposed bootstrap resampling procedure will not depend on $\Sigma_{k}^{-1}$ or the training sample. We propose the following bootstrap estimator of $\var_{\theta} [\hat{\IIFF}_{22, k}]$: choose some large integer $M$ as the number of bootstrap resamples, and define 
\begin{equation}  \label{eq:if22_bvar}
\hat{\var} [\hat{\IIFF}_{22, k}] \coloneqq \underbrace{\frac{1}{M - 1} \sum_{m = 1}^{M} \left( \hat{\IIFF}_{22, k}^{(m)} - \frac{1}{M} \sum_{m = 1}^{M} \hat{\IIFF}_{22, k}^{(m)} \right)^{2}}_{T_{1}} - \underbrace{\frac{2}{M - 1} \sum_{m = 1}^{M} \left( \hat{\IIFF}_{22, k}^{(m), c} - \frac{1}{M} \sum_{m = 1}^{M} \hat{\IIFF}_{22, k}^{(m), c} \right)^{2}}_{T_{2}}
\end{equation}
where 
\begin{align*}
\hat{\IIFF}_{22, k}^{(m)} & = \frac{1}{n (n - 1)} \sum_{1 \leq i_{1} \neq i_{2} \leq n} W_{i_{1}}^{(m)} W_{i_{2}}^{(m)} \hat{\IF}_{22, k, \bar{i}_{2}} \\
\hat{\IIFF}_{22, k}^{(m), c} & = \frac{1}{n (n - 1)} \sum_{1 \leq i_{1} \neq i_{2} \leq n} (W_{i_{1}}^{(m)} - 1) (W_{i_{2}}^{(m)} - 1) \hat{\IF}_{22, k, \bar{i}_{2}}
\end{align*}
and $\bm{W}^{(m)} = (W_{1}^{(m)}, \ldots, W_{n}^{(m)}) \overset{i.i.d.}{\sim} \mathsf{Multinom} (n, \underbrace{n^{-1}, \ldots, n^{-1}}_{(n - 1)\text{'s}})$ for $m = 1, \ldots, M$ are $M$ multinomial weights independent of the observed data $\{O_{i}, i = 1, \ldots, N\}$.

\begin{remark}
\label{rem:boot_if22} 
We now explain why $\hat{\var} [\hat{\IIFF}_{22, k}]$ is proposed as an estimator of $\var_{\theta} [\hat{\IIFF}_{22, k}]$. Conditional on the observed data (including both the training and estimation samples), the expectation over the random multinomial weights $\bm{W}$ of the term $T_{1}$ is: 
\begin{align*}
& \ \E_{\bm{W}} [T_{1} \vert \{O_{i}\}_{i = 1}^{N}] \\
= & \ \E_{\bm{W}} \left[ \left( \frac{1}{n (n - 1)} \sum_{1 \leq i_{1} \neq i_{2} \leq n} W_{i_{1}}^{(m)} W_{i_{2}}^{(m)} \hat{\IF}_{22, k, i_{1}, i_{2}} \right)^{2} \vert \{O_{i}\}_{i = 1}^{N} \right] \\
& - \left( \E_{\bm{W}} \left[ \frac{1}{n (n - 1)} \sum_{1 \leq i_{1} \neq i_{2} \leq n} W_{i_{1}}^{(m)} W_{i_{2}}^{(m)} \hat{\IF}_{22, k, i_{1}, i_{2}} \vert \{O_{i}\}_{i = 1}^{N} \right] \right)^{2} \\
= & \ \frac{1}{n^{2} (n - 1)^{2}} \sum_{1 \leq i_{1} \neq i_{2} \leq n} \{\hat{\IF}_{22, k, i_{1}, i_{2}} \hat{\IF}_{22, k, i_{1}, i_{2}} + \hat{\IF}_{22, k, i_{1}, i_{2}} \hat{\IF}_{22, k, i_{2}, i_{1}}\} \{\E_{\bm{W}} [W_{i_{1}}^{2} W_{i_{2}}^{2}] - \left( \E_{\bm{W}} [W_{i_{1}} W_{i_{2}}] \right)^{2}\} \\
& + \frac{1}{n^{2} (n - 1)^{2}} \sum_{1 \leq i_{1} \neq i_{2} \neq i_{3} \leq n} \left\{ \begin{array}{c}
\hat{\IF}_{22, k, i_{1}, i_{2}} \hat{\IF}_{22, k, i_{1}, i_{3}} \{\E_{\bm{W}} [W_{i_{1}}^{2} W_{i_{2}} W_{i_{3}}] - \E_{\bm{W}} [W_{i_{1}} W_{i_{2}}] \E_{\bm{W}} [W_{i_{1}} W_{i_{3}}]\} \\ 
+ \ \hat{\IF}_{22, k, i_{1}, i_{2}} \hat{\IF}_{22, k, i_{3}, i_{1}} \{\E_{\bm{W}} [W_{i_{1}}^{2} W_{i_{2}} W_{i_{3}}] - \E_{\bm{W}} [W_{i_{1}} W_{i_{2}}] \E_{\bm{W}} [W_{i_{3}} W_{i_{1}}]\} \\ 
+ \ \hat{\IF}_{22, k, i_{1}, i_{2}} \hat{\IF}_{22, k, i_{2}, i_{3}} \{\E_{\bm{W}} [W_{i_{1}} W_{i_{2}}^{2} W_{i_{3}}] - \E_{\bm{W}} [W_{i_{1}} W_{i_{2}}] \E_{\bm{W}} [W_{i_{2}} W_{i_{3}}]\} \\ 
+ \ \hat{\IF}_{22, k, i_{1}, i_{2}} \hat{\IF}_{22, k, i_{3}, i_{2}} \{\E_{\bm{W}} [W_{i_{1}} W_{i_{2}}^{2} W_{i_{3}}] - \E_{\bm{W}} [W_{i_{1}} W_{i_{2}}] \E_{\bm{W}} [W_{i_{3}} W_{i_{2}}]\}
\end{array} \right\} \\
& + \frac{1}{n^{2} (n - 1)^{2}} \sum_{1 \leq i_{1} \neq i_{2} \neq i_{3} \neq i_{4} \leq n} \hat{\IF}_{22, k, i_{1}, i_{2}} \hat{\IF}_{22, k, i_{3}, i_{4}} \{\E_{\bm{W}} [W_{i_{1}} W_{i_{2}} W_{i_{3}} W_{i_{4}}] - \E_{\bm{W}} [W_{i_{1}} W_{i_{2}}] \E_{\bm{W}} [W_{i_{3}} W_{i_{4}}]\}.
\end{align*}
Plugging in the following higher order moments of $\mathsf{Multinom} (n, \underbrace{n^{-1}, \ldots, n^{-1}}_{(n - 1)\text{'s}})$ \citep{newcomer2008computation}: 
\begin{align*}
\E_{\bm{W}} [W_{1} W_{2}] & = \frac{n - 1}{n}, \E_{\bm{W}} [W_{1} W_{2} W_{3}] = \frac{(n - 1) (n - 2)}{n^{2}}, \E_{\bm{W}} [W_{1} W_{2} W_{3} W_{4}] = \frac{(n - 1) (n - 2) (n - 3)}{n^{3}}, \\
\E_{\bm{W}} [W_{1}^{2} W_{2}] & = \frac{2 (n - 1)^{2}}{n^{2}}, \E_{\bm{W}} [W_{1}^{2}] = \frac{2 n - 1}{n}, \\
\E_{\bm{W}} [W_{1}^{2} W_{2} W_{3}] & = \frac{(n - 1) (n - 2) (n - 3) + n (n - 1) (n - 2)}{n^{3}} \\
& = \frac{(n - 1) (2 n^{2} - 7 n + 6)}{n^{3}}, \\
\E_{\bm{W}} [W_{1}^{2} W_{2}^{2}] & = \frac{(n - 1) (n - 2) (n - 3) + 2 n (n - 1) (n - 2) + n^{2} (n - 1)}{n^{3}} \\
& = \frac{(n - 1) (4 n^{2} - 9 n + 6)}{n^{3}},
\end{align*}
we have 
\begin{align}
& \ \E_{\bm{W}} [T_{1} \vert \{O_{i}\}_{i = 1}^{N}]  \notag \\
= & \ \frac{1}{n^{2} (n - 1)^{2}} \sum_{1 \leq i_{1} \neq i_{2} \leq n} \{\hat{\IF}_{22, k, i_{1}, i_{2}} \hat{\IF}_{22, k, i_{1}, i_{2}} + \hat{\IF}_{22, k, i_{1}, i_{2}} \hat{\IF}_{22, k, i_{2}, i_{1}}\} \left( 3 - \frac{11}{n} + \frac{14}{n^{2}} - \frac{6}{n^{3}} \right) \label{inflation} \\
& + \frac{1}{n^{2} (n - 1)^{2}} \sum_{1 \leq i_{1} \neq i_{2} \neq i_{3} \leq n} \left\{ \begin{array}{c}
\hat{\IF}_{22, k, i_{1}, i_{2}} \hat{\IF}_{22, k, i_{1}, i_{3}} + \hat{\IF}_{22, k, i_{1}, i_{2}} \hat{\IF}_{22, k, i_{3}, i_{1}} \nonumber \\ 
+ \ \hat{\IF}_{22, k, i_{1}, i_{2}} \hat{\IF}_{22, k, i_{2}, i_{3}} + \hat{\IF}_{22, k, i_{1}, i_{2}} \hat{\IF}_{22, k, i_{3}, i_{2}}
\end{array} \right\} \left( 1 - \frac{7}{n} + \frac{12}{n^{2}} - \frac{6}{n^{3}} \right) \notag \\
& + \frac{1}{n^{2} (n - 1)^{2}} \sum_{1 \leq i_{1} \neq i_{2} \neq i_{3} \neq i_{4} \leq n} \hat{\IF}_{22, k, i_{1}, i_{2}} \hat{\IF}_{22, k, i_{3}, i_{4}} \left( - \ \frac{4}{n} + \frac{10}{n^{2}} - \frac{6}{n^{3}} \right) \notag
\end{align}

The term $T_{2}$ is the bootstrap variance estimator for second order degenerate U-statistics and it has been proposed previously in \citet{arcones1992bootstrap, huskova1993consistency}. Similarly, conditional on the observed data (including both the training and estimation samples), the expectation over the random multinomial weights $\bm{W}$ of the term $T_{2}$ is: 
\begin{align*}
& \E_{\bm{W}} [T_{2} \vert \{O_{i}\}_{i = 1}^{N}] \\
& = \frac{2}{n^{2} (n - 1)^{2}} \sum_{1 \leq i_{1} \neq i_{2} \leq n} \{\hat{\IF}_{22, k, i_{1}, i_{2}} \hat{\IF}_{22, k, i_{1}, i_{2}} + \hat{\IF}_{22, k, i_{1}, i_{2}} \hat{\IF}_{22, k, i_{2}, i_{1}}\} \{\E_{\bm{W}} [(W_{i_{1}} - 1)^{2} (W_{i_{2}} - 1)^{2}] - (\E_{\bm{W}} [(W_{i_{1}} - 1) (W_{i_{2}} - 1)])^{2}\} \\
& + \frac{2}{n^{2} (n - 1)^{2}} \sum_{1 \leq i_{1} \neq i_{2} \neq i_{3} \leq n} \left\{ \begin{array}{c}
\hat{\IF}_{22, k, i_{1}, i_{2}} \hat{\IF}_{22, k, i_{1}, i_{3}} \{\E_{\bm{W}} [(W_{i_{1}} - 1)^{2} (W_{i_{2}} - 1) (W_{i_{3}} - 1)] - \E_{\bm{W}} [(W_{i_{1}} - 1) (W_{i_{2}} - 1)] \E_{\bm{W}} [(W_{i_{1}} - 1) (W_{i_{3}} - 1)]\} \\ 
+ \ \hat{\IF}_{22, k, i_{1}, i_{2}} \hat{\IF}_{22, k, i_{3}, i_{1}} \{\E_{\bm{W}} [(W_{i_{1}} - 1)^{2} (W_{i_{2}} - 1) (W_{i_{3}} - 1)] - \E_{\bm{W}} [(W_{i_{1}} - 1) (W_{i_{2}} - 1)] \E_{\bm{W}} [(W_{i_{3}} - 1) (W_{i_{1}} - 1)]\} \\ 
+ \ \hat{\IF}_{22, k, i_{1}, i_{2}} \hat{\IF}_{22, k, i_{2}, i_{3}} \{\E_{\bm{W}} [(W_{i_{1}} - 1) (W_{i_{2}} - 1)^{2} (W_{i_{3}} - 1)] - \E_{\bm{W}} [(W_{i_{1}} - 1) (W_{i_{2}} - 1)] \E_{\bm{W}} [(W_{i_{2}} - 1) (W_{i_{3}} - 1)]\} \\ 
+ \ \hat{\IF}_{22, k, i_{1}, i_{2}} \hat{\IF}_{22, k, i_{3}, i_{2}} \{\E_{\bm{W}} [(W_{i_{1}} - 1) (W_{i_{2}} - 1)^{2} (W_{i_{3}} - 1)] - \E_{\bm{W}} [(W_{i_{1}} - 1) (W_{i_{2}} - 1)] \E_{\bm{W}} [(W_{i_{3}} - 1) (W_{i_{2}} - 1)]\}
\end{array} \right\} \\
& + \frac{2}{n^{2} (n - 1)^{2}} \sum_{1 \leq i_{1} \neq i_{2} \neq i_{3} \neq i_{4} \leq n} \hat{\IF}_{22, k, i_{1}, i_{2}} \hat{\IF}_{22, k, i_{3}, i_{4}} \{\E_{\bm{W}} [(W_{i_{1}} - 1) (W_{i_{2}} - 1) (W_{i_{3}} - 1) (W_{i_{4}} - 1)] - \E_{\bm{W}} [(W_{i_{1}} - 1) (W_{i_{2}} - 1)] \E_{\bm{W}}[(W_{i_{3}} - 1) (W_{i_{4}} - 1)]\} \\
& = \frac{2}{n^{2} (n - 1)^{2}} \sum_{1 \leq i_{1} \neq i_{2} \leq n} \{\hat{\IF}_{22, k, i_{1}, i_{2}} \hat{\IF}_{22, k, i_{1}, i_{2}} + \hat{\IF}_{22, k, i_{1}, i_{2}} \hat{\IF}_{22, k, i_{2}, i_{1}}\} \{\E_{\bm{W}} [(W_{1}^{2} - 2 W_{1} + 1) (W_{2}^{2} - 2 W_{2} + 1)] - (\E_{\bm{W}} [(W_{1} - 1) (W_{2} - 1)])^{2}\} \\
& + \frac{2}{n^{2} (n - 1)^{2}} \sum_{1 \leq i_{1} \neq i_{2} \neq i_{3} \leq n} \left\{ \begin{array}{c}
\hat{\IF}_{22, k, i_{1}, i_{2}} \hat{\IF}_{22, k, i_{1}, i_{3}} + \hat{\IF}_{22, k, i_{1}, i_{2}} \hat{\IF}_{22, k, i_{3}, i_{1}} \\ 
+ \ \hat{\IF}_{22, k, i_{1}, i_{2}} \hat{\IF}_{22, k, i_{2}, i_{3}} + \hat{\IF}_{22, k, i_{1}, i_{2}} \hat{\IF}_{22, k, i_{3}, i_{2}}
\end{array} \right\} \{\E_{\bm{W}} [(W_{1} - 1)^{2} (W_{2} - 1) (W_{3} - 1)] - (\E_{\bm{W}} [(W_{1} - 1) (W_{2} - 1)])^{2}\} \\
& + \frac{2}{n^{2} (n - 1)^{2}} \sum_{1 \leq i_{1} \neq i_{2} \neq i_{3} \neq i_{4} \leq n} \hat{\IF}_{22, k, i_{1}, i_{2}} \hat{\IF}_{22, k, i_{3}, i_{4}} \{\E_{\bm{W}} [(W_{1} - 1) (W_{2} - 1) (W_{3} - 1) (W_{4} - 1)] - (\E_{\bm{W}} [(W_{1} - 1) (W_{2} - 1)])^{2}\} \\
& = \frac{2}{n^{2} (n - 1)^{2}} \sum_{1 \leq i_{1} \neq i_{2} \leq n} \{\hat{\IF}_{22, k, i_{1}, i_{2}} \hat{\IF}_{22, k, i_{1}, i_{2}} + \hat{\IF}_{22, k, i_{1}, i_{2}} \hat{\IF}_{22, k, i_{2}, i_{1}}\} \left( 1 - \frac{3}{n} + \frac{7}{n^{2}} - \frac{6}{n^{3}} - \frac{1}{n^{2}} \right) \\
& + \frac{2}{n^{2} (n - 1)^{2}} \sum_{1 \leq i_{1} \neq i_{2} \neq i_{3} \leq n} \left\{ \begin{array}{c}
\hat{\IF}_{22, k, i_{1}, i_{2}} \hat{\IF}_{22, k, i_{1}, i_{3}} + \hat{\IF}_{22, k, i_{1}, i_{2}} \hat{\IF}_{22, k, i_{3}, i_{1}} \\ 
+ \ \hat{\IF}_{22, k, i_{1}, i_{2}} \hat{\IF}_{22, k, i_{2}, i_{3}} + \hat{\IF}_{22, k, i_{1}, i_{2}} \hat{\IF}_{22, k, i_{3}, i_{2}}
\end{array} \right\} \left( - \ \frac{1}{n} + \frac{5}{n^{2}} - \frac{6}{n^{3}} - \frac{1}{n^{2}} \right) \\
& + \frac{2}{n^{2} (n - 1)^{2}} \sum_{1 \leq i_{1} \neq i_{2} \neq i_{3} \neq i_{4} \leq n} \hat{\IF}_{22, k, i_{1}, i_{2}} \hat{\IF}_{22, k, i_{3}, i_{4}} \left( \frac{3}{n^{2}} - \frac{6}{n^{3}} - \frac{1}{n^{2}} \right).
\end{align*}
Hence $T_{2}$ serves as a correction term of the inflated factor 3 appeared in equation \eqref{inflation}.

Combining the above computations, we have 
\begin{align*}
& \ \E_{\bm{W}} [\hat{\var} [\hat{\IIFF}_{22, k}] \vert \{O_{i}\}_{i = 1}^{N}] = \E_{\bm{W}} [T_{1} \vert \{O_{i}\}_{i = 1}^{N}] - \E_{\bm{W}} [T_{2} \vert \{O_{i}\}_{i = 1}^{N}] \\
= & \ \frac{1}{n^{2} (n - 1)^{2}} \sum_{1 \leq i_{1} \neq i_{2} \leq n} \{\hat{\IF}_{22, k, i_{1}, i_{2}} \hat{\IF}_{22, k, i_{1}, i_{2}} + \hat{\IF}_{22, k, i_{1}, i_{2}} \hat{\IF}_{22, k, i_{2}, i_{1}}\} \left( 1 - \frac{5}{n} + \frac{2}{n^{2}} + \frac{6}{n^{3}} \right) \\
& + \frac{1}{n^{2} (n - 1)^{2}} \sum_{1 \leq i_{1} \neq i_{2} \neq i_{3} \leq n} \left\{ \begin{array}{c}
\hat{\IF}_{22, k, i_{1}, i_{2}} \hat{\IF}_{22, k, i_{1}, i_{3}} + \hat{\IF}_{22, k, i_{1}, i_{2}} \hat{\IF}_{22, k, i_{3}, i_{1}} \\ 
+ \ \hat{\IF}_{22, k, i_{1}, i_{2}} \hat{\IF}_{22, k, i_{2}, i_{3}} + \hat{\IF}_{22, k, i_{1}, i_{2}} \hat{\IF}_{22, k, i_{3}, i_{2}}
\end{array} \right\} \left( 1 - \frac{5}{n} + \frac{6}{n^{3}} \right) \\
& - \frac{4}{n} \frac{1}{n^{2} (n - 1)^{2}} \sum_{1 \leq i_{1} \neq i_{2} \neq i_{3} \neq i_{4} \leq n} \hat{\IF}_{22, k, i_{1}, i_{2}} \hat{\IF}_{22, k, i_{3}, i_{4}} \left( 1 - \frac{1.5}{n} - \frac{1.5}{n^{2}} \right).
\end{align*}
Finally, it is not difficult to see that 
\begin{equation*}
\E_{\theta} [\E_{\bm{W}} [\hat{\var} [\hat{\IIFF}_{22, k}] \vert \{O_{i}\}_{i = 1}^{N}]] = \var_{\theta} [\hat{\IIFF}_{22, k}] \left( 1 + O (n^{-1}) \right).
\end{equation*}
\end{remark}

\begin{proposition}
\label{prop:boot_if22} 
Under Condition \ref{cond:sw}, as $n \rightarrow \infty$ and $M \rightarrow \infty$, 
\begin{equation*}
\frac{\tilde{\var} [\hat{\IIFF}_{22, k}]}{\var_{\theta} [\hat{\IIFF}_{22, k}]} = 1 + o_{\P_{\theta}} (1).
\end{equation*}
\end{proposition}

\begin{proof}
Let the $\tilde{\var} [\hat{\IIFF}_{22, k}]$ be the limit (in $\P_{\bm{W}}$-probability) of $\hat{\var} [\hat{\IIFF}_{22, k}]$ as $M \rightarrow \infty$ given all the observed data. Using the moments of multinomial distributions \citep{newcomer2008computation}, together with the explanation in Comment \ref{rem:boot_if22}, it is easy to show that $\E_{\theta} \left[ \frac{\tilde{\var} [\hat{\IIFF}_{22, k}]}{\var_{\theta} [\hat{\IIFF}_{22, k}]} \right] = 1 + o(1)$ and $\var_{\theta} \left[ \frac{\tilde{\var} [\hat{\IIFF}_{22, k}]}{\var_{\theta} [\hat{\IIFF}_{22, k}]} \right] = o(1)$. Combining the above two claims, we have $\frac{\tilde{\var} [\hat{\IIFF}_{22, k}]}{\var_{\theta} [\hat{\IIFF}_{22, k}]} = 1 + o_{\P_{\theta}} (1)$.
\end{proof}

Following computations similar in spirit to but more tedious than the computations above, we propose the following estimator of $\var_{\theta} [\hat{\IIFF}_{33, k}]$: 
\begin{equation}  \label{eq:if33_bvar}
\begin{split}
\hat{\var} [\hat{\IIFF}_{33, k}] \coloneqq & \ \underbrace{\frac{1}{M - 1} \sum_{m = 1}^{M} \left( \hat{\IIFF}_{33, k}^{(m)} - \frac{1}{M} \sum_{m = 1}^{M} \hat{\IIFF}_{33, k}^{(m)} \right)^{2}}_{S_{1}} \\
& - \underbrace{\frac{1}{M - 1} \sum_{m = 1}^{M} \left( \hat{\IIFF}_{33, k}^{(m), c(b, p)} - \frac{1}{M} \sum_{m = 1}^{M} \hat{\IIFF}_{33, k}^{(m), c(b, p)} \right)^{2}}_{S_{2}} \\
& - \underbrace{\frac{1}{M - 1} \sum_{m = 1}^{M} \left( \hat{\IIFF}_{33, k}^{(m), c(b, \Sigma)} - \frac{1}{M} \sum_{m = 1}^{M} \hat{\IIFF}_{33, k}^{(m), c(b, \Sigma)} \right)^{2}}_{S_{3}} \\
& - \underbrace{\frac{1}{M - 1} \sum_{m = 1}^{M} \left( \hat{\IIFF}_{33, k}^{(m), c(p, \Sigma)} - \frac{1}{M} \sum_{m = 1}^{M} \hat{\IIFF}_{33, k}^{(m), c(p, \Sigma)} \right)^{2}}_{S_{4}}
\end{split}
\end{equation}
where 
\begin{align*}
\hat{\IIFF}_{33, k}^{(m)} & = \frac{1}{n (n - 1) (n - 2)} \sum_{1 \leq i_{1} \neq i_{2} \neq i_{3} \leq n} W_{i_{1}}^{(m)} W_{i_{2}}^{(m)} W_{i_{3}}^{(m)} \hat{\IF}_{33, k, \bar{i}_{3}} \\
\hat{\IIFF}_{33, k}^{(m), c(b, p)} & = \frac{1}{n (n - 1) (n - 2)} \sum_{1 \leq i_{1} \neq i_{2} \neq i_{3} \leq n} (W_{i_{1}}^{(m)} - 1) W_{i_{2}}^{(m)} (W_{i_{3}}^{(m)} - 1) \hat{\IF}_{33, k, \bar{i}_{3}} \\
\hat{\IIFF}_{33, k}^{(m), c(b, \Sigma)} & = \frac{1}{n (n - 1) (n - 2)} \sum_{1 \leq i_{1} \neq i_{2} \neq i_{3} \leq n} (W_{i_{1}}^{(m)} - 1) (W_{i_{2}}^{(m)} - 1) W_{i_{3}}^{(m)} \hat{\IF}_{33, k, \bar{i}_{3}} \\
\hat{\IIFF}_{33, k}^{(m), c(p, \Sigma)} & = \frac{1}{n (n - 1) (n - 2)} \sum_{1 \leq i_{1} \neq i_{2} \neq i_{3} \leq n} W_{i_{1}}^{(m)} (W_{i_{2}}^{(m)} - 1) (W_{i_{3}}^{(m)} - 1) \hat{\IF}_{33, k, \bar{i}_{3}}.
\end{align*}

We then have:
\begin{proposition}
\label{prop:boot_if33} 
Under Condition \ref{cond:sw}, as $n \rightarrow \infty$ and $M \rightarrow \infty$, 
\begin{equation*}
\frac{\hat{\var} [\hat{\IIFF}_{33, k}]}{\var_{\theta} [\hat{\IIFF}_{33, k}]} = 1 + o_{\P_{\theta}}(1).
\end{equation*}
\end{proposition}

Finally, we consider estimating $\var_{\theta} [\hat{\IIFF}_{22 \rightarrow 33, k}]$ through nonparametric bootstrap. It is not difficult to see that the following estimator has the desired property $\frac{\hat{\var} [\hat{\IIFF}_{22 \rightarrow 33, k}]}{\var_{\theta} [\hat{\IIFF}_{22 \rightarrow 33, k}]} = 1 + o_{\P_{\theta}}(1)$ as $M \rightarrow \infty$: 
\begin{equation}
\label{bvar:if2233}
\hat{\var} [\hat{\IIFF}_{22 \rightarrow 33, k}] = \hat{\var} [\hat{\IIFF}_{22, k}] + \hat{\var} [\hat{\IIFF}_{33, k}] + 2 \hat{\cov} [\hat{\IIFF}_{22, k}, \hat{\IIFF}_{33, k}]
\end{equation}
where 
\begin{align*}
\hat{\cov} [\hat{\IIFF}_{22, k}, \hat{\IIFF}_{33, k}] = & \ \frac{1}{M - 1} \sum_{m = 1}^{M} \left( \hat{\IIFF}_{22, k}^{(m)} - \frac{1}{M} \sum_{m^{\prime} = 1}^{M} \hat{\IIFF}_{22, k}^{(m^{\prime})} \right) \left( \hat{\IIFF}_{33, k}^{(m)} - \frac{1}{M} \sum_{m^{\prime} = 1}^{M} \hat{\IIFF}_{33, k}^{(m^{\prime})} \right) \\
& - \frac{2}{M - 1} \sum_{m = 1}^{M - 1} \left( \hat{\IIFF}_{22, k}^{(m), c} - \frac{1}{M} \sum_{m^{\prime} = 1}^{M} \hat{\IIFF}_{22, k}^{(m^{\prime}), c} \right) \left( \hat{\IIFF}_{33, k}^{(m), c(b, p)} - \frac{1}{M} \sum_{m^{\prime} = 1}^{M} \hat{\IIFF}_{33, k}^{(m^{\prime}), c(b, p)} \right).
\end{align*}

It is straightforward but tedious to generalize the above arguments to construct bootstrapped estimators of $\var_{\theta} [\hat{\IIFF}_{mm, k}]$ for general $m \geq 2$. We plan to report the general construction in a separate paper. But when estimating $\var_{\theta} [\hat{\IIFF}_{22 \rightarrow mm, k}]$, it is sufficient to use $\hat{\var} [\hat{\IIFF}_{22 \rightarrow 33, k}]$ as it dominates the variance of higher-order terms.

\subsection{Proof of Theorem \ref{thm:cs} and Theorem \ref{prop:hoif_test}}
\label{app:sigma_test} 
As discussed in Sections \ref{sec:main} and \ref{sec:hierarchy}, since $\Sigma_{k}^{-1}$ is generally unknown, $\hat{\IIFF}_{22, k} (\Sigma_{k}^{-1})$ needs to be replaced by $\hat{\IIFF}_{22 \rightarrow mm, k} (\hat{\Sigma}_{k}^{-1})$. We consider the following statistic used in the test $\hat{\chi}_{m, k} (\hat{\Sigma}_{k}^{-1}; z_{\alpha^{\dag} / 2}, \delta)$ after standardization: 
\begin{equation} \label{eq:effect}
\begin{split}
& \ \frac{\hat{\se} [\hat{\psi}_{1}]}{\hat{\se} [\hat{\IIFF}_{22 \rightarrow mm, k} (\hat{\Sigma}_{k}^{-1})]} \left( \frac{\hat{\IIFF}_{22 \rightarrow mm, k} (\hat{\Sigma}_{k}^{-1})}{\hat{\se} [\hat{\psi}_{1}]} - \delta \right) \\
= & \ \left( \frac{\hat{\IIFF}_{22 \rightarrow mm, k} (\hat{\Sigma}_{k}^{-1}) - \Bias_{\theta, k} (\hat{\psi}_{1})}{\hat{\se} [\hat{\psi}_{1}]} \frac{\hat{\se} [\hat{\psi}_{1}]}{\hat{\se} [\hat{\IIFF}_{22 \rightarrow mm, k} (\hat{\Sigma}_{k}^{-1})]} \right) + \frac{\hat{\se} [\hat{\psi}_{1}]}{\hat{\se} [\hat{\IIFF}_{22 \rightarrow mm, k} (\hat{\Sigma}_{k}^{-1})]} \left( \frac{\Bias_{\theta, k} (\hat{\psi}_{1})}{\hat{\se} [\hat{\psi}_{1}]} - \delta \right) \\
= & \ \left( \frac{\hat{\IIFF}_{22 \rightarrow mm, k} (\hat{\Sigma}_{k}^{-1}) - \E_{\theta} [\hat{\IIFF}_{22 \rightarrow mm, k} (\hat{\Sigma}_{k}^{-1})]}{\hat{\se} [\hat{\psi}_{1}]} \frac{\hat{\se} [\hat{\psi}_{1}]}{\hat{\se} [\hat{\IIFF}_{22 \rightarrow mm, k} (\hat{\Sigma}_{k}^{-1})]} \right) + \frac{\EB_{\theta, m, k}}{\hat{\se} [\hat{\psi}_{1}]} \frac{\hat{\se} [\hat{\psi}_{1}]}{\hat{\se} [\hat{\IIFF}_{22 \rightarrow mm, k} (\hat{\Sigma}_{k}^{-1})]} \\
& + \frac{\hat{\se} [\hat{\psi}_{1}]}{\hat{\se} [\hat{\IIFF}_{22 \rightarrow mm, k} (\hat{\Sigma}_{k}^{-1})]} \left( \frac{\Bias_{\theta, k} (\hat{\psi}_{1})}{\hat{\se} [\hat{\psi}_{1}]} - \delta \right) \\
= & \; \left\{ \left( \frac{\hat{\IIFF}_{22 \rightarrow mm, k} (\hat{\Sigma}_{k}^{-1}) - \E_{\theta} [\hat{\IIFF}_{22 \rightarrow mm, k} (\hat{\Sigma}_{k}^{-1})]}{\se_{\theta} [\hat{\IIFF}_{22 \rightarrow mm, k} (\hat{\Sigma}_{k}^{-1})]} \right) + \left( \frac{\Bias_{\theta, k} (\hat{\psi}_{1}) + \EB_{\theta, m, k}}{\se_{\theta} [\hat{\psi}_{1}]} - \delta \right) \frac{\se_{\theta} [\hat{\psi}_{1}]}{\se_{\theta} [\hat{\IIFF}_{22 \rightarrow mm, k} (\hat{\Sigma}_{k}^{-1})]} \right\} (1 + o_{\P_{\theta}} (1)) \\
= & \ \left\{ \left( \frac{\hat{\IIFF}_{22 \rightarrow mm, k} (\hat{\Sigma}_{k}^{-1}) - \E_{\theta} [\hat{\IIFF}_{22 \rightarrow mm, k} (\hat{\Sigma}_{k}^{-1})]}{\se_{\theta} [\hat{\IIFF}_{22 \rightarrow mm, k} (\hat{\Sigma}_{k}^{-1})]} \right) + \left( \gamma + \frac{\EB_{\theta, m, k}}{\se_{\theta} [\hat{\psi}_{1}]} - \delta \right) \frac{\se_{\theta} [\hat{\psi}_{1}]}{\se_{\theta} [\hat{\IIFF}_{22 \rightarrow mm, k} (\hat{\Sigma}_{k}^{-1})]} \right\} (1 + o_{\P_{\theta}} (1)) \\
= & \ \left\{ \underbrace{\left( \frac{\hat{\IIFF}_{22 \rightarrow mm, k} (\hat{\Sigma}_{k}^{-1}) - \E_{\theta} [\hat{\IIFF}_{22 \rightarrow mm, k} (\hat{\Sigma}_{k}^{-1})]}{\se_{\theta} [\hat{\IIFF}_{22 \rightarrow mm, k} (\hat{\Sigma}_{k}^{-1})]} \right) - (\delta - \gamma) \frac{\se_{\theta} [\hat{\psi}_{1}]}{\se_{\theta} [\hat{\IIFF}_{22 \rightarrow mm, k} (\hat{\Sigma}_{k}^{-1})]}}_{A} + \underbrace{\frac{\EB_{\theta, m, k}}{\se_{\theta} [\hat{\IIFF}_{22 \rightarrow mm, k} (\hat{\Sigma}_{k}^{-1})]}}_{B} \right\} (1 + o_{\P_{\theta}} (1)).
\end{split}
\end{equation}

The effect of estimating $\Sigma_{k}^{-1}$ on the asymptotic validity of the test $\hat{\chi}_{m, k} (\hat{\Sigma}_{k}^{-1}; z_{\alpha^{\dag} / 2}, \delta)$ of $\H_{0, k} (\delta)$ thus depends on the orders of terms A and B. A has variance 1 and mean $- (\delta - \gamma) \frac{\se_{\theta} [\hat{\psi}_{1}]}{\se_{\theta} [\hat{\IIFF}_{22 \rightarrow mm, k} (\hat{\Sigma}_{k}^{-1})]}$. B depends on the estimation bias due to estimating $\Sigma_{k}^{-1}$ by $\hat{\Sigma}_{k}^{-1}$. Hence if we have:

\begin{enumerate}[label=(\arabic*)]
\item $\frac{\hat{\IIFF}_{22 \rightarrow mm, k} (\hat{\Sigma}_{k}^{-1}) - \E_{\theta} [\hat{\IIFF}_{22 \rightarrow mm, k} (\hat{\Sigma}_{k}^{-1})]}{\se_{\theta} [\hat{\IIFF}_{22 \rightarrow mm, k} (\hat{\Sigma}_{k}^{-1})]}$ is asymptotically $N(0, 1)$ conditional on the training sample;

\item $\frac{\se_{\theta} [\hat{\psi}_{1}]}{\se_{\theta} [\hat{\IIFF}_{22 \rightarrow mm, k} (\hat{\Sigma}_{k}^{-1})]} / \frac{\se_{\theta} [\hat{\psi}_{1}]}{\se_{\theta} [\hat{\IIFF}_{22 \rightarrow mm, k} (\hat{\Sigma}_{k}^{-1})]} \rightarrow 1$;

\item $B = o(1)$ under $\H_{0, k} (\delta)$ or fixed alternatives to $\H_{0, k} (\delta)$;
\end{enumerate}
\begin{enumerate}[label=(\arabic*')] 
\setcounter{enumi}{2}
\item $B \ll - (\delta - \gamma) \frac{\se_{\theta} [\hat{\psi}_{1}]}{\se_{\theta} [\hat{\IIFF}_{22 \rightarrow mm, k} (\hat{\Sigma}_{k}^{-1})]}$ under diverging alternatives to $\H_{0, k} (\delta)$ i.e. $\delta - \gamma = c$ for some $c \rightarrow \infty$ (at any rate).
\end{enumerate}

$\hat{\chi}_{m, k} (\hat{\Sigma}_{k}^{-1}; z_{\alpha^{\dag} / 2}, \delta)$ is an asymptotically level $\alpha^{\dag}$ two-sided test for $\H_{0, k} (\delta)$ and rejects the null with probability approaching 1 under diverging alternatives to $\H_{0, k} (\delta)$.

According to Theorem \ref{thm:hoif_stats}, both (1) and (2) hold for $\hat{\IIFF}_{22 \rightarrow mm, k} (\hat{\Sigma}_{k}^{-1})$. In terms of (3) and (3'), under the conditions of Theorem \ref{prop:hoif_test}:

\begin{itemize}
\item Under $\H_{0, k} (\delta)$ or fixed alternatives to $\H_{0, k} (\delta)$, $\EB_{\theta, m, k} (\hat{\Sigma}_{k}^{-1}) \lesssim \frac{(k \log k)^{(m - 1) / 2}}{n^{m / 2}} \ll \frac{\sqrt{k}}{n} + \frac{1}{\sqrt{n}}$ and hence $B = o(1)$. Thus (3) is satisfied.

\item Under diverging alternatives to $\H_{0, k} (\delta)$ i.e. $\gamma - \delta = c \rightarrow \infty$, $(\gamma - \delta) \se_{\theta} (\hat{\psi}_{1}) \asymp \BL_{\theta, 2, \hat{b}, k} \BL_{\theta, 2, \hat{p}, k}$, 
\begin{align*}
B & \lesssim \frac{\BL_{\theta, 2, \hat{b}, k} \BL_{\theta, 2, \hat{p}, k}}{\se_{\theta} [\hat{\IIFF}_{22 \rightarrow mm, k} (\hat{\Sigma}_{k}^{-1})]} \( \frac{k \log k}{n} \)^{(m - 1) / 2} \ll \frac{\BL_{\theta, 2, \hat{b}, k} \BL_{\theta, 2, \hat{p}, k}}{\se_{\theta} [\hat{\IIFF}_{22 \rightarrow mm, k} (\hat{\Sigma}_{k}^{-1})]} \\
& \asymp - \left( \delta - \gamma \right) \frac{\se_{\theta} [\hat{\psi}_{1}]}{\se_{\theta} [\hat{\IIFF}_{22 \rightarrow mm, k} (\hat{\Sigma}_{k}^{-1})]}.
\end{align*}
Thus (3') is satisfied.
\end{itemize}

In summary, $\hat{\chi}_{m, k} (\hat{\Sigma}_{k}^{-1}; z_{\alpha^{\dag} / 2}, \delta)$ is an asymptotically valid level $\alpha^{\dag}$ two-sided test for $\H_{0, k} (\delta)$ and rejects the null with probability approaching 1 under diverging alternatives to $\H_{0, k} (\delta)$.


\section{Supplementary information for simulation studies}
\label{app:simulations} 
The functions $h_{f}$, $h_{b}$ and $h_{p}$ appeared in Section \ref{sec:simulation} are of the following forms: 
\begin{align}
h_{f} (x; s_{f}) & \propto 1 + \text{exp} \left\{ \frac{1}{2} \sum_{j \in \mathcal{J}, \ell \in \mathbb{Z}} 2^{- j (s_{f} + 0.5)} \omega_{j, \ell} (x) \right\},  \label{f} \\
h_{b} (x; s_{b}) & = \sum_{j \in \mathcal{J}, \ell \in \mathbb{Z}} 2^{- j (s_{b} + 0.5)} \omega_{j, \ell} (x),  \label{b} \\
h_{p} (x; s_{p}) & = - 2 \sum_{j \in \mathcal{J}, \ell \in \mathbb{Z}} 2^{- j (s_{p} + 0.5)} \omega_{j, \ell} (x)
\label{p}
\end{align}
where $\mathcal{J} = \{ 0, 3, 6, 9, 10, 16 \}$ and $\omega_{j, \ell} (\cdot)$ is the D12 (or equivalently db6) father wavelets function dilated at resolution $j$, shifted by $\ell$. By the equivalent characterization of Besov-Triebel spaces by the corresponding wavelet coefficients in the frequency domain \citep[page 331]{gine2016mathematical}, $h_{f} (\cdot; s_{f}) \in \Holder (s_{f})$, $h_{b} (\cdot; s_{b}) \in \Holder (s_{b})$ and $h_{p} (\cdot; s_{p}) \in \Holder (s_{p})$. We fix $s_{f} = 0.1$. In simulation setup I we choose $s_{b} = s_{p} = 0.25$ whereas in simulation setup II we choose $s_{b} = s_{p} = 0.6$.

\begin{table}[!htb]
\begin{tabular}{c|c|c}
\hline
$j$ & $\tau_{b, j}$ & $\tau_{p, j}$ \\ 
\hline
1 & -0.2819 & 0.09789 \\ 
2 & 0.4876 & 0.08800 \\ 
3 & -0.1515 & -0.4823 \\ 
4 & -0.1190 & 0.4588 \\ 
\hline
\end{tabular}
\caption{Coefficients used in constructing $b$ and $\pi$ in Section \ref{sec:simulation}.}
\label{tab:s1}
\end{table}
In Table \ref{tab:s1}, we provide the numerical values for $\left( \tau_{b, j}, \tau_{p, j} \right)_{j = 1}^{8}$ used in generating the simulation experiments in Section \ref{sec:simulation}.

\subsection{Generating correlated multidimensional covariates $X$ with fixed non-smooth marginal densities}
\label{app:multiX} 
In the simulation study conducted in Section \ref{sec:simulation}, one key step of generating the simulated datasets is to draw correlated multidimensional covariates $X \in [0, 1]^{d}$ with fixed non-smooth marginal densities. First, we fix the marginal densities of $X$ in each dimension proportional to $h_{f} (\cdot)$ (equation \eqref{f}). Then we make $2K$ independent draws of $\tilde{X}_{i, j}$, $i = 1, \ldots, 2K$, from $h_{f}$ for every $j = 1, \ldots, d$ so $\tilde{X} = (\tilde{X}_{1, \cdot}, \ldots, \tilde{X}_{2K, \cdot})^{\top} \in [0, 1]^{2K \times d}$. Next, to create correlations between different dimensions, we follow the strategy proposed in \citet{baker2008order}. First we group every two consecutive draws: 
\begin{equation*}
(\tilde{X}_{1, \cdot}, \tilde{X}_{2, \cdot})^{\top}, (\tilde{X}_{3, \cdot}, \tilde{X}_{4, \cdot})^{\top}, \ldots, (\tilde{X}_{2K - 1, \cdot}, \tilde{X}_{2K, \cdot})^{\top}.
\end{equation*}
Then for each pair $(\tilde{X}_{2 i - 1, \cdot}, \tilde{X}_{2 i, \cdot})^{\top}$ for $i = 1, \ldots, K$, we form the following $d$-dimensional random vectors 
\begin{align*}
U_{i} \coloneqq (\max (\tilde{X}_{2 i - 1, 1}, \tilde{X}_{2 i, 1}), \ldots, \max (\tilde{X}_{2 i - 1, d}, \tilde{X}_{2 i, d}))^{\top}, \\
V_{i} \coloneqq (\min (\tilde{X}_{2 i - 1, 1}, \tilde{X}_{2 i, 1}), \ldots, \min (\tilde{X}_{2 i - 1, d}, \tilde{X}_{2 i, d}))^{\top}.
\end{align*}
Lastly, we construct $K$ independent $d$-dimensional vectors $X$ by the following rule: for each $i = 1, \ldots, K$, we draw a Bernoulli random variable $B_{i}$ with probability 1 / 2, and if $B_{i} = 0$, $X_{i, \cdot} = U_{i}$, otherwise $X_{i, \cdot} = V_{i}$. Following the above strategy, we preserve the marginal density of $X_{\cdot, j}$ as that of $\tilde{X}_{\cdot, j}$ but create dependence between different dimensions.

\section{Proximal higher-order influence functions}
\label{app:prox_hoif}
In this section, we further generalize the derivations in Appendix \ref{app:hoif} for $\psi (\theta) = - \E_{\theta} [Y (a)]$ to the case under the proximal causal inference framework considered in \citet{miao2018confounding, tchetgen2020introduction, cui2023semiparametric}. The purpose of this section is to show that our framework can be extended to the endogeneity settings which economists are extremely interested in; e.g. \citet{newey1991uniform, ai2003efficient, ai2007estimation, newey2003instrumental}. We only consider $\psi (\theta) = - \E_{\theta} [Y (a)]$, since it is yet unclear what is the ``right'' generalization under endogeneity for the general DR functionals considered in \citet{rotnitzky2021characterization}. We leave such generalization to future work.

We immediately begin with the formal setup, followed with a theorem on the forms of HOIFs of $\psi (\theta)$. Instead of observing i.i.d. $O_{i} = (X_{i}, A_{i}, Y_{i})_{i = 1}^{n}$ under strong ignorability, there might exist latent confounding variable $U$, so $\psi (\theta)$ is not point identified in general. But suppose we observe $O_{i} = (X_{i}, Z_{i}, A_{i}, W_{i}, Y_{i})_{i = 1}^{n}$, where $W$ and $Z$ are ``proxy variables'' of $Y$ and $A$ respectively, satisfying the following proximal causal identification conditions given in \citet{tchetgen2020introduction}:
\begin{condition}\leavevmode
\label{cond:prox}
\begin{enumerate}
\item Consistency: $Y = Y_{A}$ almost surely;
\item No direct effect from the treatment $A$ and treatment proxy $Z$ to the outcome proxy $W$: $W_{a, z} = W$ for all $a$ and $z$ almost surely;
\item No direct effect from the treatment proxy $Z$ to the outcome $Y$: $Y_{a, z} = Y_{a}$ for all $a$ and $z$ almost surely;
\item $(Z, A) \indep (Y_{a}, W) | (U, X)$ for $a = 0, 1$;
\item $0 < \Pr (A = a | U, X) < 1$ almost surely for $a = 0, 1$;
\item For any $a, x$, if $\E_{\theta} [g (U) | W, A = a, X = x] = 0$ almost surely, then $g (U) = 0$ almost surely;
\item For any $a, x$, if $\E_{\theta} [g (W) | Z, A = a, X = x] = 0$ almost surely, then $g (W) = 0$ almost surely;
\end{enumerate}
\end{condition}
For more details of proximal causal inference and the intuition behind the above regularity conditions, we refer interested readers to many recent papers on this topic \citep{miao2018confounding, tchetgen2020introduction, cui2023semiparametric, kallus2021causal, shpitser2023proximal}. 

Then we have
\begin{theorem}
\label{thm:prox_hoif} 
Under the above Condition \ref{cond:prox} (Assumptions 1, and 4-9 of \citet{cui2023semiparametric}), together with the following conditions on the so-called confounding bridge functions $r$ and $q$ (Assumption 10 of \citet{cui2023semiparametric}): Denote $\Psi (z, a, x) \equiv \E_{\theta} [(Y - r (W, A = a, X))^{2} | Z = z, A = a, X = x]$ and $\Gamma (w, a, x) \equiv \E_{\theta} \[ \( q (Z, A, X) - \frac{1}{f (A | W, X)} \)^{2} | W = w, A = a, X = x \]$:
\begin{align*}
(1) & \ 0 < \inf_{z, a, x} \Psi (z, a, x) \leq \sup_{z, a, x} \Psi (z, a, x) < \infty; \\
(2) & \ 0 < \inf_{w, a, x} \Gamma (w, a, x) \leq \sup_{w, a, x} \Gamma (w, a, x) < \infty; \\
(3) & \ \text{Let $T: L_{2} (W, A, X) \rightarrow L_{2} (Z, A, X)$ be the conditional expectation operator given by $T (g) \equiv \E_{\theta} [g (W, A, X) | Z, A, X]$} \\
& \ \text{and the adjoint $T': L_{2} (Z, A, X) \rightarrow L_{2} (W, A, X)$ be $T' (g) \equiv \E_{\theta} [g (Z, A, X) | W, A, X]$.} \\
& \ \text{Suppose that the inverse mapping $\{T' \Psi^{-1} T\}^{-1}$ exists and is uniquely defined and that} \\
& \ T' \Psi^{-1} \E_{\theta} [(Y - r (W, A, X)) (r (W, A = a, X) - \psi (\theta)) | Z, A, X] \in \text{Domain} (\{T' \Psi^{-1} T\}^{-1}) \\
& \ \mathbbm{1} \{A = a\} / f (A | W, X) \in \text{Domain} (\{T' \Psi^{-1} T\}^{-1}).
\end{align*}

Then the confounding bridge functions $r$ and $q$ that solve integral equations
\begin{equation}
\begin{split}
\E_{\theta} [Y | X, A = a, Z] = \int r (w, A = a, X) \mathrm{d} F_{\theta} (w | X, Z), \\
\frac{1}{\Pr_{\theta} (A = a | X, W)} = \int q (z, A = a, X) \mathrm{d} F_{\theta} (z | X, W, A = a).
\end{split}
\end{equation}
are uniquely identified. The nonparametric first order influence function of $\psi (\theta) = - \E_{\theta} [Y (a)]$ is given by
\begin{equation}
\IF_{1, \psi} (\theta) = \mathbbm{1} \{A = a\} q (Z, A = a, X) r (W, A = a, X) - \mathbbm{1} \{A = a\} Y q (Z, A = a, X) - r (W, A = a, X) - \psi (\theta).
\end{equation}
In the main text, all the notations related to $b$ and $p$ are extended to $h$ and $q$. For the truncated nuisance functions $\tilde{r}_{k, \theta}$ and $\tilde{q}_{k, \theta}$, we use $k$-dimensional basis $\bar{\mathfrak{z}}_{k} (W, A, X)$ and $\zbar_{k} (Z, A, X)$, respectively.

Furthermore, the HOIFs of order $m$ of $\zkpsi = \E_{\theta} \[ \calH (\tilde{r}_{k, \theta}, \tilde{q}_{k, \theta}) \]$ are $\IIFF_{m, k} (\Sigma_{k}^{-1}) = \IIFF_{1} - \IIFF_{22 \rightarrow mm, k} (\Sigma_{k}^{-1})$, where $\IIFF_{\ell \ell \rightarrow mm, k} (\Sigma_{k}^{-1}) \coloneqq \sum_{j = \ell}^{m} \IIFF_{jj, k} (\Sigma_{k}^{-1})$, $\Sigma_{k} \coloneqq \E_{\theta} [\mathbbm{1} \{A = a\} \bar{\mathfrak{z}}_{k} (W, A = a, X) \zbar_{k} (Z, A = a, X)^{\top}]$ which is assumed to be invertible,
\begin{align}
\IIFF_{22, k} (\Sigma_{k}^{-1}) \coloneqq \frac{1}{n (n - 1)} \sum_{1 \leq i_{1} \neq i_{2} \leq n} \IF_{22, k, \bar{i}_{2}} (\Sigma_{k}^{-1})
\end{align}
and for $j > 2$,
\begin{align}
\IIFF_{jj, k} (\Sigma_{k}^{-1}) \coloneqq \frac{(n - j)!}{n!} \sum_{1 \leq i_{1} \neq \ldots \neq i_{j} \leq n} \IF_{jj, k, \bar{i}_{j}} (\Sigma_{k}^{-1}).
\end{align}
Here
\begin{align}
\IF_{22, k, \bar{i}_{2}} (\Sigma_{k}^{-1}) \equiv \IF_{22, \tilde{\psi}_{k}, \bar{i}_{2}} (\theta) = & \ \left[ \mathbbm{1} \{A = a\} (Y - \tilde{r}_{k, \theta} (W, A = a, X)) \zbar_{k} (Z, A = a, X) \right]^{\top}_{i_{1}} \Sigma_{k}^{-1} \left[ \bar{\mathfrak{z}}_{k} (W, A = a, X) (\mathbbm{1} \{A = a\} \tilde{q}_{k, \theta} (Z, A = a, X) - 1) \right]_{i_{2}},
\end{align}
and
\begin{align*}
& \; \IF_{jj, k, \bar{i}_{j}} (\Sigma_{k}^{-1}) \equiv \IF_{jj, \tilde{\psi}_{k}, \bar{i}_{j}} (\theta)  \nonumber \\
= & \; (-1)^{j}  \left[ \mathbbm{1} \{A = a\} (Y - \tilde{r}_{k, \theta} (W, A = a, X)) \zbar_{k} (Z, A = a, X) \right]^{\top}_{i_{1}} \left\{ \prod\limits_{s = 3}^{j} \Sigma_{k}^{-1} \left( \left[ \mathbbm{1} \{A = a\} \bar{\mathfrak{z}}_{k} (W, A = a, X) \zbar_{k} (Z, A = a, X)^{\top} \right]_{i_{s}} - \Sigma_{k} \right) \right\} \\
& \times \Sigma_{k}^{-1} \left[ \bar{\mathfrak{z}}_{k} (W, A = a, X) (\mathbbm{1} \{A = a\} \tilde{q}_{k, \theta} (Z, A = a, X) - 1) \right]_{i_{2}}. \label{eq:ifjj}
\end{align*}
are the kernels of second order influence function $\IIFF_{22, k} (\Sigma_{k}^{-1})$ and $j$-th order influence function $\IIFF_{jj, k} (\Sigma_{k}^{-1})$ respectively, where $\bar{i}_{j} \coloneqq \{ i_{1}, \ldots, i_{j} \}$.
\end{theorem}

Note that the first order influence function in the above theorem has appeared in \citet{cui2023semiparametric}. We state it in the above theorem for the sake of completeness. The forms of HOIFs are new, but are straightforward applications of the derivation given in Appendix \ref{app:hoif} for DR functionals.

\newpage
\section{Supplementary figures}
The figures mentioned in Section \ref{sec:simulation} are collected here.

\printbibliography[title = {Appendix References}]
\end{document}